\documentclass[iop]{emulateapj_pdftex}
\usepackage{amsmath,dsnr,epsfig,ifthen,natbib}
\newcommand{\hst}{{\it HST}}
\newcommand{\vfifty}{\ifmmode v_{50\%}\else $v_{50\%}$\fi}
\newcommand{\vfiftyave}{\ifmmode \langle v_{50\%}\rangle \else $\langle
  v_{50\%}\rangle$\fi}
\newcommand{\vtsig}{\ifmmode v_{98\%}\else $v_{98\%}$\fi}
\newcommand{\vtsigave}{\ifmmode \langle v_{98\%}\rangle \else $\langle
  v_{98\%}\rangle$\fi}

\shorttitle{Galactic Winds in Mergers}
\shortauthors{Rupke \& Veilleux}

\begin{document}

\slugcomment{Submitted to ApJ 31 Aug 2012; Accepted 22 Mar 2013}

\title{The Multiphase Structure and Power Sources of Galactic Winds in
  Major Mergers}

\author{David S. N. Rupke$^1$} \affil{Department of Physics, Rhodes
  College, Memphis, TN 38112; \\ Institute for Astronomy, University of
  Hawaii, Honolulu, HI 96822} \email{$^1$drupke@gmail.com}

\author{and Sylvain Veilleux} \affil{Department of Astronomy,
  University of Maryland, College Park, MD 20742}

\begin{abstract}
  Massive, galaxy-scale outflows are known to be ubiquitous in major
  mergers of disk galaxies in the local universe. In this paper, we
  explore the multiphase structure and power sources of galactic winds
  in six ultraluminous infrared galaxies (ULIRGs) at $z < 0.06$ using
  deep integral field spectroscopy with the Gemini Multi-Object
  Spectrograph (GMOS) on Gemini North. We probe the neutral, ionized,
  and dusty gas phases using \nad, strong emission lines (\oo, \ha,
  and \nt), and continuum colors, respectively. We separate outflow
  motions from those due to rotation and tidal perturbations, and find
  that all of the galaxies in our sample host high-velocity flows on
  kiloparsec scales. The properties of these outflows are consistent
  with multiphase (ionized, neutral, and dusty) collimated bipolar
  winds emerging along the minor axis of the nuclear disk to scales of
  $1-2$~kpc. In two cases, these collimated winds take the form of
  bipolar superbubbles, identified by clear kinematic signatures. Less
  collimated (but still high-velocity) flows are also present on
  scales up to 5~kpc in most systems. The three galaxies in our sample
  with obscured QSOs host higher velocity outflows than those in the
  three galaxies with no evidence for an AGN. The peak outflow
  velocity in each of the QSOs is in the range $1450-3350$~\kms, and
  the highest velocities ($2000-3000$~\kms) are seen only in ionized
  gas. The outflow energy and momentum in the QSOs are difficult to
  produce from a starburst alone, but are consistent with the QSO
  contributing significantly to the driving of the flow. Finally, when
  all gas phases are accounted for, the outflows are massive enough to
  provide negative feedback to star formation.
\end{abstract}

\keywords{galaxies: evolution --- galaxies: interactions --- galaxies:
  ISM --- galaxies: kinematics and dynamics --- ISM: jets and outflows
  --- galaxies: quasars}

%%%%%%%%%%%%%%%%%%%%%%%%

\section{INTRODUCTION} \label{sec:introduction}

Massive, high-velocity outflows are present in almost all major galaxy
mergers in the local universe
\citep{heckman90a,heckman00a,rupke02a,rupke05c,rupke05a,rupke05b,martin05a,martin06a,westmoquette12a,soto12a,soto12b}. These
kpc-scale outflows exist in both the neutral and ionized gas
phases. They have high velocities (typically a few hundred \kms,
though reaching $>$1000~\kms\ in some cases); large mass outflow rates
(estimates suggest a few tens of percent of the star formation rate);
and large extents ($\ga$5~kpc). They have been thought to be powered
primarily by star formation \citep{rupke05c,krug10a}. Their properties
are consistent with requirements for negative feedback on star
formation and black hole accretion in mergers, as suggested by
numerical simulations \citep[e.g.,][]{springel05a,hopkins05a}.

However, these conclusions rest on studies using long-slit optical
spectroscopy of a single gas phase. Such studies do not probe the
detailed structures of winds, and have not compared in detail the
properties of different gas phases. In particular, disentangling
motions in the ionized gas (as probed by strong emission lines) due to
rotation, tidal motions, and outflows is difficult without
spatially-resolved data. The present ubiquity of sensitive optical
integral field spectrographs on large telescopes has made it possible
to study the winds in these systems in much greater detail.

Late-stage, major mergers of massive disk galaxies in the local
universe are synonymous with ultraluminous infrared galaxies (ULIRGs;
e.g., \citealt{veilleux02a}). ULIRGs have been studied for many years
using integral field spectroscopy (IFS; e.g., \citealt{colina05a}, and
references therein). However, these early studies used 4m class
telescopes at relatively low spectral resolution, and did not
typically address the outflows in these systems (though see
\citealt{colina99a}).

Recent deep observations on $8-10$m class telescopes at high spatial
resolution have made possible the detailed study of the multiphase
wind structure and power sources in ULIRGs
\citep{arribas08a,shih10a,rupke11a,westmoquette12a}. \citet{shih10a}
confirmed the outflow velocities, mass, and radial extent measured
previously in F10565$+$2448 \citep{rupke05b,martin06a}, but also
showed that the outflowing gas has a complicated structure that is
inconsistent with lifting of rotating gas from the disk
\citep{martin06a}. \citet{rupke11a} confirmed the detection of an
extended outflow in Mrk~231 \citep{rupke05c} and showed that this
outflow emerges in all projected directions and is likely powered by
the central QSO. This galaxy may host the first known example of
QSO-mode feedback in a major merger. (See also
\citealt{fischer10a,feruglio10a} for molecular gas detections of the
outflow.) \citet{westmoquette12a} presented the first large-scale IFS
survey of outflows in ULIRGs. These authors focused on the ionized gas
phase, and detected ubiquitous kiloparsec-scale outflows through the
presence of broad, blueshifted velocity components. The high
velocities they detected in some systems (reaching several thousand
\kms) implicate an AGN in powering the outflows.

In this paper, we present an IFS study of six ULIRGs. This study
differs from early ULIRG studies in its high depth, spatial
resolution, and spectral resolution. It also differs from the recent
study of \citet{westmoquette12a} by our ability to probe multiple gas
phases (neutral and ionized) at higher spatial resolution (enabled by
0\farcs2 spatial sampling) and by our modeling of the resulting wind
structure. The data on two of these systems were first presented in
\citet{shih10a} and \citet{rupke11a}, and we re-reduce and re-analyze
these data along with four new systems. Five of these systems were
first surveyed for outflows using long-slit spectroscopy
\citep{heckman00a,rupke02a,rupke05c,rupke05b,martin05a}.

In Section \ref{sec:obs}, we discuss our observations and data
analysis.  We discuss our analysis procedures and wind modeling in
Section\,\ref{sec:analysis-modeling}. We then detail the properties of
individual objects in Section\,\ref{sec:systems}. The results on
individual systems are synthesized to form a sample-wide picture in
Section\,\ref{sec:sample}, with the crucial results found in
Figures~\ref{fig:vel_v_fwhm}$-$~\ref{fig:vel_v_lir}. We summarize
in Section\,\ref{sec:summary}.

\section{SAMPLE, OBSERVATIONS, AND REDUCTION} \label{sec:obs}

The six systems in our sample (Table \ref{tab:sample}) were selected
for their proximity and high infrared luminosity: $z < 0.06$ and
log(\lir/\lsun) $\ga12$. All six are members of the Revised Bright
Galaxy Sample (RBGS; \citealt{sanders03a}), which was assembled based
on {\it Infrared Astronomical Satellite} data. They span a range of
nuclear properties and merger classes, though all are thought to be in
the late stages of a major merger (mass ratio $\la 3$;
\citealt{veilleux02a}).

The galaxies in our study were observed with the Integral Field Unit
in the Gemini Multi-Object Spectrograph (GMOS;
\citealt{allington-smith02a,hook04a}) on the Gemini North
telescope. In each case we used the integral field unit in 2-slit
mode, with the B600 grating and $r^\prime$ filter, yielding wavelength
coverage from 5600 to 7000~\AA\ and a spectral resolution averaging
1.6~\AA\ at the central wavelength of 6300~\AA. (This spectral
resolution is measured directly from night sky emission lines in each
spaxel.)

To assess image quality and relative photometry in galaxies with an
identifiable nucleus, we fit a broad Gaussian to the nucleus in each
exposure. Variations in image quality and cloud cover between
exposures led to at most a 30\%\ difference in this peak flux between
the highest and lowest fluxes. A small number of exposures with larger
variations were discarded. Delivered image quality through the IFS was
difficult to assess for most sources, since there are no point sources
in the field of view of the integral field unit (except for the QSO in
Mrk~231, which has a full width at half maximum, or FWHM, of
$0\farcs5-0\farcs6$). However, fits to multiple field stars in 12 of
the 13 acquisition images in our dataset yield a mean FWHM of
0\farcs59, with a standard deviation of 0\farcs16 and a range of
$0\farcs4-0\farcs9$.

The native FOV of GMOS in 2-slit mode is $5\arcsec\times7\arcsec$. All
sources were observed in two pointings. In four systems, GMOS was
dithered along the short axis, with some overlap between the two
pointings. In the other two (F08572$+$3915 and VV~705), it was
dithered between two nuclei.

Table \ref{tab:obs} lists further details of the observations of each
source.

We reduced the data using IRAF v1.14 and v1.10 of the data reduction
package provided by the Gemini Observatory. We modified the Gemini
software slightly to accommodate error propagation, and supplemented it
with custom IDL routines. These IDL routines perform the following
steps: (1) correcting single-wavelength images to a common center (due
to differential atmospheric refraction); (2) applying a small shift to
the wavelength calibration using the line centers of atmospheric
emission lines (\ooll\ or \nadl); (3) telluric correction (using a
scaled version of the high SNR Mrk~231 nuclear spectrum as a
template); (4) spatial rebinning of spaxels; and (5) mosaicing of
dithered positions.

The native 0\farcs2-diameter hexagonal spaxels of GMOS are resampled
to 0\farcs1 square spaxels by the GMOS IRAF pipeline. To Nyquist
sample the seeing disk while maximizing sensitivity, we binned the
0\farcs1 square pixels into 0\farcs3 square pixels.

The relative flux calibration of our data is robust. However, flux
calibration uncertainties at the red filter transmission edge may
affect the \nt/\ha\ ratio at the $\la$50\%\ level in F08572$+$3915,
which has the highest redshift in our sample and thus an \nt\ line
near the filter edge.

The final result is eight data cubes (one per system dithered along
the GMOS short axis, and two per system dithered between nuclei).  The
resulting cubes contain hundreds of spectra.

\section{ANALYSIS AND MODELING} \label{sec:analysis-modeling}

\subsection{Identification of Nuclei}

The nuclei in these systems are heavily obscured, and the optical
continuum peak is often not coincident with the underlying stellar
mass peak. The stellar mass peak can be approximated by the location
of the near-infrared (NIR) luminosity peak
\citep[e.g.,][]{scoville00a}.

Archival \hst\ F814W and F160W images exist for each galaxy in our
sample. To identify the nuclei within the GMOS data cubes, we located
the NIR peak or peaks in each \hst\ F160W image. We registered the NIR
image with the F814W image by matching the positions of $1-3$
relatively unobscured nuclear star clusters, as in
\citet{iwasawa11a}. We then convolved each F814W image with a 5-pixel
Gaussian to match the average seeing in our data (0\farcs6). Finally,
we matched the convolved F814W and GMOS continuum peaks. The GMOS
continuum is slightly bluer than the F814W continuum ($R$ vs. $I$
band), but we do not expect the continuum peak to shift significantly
over this wavelength range. The resulting nuclear locations in the
GMOS data are accurate to $\la$0\farcs1, or a third of a spaxel.

The exception to this procedure is Mrk~231; the nucleus for this
galaxy is easily identifiable as the bright QSO.

\subsection{Continuum, Emission Line, and Absorption Line Fitting}

We modeled the continuum and emission lines in each spectrum using
UHSPECFIT, a suite of IDL software which optimizes continuum and
emission line fits to spectra. UHSPECFIT is described elsewhere
\citep{rupke10b}. In brief, it masks emission line regions, fits the
continuum, and then simultaneously fits all emission lines in the
continuum-subtracted spectrum. The fit is improved by repeating this
process based on the parameters derived from the first fit.

We fit simple polynomial continua (of order $3-4$) to each
spectrum. For Mrk~231, we also fit a scaled version of the extremely
bright nuclear QSO spectrum \citep{rupke11a}. The QSO point spread
function (PSF) bleeds into many spaxels and must be removed.

The continua in our galaxies are dominated by young stars, which have
weak absorption lines in the wavelength range of our data
($>$5200~\AA\ rest frame). The stellar absorption line that would most
affect our emission line fits is \ha; because our spectra do not cover
other Balmer lines, we cannot constrain the strength of \ha\
absorption using other lines. However, observations of ULIRGs show
that the equivalent width of \hb\ absorption is typically in the range
$3-7$~\AA\ within 2~kpc of the nuclei of ULIRGs, and only 1 or 2~\AA\
higher at larger radii \citep{soto10a}.

The \ha\ emission line equivalent widths in our data cover a broad
range (from $-$1 to $-$300 \AA), but typically dip below $-$10~\AA\
only at radii $\ga$2~kpc, and sometimes not at all. Some systems
(Mrk~273 and VV~705:NW) have \ha\ emission line equivalent widths of
$\la-20$~\AA\ in virtually every spaxel. The equivalent widths vary
spatially in a complex way, but generally increase with increasing
radius. Furthermore, the Balmer absorption lines in ULIRGs are
typically much broader than the emission lines \citep{soto10a},
minimizing their impact on the emission line fitting. The most likely
effect of the lack of Balmer absorption correction in our data would
be an overestimate of the \ntl/\ha\ and \ool/\ha\ emission line ratios
$2-3$~kpc from the nucleus in a few systems. Conservatively, this
overestimate is no worse than $50\%$ in those spaxels with emission
line equivalent widths near $-$10~\AA. Given the relative widths of
the absorption and emission lines, the likely effect is smaller than
this. This error does not affect the conclusions of
\S\,\ref{sec:gas-excitation}.

The emission lines visible in the wavelength range of our data are
\ha, \ntll, and \ooll. We fit $1-2$ Gaussian velocity components to
these emission lines in each spectrum in all cases but Mrk~273, for
which we allowed up to 3 components. We fitted a common velocity and
FWHM to all emission lines in a given velocity component. The broad
emission line component in F08572$+$3915:NW required a more
constrained fit, which is described in \S\,\ref{sec:f08572_kin}. The
required number of Gaussian components in each spaxel was determined
by a combination of software automation and manual
inspection. UHSPECFIT discarded all components that were narrower than
the spectral resolution, too broad to be robustly distinguished from
the continuum, or fell below SNR $=3$ in the \ha\ and \ntl\ lines. The
exception is Mrk~231, for which we required SNR $>7$ in \ha\ or \ntl.

For spaxels with multiple velocity components, we organize them
principally by FWHM into a (1) narrow and (2) broad component (except
for Mrk~273; see \S\,\ref{sec:mrk273}). However, we adjust the
classification of some spaxels by hand to achieve continuity in the
velocity maps, particularly in cases where the FWHM of the two
components are similar.

For five of the six galaxies in our sample, we found a $>$3$\sigma$
component in \ha\ or \ntl\ in $>$90\%\ of spaxels. In the sixth
galaxy, F08572$+$3915, we detected emission in $>$60\%\ of spaxels.

We fit the \nad\ absorption lines using the method of
\citet{rupke05a}. This method robustly separates optical depth and
covering factor, even for a blended doublet and multiple velocity
components, by using a Gaussian in optical depth. The resulting values
for optical depth vary from 0.1 to 5 (the latter is the upper limit of
values to which we are sensitive), but are typically of order
unity. The covering factor can vary between 0 and 1, but is typically
0.5. These results are consistent with previous studies of nuclear
spectra of ULIRGs \citep{heckman00a,rupke02a,rupke05a}.

One velocity component in \nad\ provided good fits to spectra
throughout the sample. For most systems, we ignored the nearby
\ion{He}{1}~5876~\AA\ emission line because of its weakness. The
exceptions are VV~705 and Mrk~231; in these cases we fixed the line
width and centroid of \ion{He}{1} in each spaxel to the values
determined by fits to other emission lines in that spaxel.

The stellar contribution to the \nad\ absorption line in ULIRGs is
generally quite small \citep{rupke05a}. In previous studies, the
stellar contribution to \nad\ has been estimated by scaling the
strength of \ion{Mg}{1}~b \citep{heckman00a,rupke02a,rupke05a}. The
current data do not probe \ion{Mg}{1}~b, or indeed other strong
stellar features, so we cannot estimate this contribution on a
spaxel-by-spaxel basis. For four of the six galaxies in our sample,
measurements of nuclear spectra from previous work
\citep{kim95a,veilleux99a} or from the SDSS allow us to estimate this
contribution for at least one nucleus (F08572$+$3915:NW,
F10565$+$2448, Mrk~273, and VV~705:NW). We find that the equivalent
width of \ion{Mg}{1}~b is 1.0~\AA\ or less, yielding a stellar \nad\
equivalent width of 0.5~\AA\ or less. Compared to the measured total
\nad\ equivalent widths in these sources, then, the stellar \nad\
represents only $5-10$~\%\ of the line.

\subsection{Maps and Velocity Distributions} \label{sec:maps}

\begin{figure}
%  \plotone{f01.eps}
  \centering \includegraphics[width=6.5in]{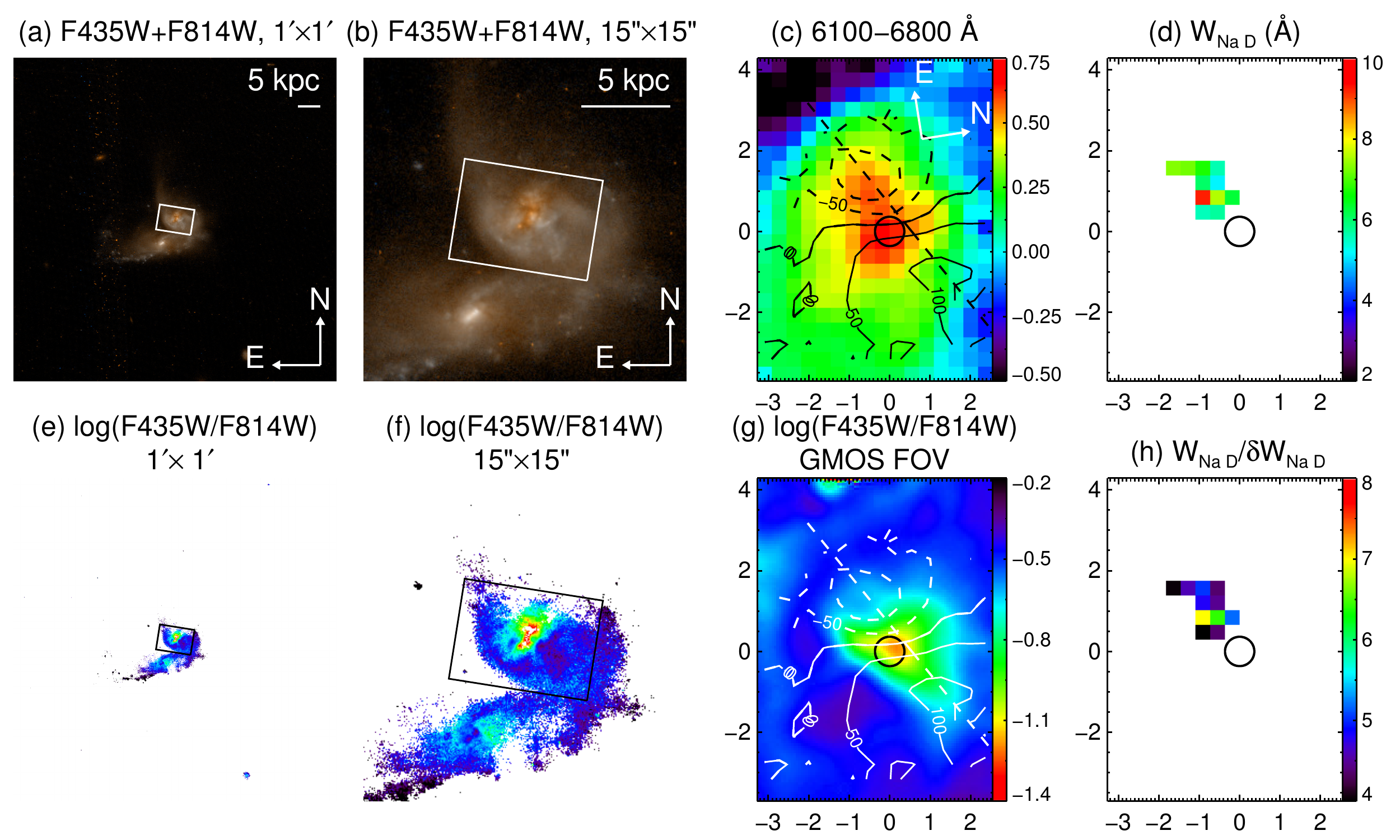}
  \caption{$(a)-(b)$: Two-color \hst\ images of F08572$+$3915:NW at
    sizes $1\arcmin\times1\arcmin$ and $15\arcsec\times15\arcsec$. Red
    is $F814W$, blue is $F435W$, and green is their average. The image
    orientation is indicated by the compass rose. The FOV of the GMOS
    observations is indicated by the box. $(c)$ Logarithm of total
    GMOS flux integrated from 6100 to 6800 \AA, in arbitrary
    units. The image orientation is indicated by the compass rose, and
    the axis labels are in units of kpc. The near-infrared (NIR)
    continuum peak is circled. Isovelocity contours of the narrow
    ionized gas component are overplotted and labeled with velocity in
    \kms. Velocities are relative to $z_{sys} = 0.0584$. The dashed
    line intersects the peaks in rotational velocity and has PA $=$
    120\arcdeg\ E of N. $(d)$ Rest-frame equivalent width of \nad,
    computed from line fits, in \AA. $(e)-(f)$ Maps of the \hst\
    continuum flux ratio, log(F435W/F814W), with sizes and
    orientations identical to $(a)$ and $(b)$, respectively. $(g)$ Map
    of \hst\ color within the GMOS FOV, rotated and smoothed with a
    Gaussian kernel of $\mathrm{FWHM}=1\arcsec$. $(h)$ Signal-to-noise
    ratio of \nad\ equivalent width.}
  \label{fig:map_cont_f08572nw}
\end{figure}

The immediate results of our emission- and absorption-line fits to the
data are maps of physical properties in each system. The first set of
maps for each system (e.g., Figure \ref{fig:map_cont_f08572nw})
displays continuum properties: (1) two-color {\it Hubble Space
  Telescope} images (using Advanced Camera for Surveys Wide Field
Camera $F814W$ and $F435W$ data in most cases); (2) continuum flux
ratio (color) images using \hst\ data; (3) continuum images from our
GMOS data; and (4) the rest-frame equivalent width of \nad\ in each
spaxel.

\begin{figure}
%  \plotone{f03.eps}
  \centering
  \centering \includegraphics[width=6.5in]{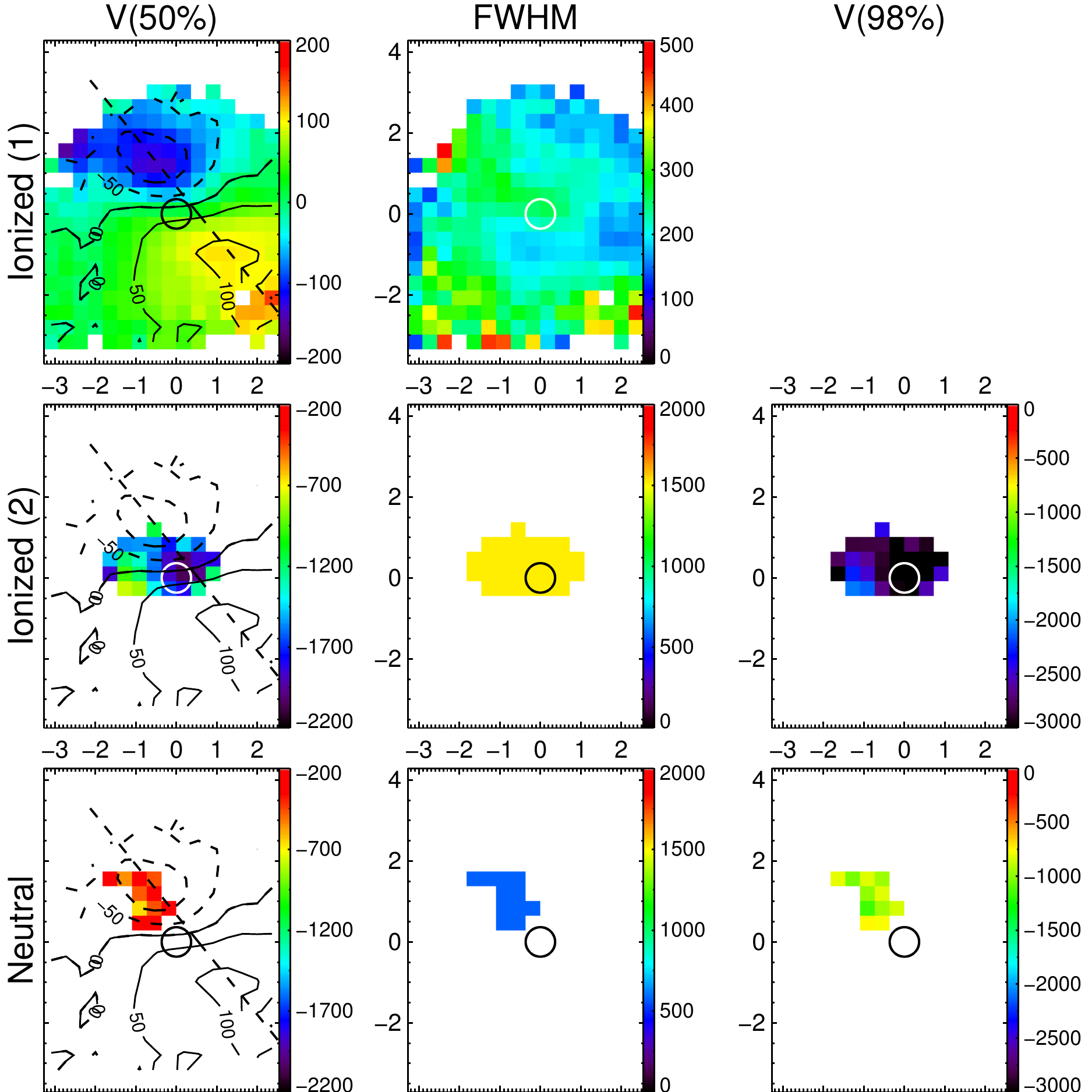}
  \caption{For F08572$+$3915:NW, maps of the centroid velocity
    (\vfifty), FWHM, and most blueshifted velocity
    ($\vtsig\equiv\vfifty-2\sigma$) in each velocity component in the
    ionized gas (based on emission lines) and neutral gas (based on
    \nad\ absorption). Each column shows a different velocity measure,
    and each row a different gas phase or velocity component. The
    ionized gas velocity component is labeled in parentheses (1 $=$
    narrow, rotating component; 2 $=$ broad, outflowing
    component). Colors indicate velocity in \kms. FWHM is corrected in
    quadrature for the instrumental profile. The contours, dashed
    lines, and open circles have the same meaning as in Figure
    \ref{fig:map_cont_f08572nw}.}
  \label{fig:map_vel_f08572nw}
\end{figure}

The second set of maps (e.g., Figure \ref{fig:map_vel_f08572nw}) shows
the line-of-sight (LOS) velocity of each emission- and absorption-line
component in each spaxel. We display the centroid velocity, the most
blueshifted velocity (for the broad components), and the FWHM (with
the spectral resolution subtracted in quadrature) of each component in
each spaxel. Our line profile modeling is based on Gaussian velocity
profiles, so the most blueshifted velocity is defined by the
properties of the normal distribution:
\begin{align}
  v_{50\%} &\equiv \mathrm{center~of~Gaussian~profile}\\
 ~ & \mathrm{(50\%~of~gas~has~lower~outflow~velocity)}\notag\\
   v_{98\%} &\equiv v_{50\%} - 2\sigma\\
 ~ & \mathrm{(98\%~of~gas~has~lower~outflow~velocity)}\notag
\end{align}
These definitions assume that negative velocities are blueshifted, and
apply to individual Gaussian components in cases where we fit more
than one velocity component.

The sole exception to these definitions is Mrk~273. Because Mrk~273
shows significantly redshifted emission as well as blueshifted
emission, we define \vtsig\ for redshifted components in this system
by adding, instead of subtracting, $\sigma$.

\begin{figure}
  % \plotone{f04.eps}
  \centering
  \centering \includegraphics[width=6.5in]{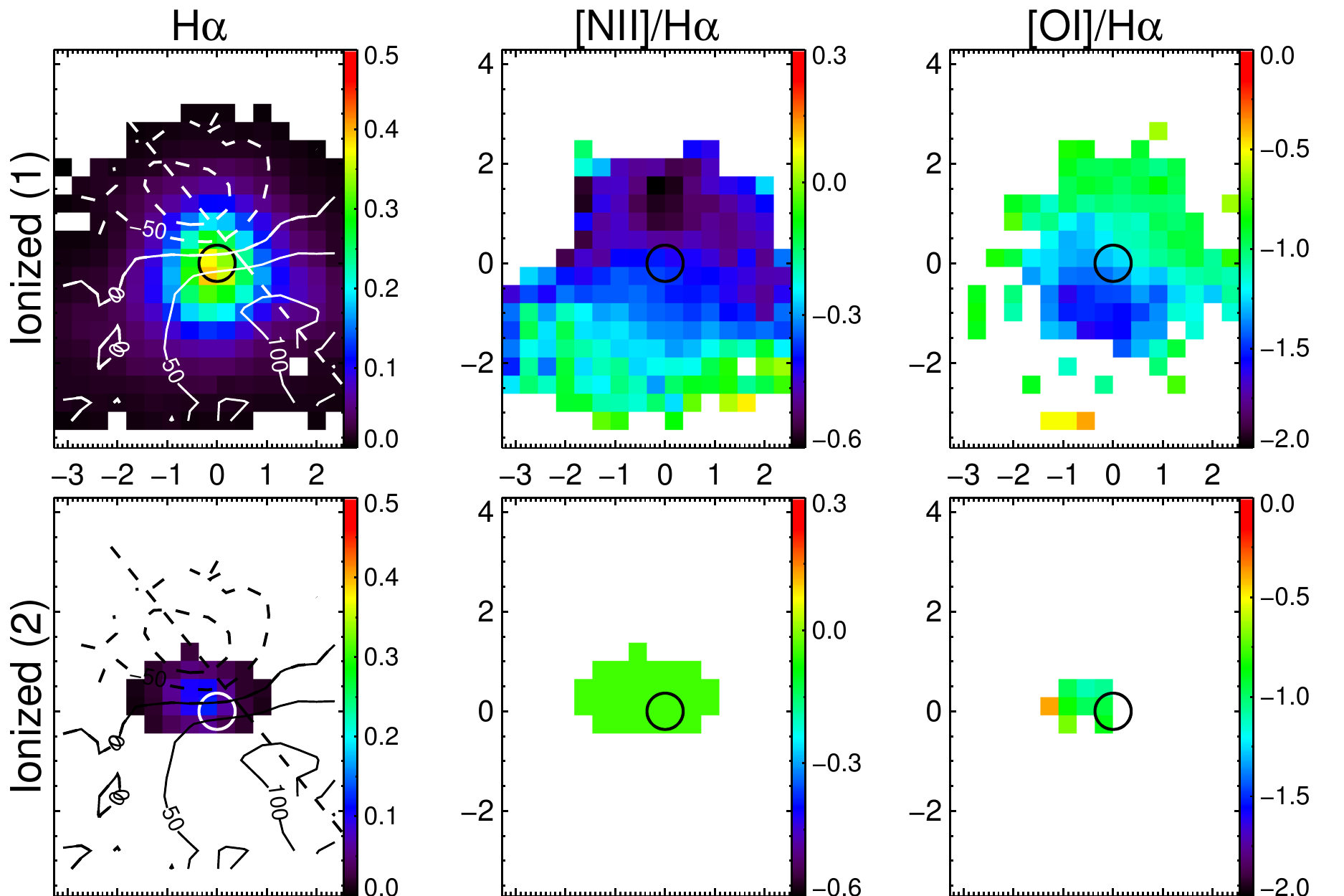}
  \caption{For F08572$+$3915:NW, maps of \ha\ flux in units of
    $1.0\times10^{-15}$ erg s$^{-1}$ cm$^{-2}$ per spaxel, or
    $1.1\times10^{-14}$ erg s$^{-1}$ cm$^{-2}$ arcsec$^{-2}$ (left
    column), \ntl/\ha\ flux ratio (central column) and \ool/\ha\ flux
    ratio (right column) in each ionized gas velocity component. The
    ionized gas velocity component is labeled in parentheses (1 $=$
    narrow, rotating component; 2 $=$ broad, outflowing
    component). The contours, dashed lines, and open circles have the
    same meaning as in Figure \ref{fig:map_cont_f08572nw}.}
  \label{fig:map_lrat_f08572nw}
\end{figure}

The third set of maps (e.g., Figure \ref{fig:map_lrat_f08572nw})
presents \ha\ flux and two emission line flux ratios (\ntl/\ha\ and
\ool/\ha) of each component in each spaxel.

\begin{figure}
  % \plotone{f05.eps}
  \centering
  \centering \includegraphics[width=6.5in]{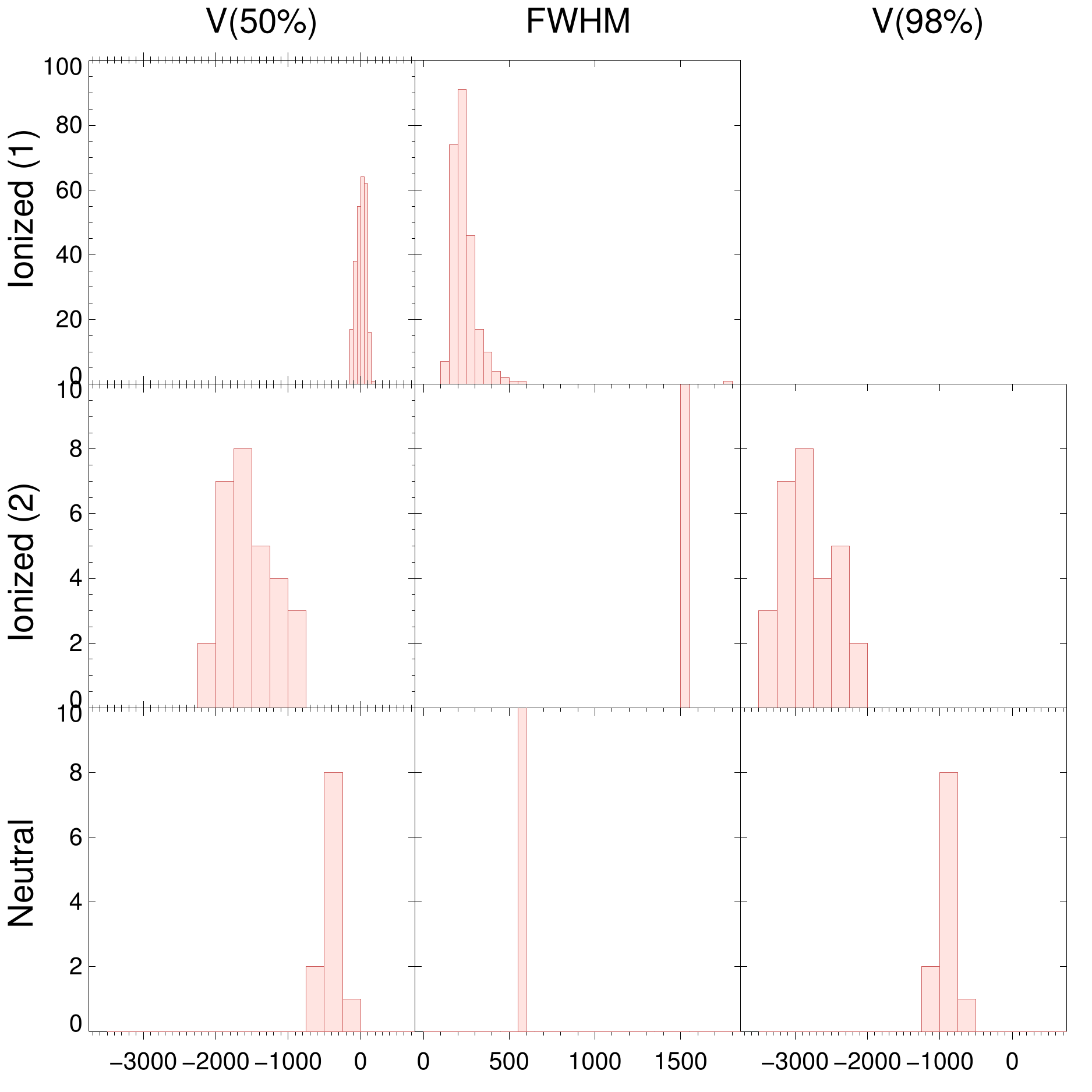}
  \caption{Distributions of \vfifty, FWHM, and \vtsig\ in each ionized
    gas component and the neutral gas in F08572$+$3915:NW. FWHM is
    corrected in quadrature for the instrumental profile. The vertical
    axis shows the number of spaxels in each bin, and the horizontal
    axis velocity in \kms.}
  \label{fig:veldist_f08572nw}
\end{figure}

The statistical properties of the derived LOS velocity fields are
presented in two ways. First, we show histograms of the distributions
of all LOS velocities in each system (e.g., Figure
\ref{fig:veldist_f08572nw}). Second, we tabulate the mean and extreme
values of FWHM, \vfifty, and \vtsig\ in each galaxy with an outflow in
Table \ref{tab:vels}. In the latter, we include only outflowing
components in each spaxel.

\subsection{Structural Modeling} \label{sec:models}

A primary goal of this study is to study the structural properties of
the outflows in individual systems. Structural models enable us to
better quantify the extent, mass, momentum, and energy in an outflow,
and thereby its impact on the galaxy.

Previous surveys of winds in ULIRGs used a spherically symmetric, mass
conserving free wind to estimate mass, momentum, and energy in
outflows \citep{heckman00a,rupke02a,rupke05b,soto12b}. These studies
made assumptions about the radial extent, radial distribution, and
variation of velocity with position of outflows in these
systems. Recent IFS studies of outflows in two ULIRGs
\citep{shih10a,rupke11a} were able to mitigate these uncertainties by
providing lower limits on the radial extent of the gas and detailed
maps of the gas mass and velocity as a function of position.

In this study, we employ two strategies to model the structure of the
outflows in each system. First, we estimate mass, momentum, energy,
and their flow rates using a single radius, mass conserving free
wind. Next, we use the detailed IFS data to motivate alternative
geometries. In particular, we use superbubble models to fit the data
for two systems (Mrk~273 and VV~705) that show strong evidence for
such structure. Other geometries are possible (e.g., biconical
outflows, filled-in geometries instead of shells), but require
modeling complexity beyond the scope of the current paper.

\subsubsection{Single Radius Free Wind
  (SRFW)} \label{sec:SRFW}

For the Single Radius Free Wind (SRFW) model, we use the time-averaged
thin shell formulae of \citet{rupke05b}, which are derived by assuming
a mass-conserving flow. These formulae were modified by
\citet{shih10a} for IFS data by letting each spaxel (corresponding to
a unique angular coordinate with respect to the wind's center) have
its own mass and velocity. These formulae are equivalent to computing
the mass, momentum, and energy at each angular location on the thin
shell, and then computing outflow rates by dividing by the dynamical
time $t_{dyn}\equiv R_{wind}/v$ of each spaxel (or angular
coordinate). We assume a single radius for the wind in each system,
estimated from the maximum extent of the wind in the plane of the
sky. Within each system, we use the same wind radius for the ionized
and neutral gas phases. We use the velocity measured in each spaxel,
deprojected using the assumed radius, as the true wind velocity at
that angular location in the wind.

Most of the formulae in \citet{rupke05b} and \citet{shih10a} are
expressed in terms of column density, and are in this form only
applicable to the neutral gas in our study. However, by computing the
ionized gas mass in each spaxel from the \ha\ luminosity, and dividing
by the dynamical time for each spaxel, we can compute momentum,
energy, and their outflow rates in each spaxel from their dependence
on mass and velocity.

The neutral gas mass in each spaxel is proportional to the H column
density and the deprojected solid angle at that location (i.e., the
solid angle subtended by the spaxel with respect to the outflow
center). To compute the H column density, we first find the Na column
density, which depends on the measured optical depth and velocity
width in a simple way for our assumption of a Gaussian velocity
distribution in each component \citep[see, e.g., the formula
in][]{rupke02a}. This is converted into H column density using typical
Milky Way dust depletion for Na ($-0.95$~dex; \citealt{savage96a}) and
metal abundance of Na ($-5.69$~dex; \citealt{savage96a}), and an
uncertain ionization fraction ($y = 1 - N$(\ion{Na}{1})/$N$(Na) = 0.9;
\citealt{rupke05a}). $N$(H) depends on $y$ as $(1-y)^{-1}$. The solid
angle subtended by each spaxel is easily computed in our spherical
model, and we then multiply this solid angle by the measured covering
factor of the neutral gas, to account for clumpiness in the
outflow. (The covering factor in each spaxel equals the fractional
area of the spaxel that is covered by absorbing gas, and is a
parameter in the absorption line fit; see \citealt{rupke05a} for more
details.)

The ionized gas mass in each spaxel is computed from the \ha\
luminosity, assuming Case B recombination and $T=10^4$~K. It depends
on electron density as $n_e^{-1}$. We assume $n_e = 10$~cm$^{-3}$,
which is consistent with limits on $n_e$ for the superbubble in
NGC~3079 ($n_e = 5-100$~cm$^{-3}$; \citealt{veilleux94a,cecil01a}).

The assumed radii in the SRFW model for each galaxy with an outflow
are listed in Table~\ref{tab:modsrfw}. This table also lists the
deprojected covering factor over the surface of the outflow, as seen
from the galaxy nucleus, assuming the given radius, and summed over
all spaxels with outflowing gas. This integrated covering factor
(which we label $\Omega/4\pi$) represents the fraction of sightlines,
for an observer at the galaxy center, that will encounter the wind. It
differs from the opening angle of the wind, which describes the solid
angle contained by the outer boundary of the wind. The integrated
covering factor is a lower limit to the wind opening angle.

\subsubsection{Bipolar Superbubble (BSB)} \label{sec:bsb}

The ionized gas kinematics in Mrk~273 (\S\,\ref{sec:mrk273-bsb}) and
neutral gas kinematics in VV~705:NW (\S\,\ref{sec:vv705-kinematics})
suggest the presence of a Bipolar SuperBubble (BSB). Following similar
treatments of other nuclear superbubbles \citep[e.g.,][]{veilleux94a},
we used two prolate ellipsoidal bubbles on either side of the galaxy
nucleus. Each bubble has semi-principal axis lengths $R_z$ (along the
axis shared by the two bubbles) and $R_{x}=R_{y}\equiv R_{xy}$
(perpendicular to the $z$ axis). We assumed that the gas lies on the
bubbles' surfaces and that each parcel of gas is moving radially with
velocity $v = v_z (r/r_z)^n$, where $r$ is the distance from the
center of the wind, and $v_z$ and $r_z = 2R_z$ are the velocity and
distance along the $z$ axis.

For Mrk~273, we optimized seven of the eight BSB parameters
($v_z$, $R_{xy}$, $R_z$, line of sight inclination, position angle,
and center coordinates) by minimizing the RMS difference between the
model and the data. We found that it was not possible to robustly
constrain the power law index $n$ beyond the statement that $n\ga1$ is
favored. The best-fit z-axis velocity $v_z$ varied with $n$, such that
larger values of $n$ favored larger $v_z$. However, the best fit
values of other parameters varied little or not at all with $n$.

For two different plausible values of $n$ based on studies of other
sources (2 and 3; see, e.g., \citealt{veilleux94a,cecil01a}), we list
in Table~\ref{tab:modbbl} the best fit BSB parameters for the
ionized gas in Mrk~273. In the best fit models, 67 Nyquist-sampled
spaxels lie completely inside the bubble edges. However, each spaxel
has two outflowing kinematic components (the near and far sides of the
bubble), leading to 134 data points in the fit. The RMS difference
between the model and data for the redshifted and blueshifted
components (with the redshifted and blueshifted RMS values combined in
quadrature) is $\sim$200~\kms. We estimated the errors in each
parameter by fixing all other parameters and varying the parameter
until the RMS increased by 50~\kms.

For VV~705:NW, we performed a similar procedure. Because this disk is
almost face-on (\S\,\ref{sec:vv705-kinematics}), any bubble emerging
along the minor axis of the disk will also be face on. The outflow in
this system was detected primarily in absorption, so we could not
detect the far side of the bubble (behind the disk). The resulting 38
distinct data points (in the best fit model) made it infeasible to fit
all 8 parameters of the BSB model. Thus, we fixed the center of the
outflow to be the NIR luminosity peak
(\S\,\ref{sec:vv705-kinematics}). We also constrained the cylindrical
radius of the outflow in the plane perpendicular to the z-axis to be
$R_{xy}=1.0$~kpc after inspecting the extent of the outflow in the
plane of the sky. Furthermore, as with Mrk~273 we could only determine
that $n\ga1$. Thus, for $n=2$ and $n=3$ we fit four of the eight BSB
parameters ($v_z$, $R_z$, line of sight inclination, position angle).

To compute the neutral gas mass in the BSB model, we measured the
column density in each spaxel and multiplied by the projected spaxel
area. The ionized gas mass was computed as in the SRFW model, and
momentum and energy followed from the deprojected velocity of each
spaxel. Outflow rates come from dividing by the dynamical time,
$2R_z/v_z$.

\subsubsection{Results}

The mass, momenta, and energies and their outflow rates derived from
the SRFW and BSB models are listed for each system and each gas phase
in Table~\ref{tab:mpe}.\footnote{We note that the mass, momentum, and
  energy outflow rates for Mrk~231 are a factor of $\sim$2 lower than
  those listed in \citet{rupke11a}. This is because in
  \citet{rupke11a} we did not multiply the neutral gas quantities in
  each spaxel by the covering factor of the absorption line gas in
  that spaxel. In this paper we do include this factor.}

Some uncertainties remain in these models. First, there is still LOS
uncertainty in the gas position. Second, we employ only ``thin shell''
models: we assume that the gas is confined to a small radial range
($\Delta R/R << 1$). Thick shell models are likely to yield smaller
masses, momenta, and energies \citep[e.g.,][]{rupke05b}. Finally,
there is uncertainty due to the ionization correction required to
convert the amount of \ion{Na}{1} to the total amount of Na
\citep[e.g.,][]{rupke02a,rupke05b}. We assume that the \ion{Na}{1}
represents 10\%\ of the total Na, based on Galactic sight lines
\citep{rupke02a}. Since Na resides primarily in gas well-shielded from
UV radiation (because of its low ionization potential), the correction
is likely not much larger than this.

\section{INDIVIDUAL SYSTEMS} \label{sec:systems}

In this section, we discuss the power source and kinematics of each
system in detail.

\subsection{F08572$+$3915} \label{sec:f08572}

\begin{figure}
%  \plotone{f02.eps}
  \centering
  \centering \includegraphics[width=5in]{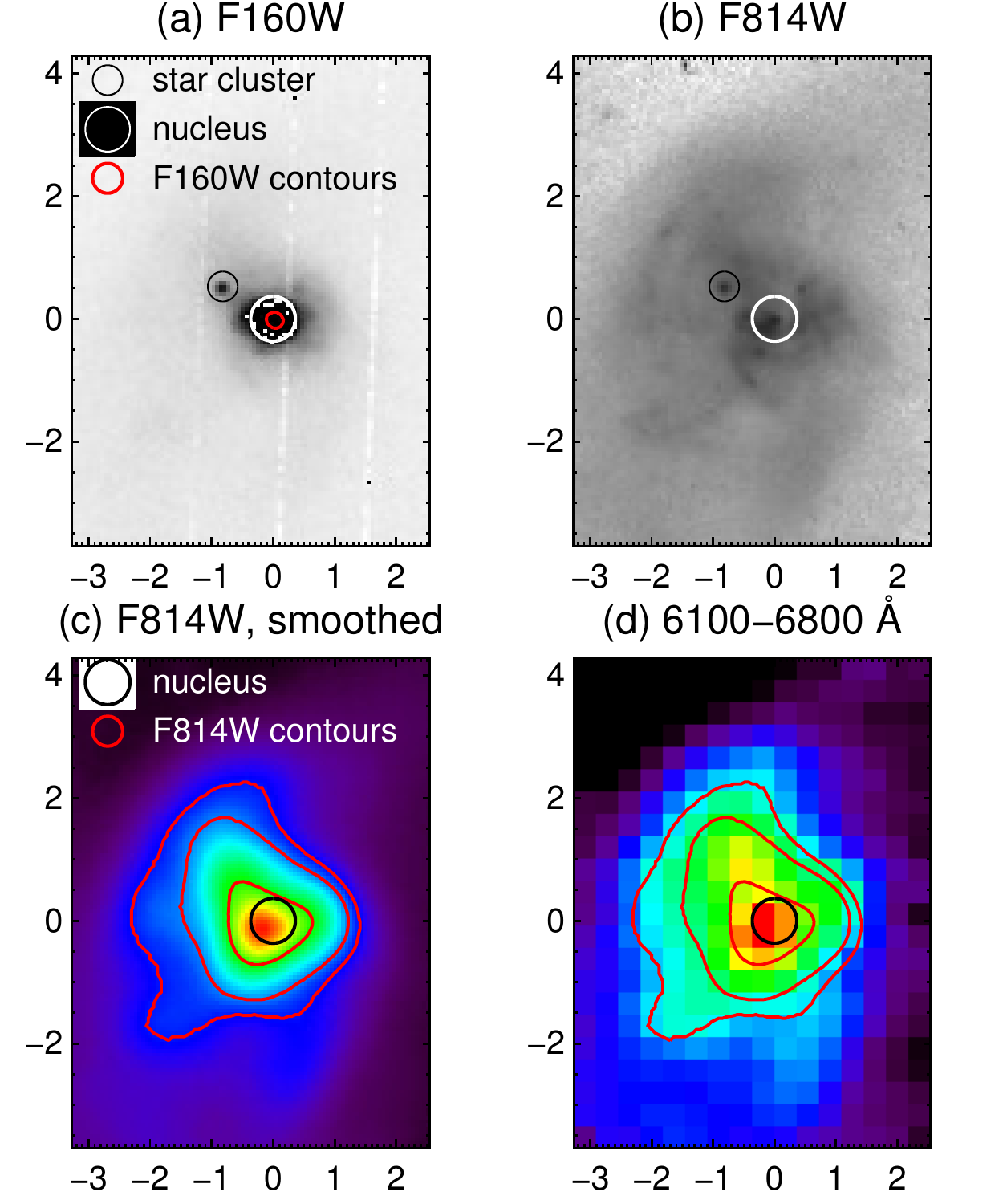}
  \caption{\scriptsize (a) \hst\ F160W image of F08572+3915:NW, within the
    GMOS FOV. A contour highlighting the nucleus is in red. The
    nuclear star cluster used to align the images is circled, as is
    the NIR nucleus. Slight defects resulting from interpolation over
    detector artifacts appear near the nucleus but do not affect
    centering. (b) \hst/ACS F814W image. Note the common alignment of
    the star cluster between this image and the F160W image, and the
    relative faintness of the nucleus in the optical. (c) F814W image,
    smoothed with a Gaussian kernel of $\sigma = 0\farcs25$ to match
    ground-based seeing. Image contours are in red, and the NIR
    nucleus is circled. (d) GMOS $6100-6800$~\AA\ image. Smoothed
    F814W contours are in red. Note that the smoothed F814W image is
    well aligned with the GMOS image. The location of the nucleus in
    the GMOS data is tied to the F160W image by common alignment with
    the F814W image. Axis labels are in kpc.}
  \label{fig:register_f08572nw}
\end{figure}

\begin{figure}
  % \plotone{f06.eps}
  \centering \includegraphics[width=6.5in]{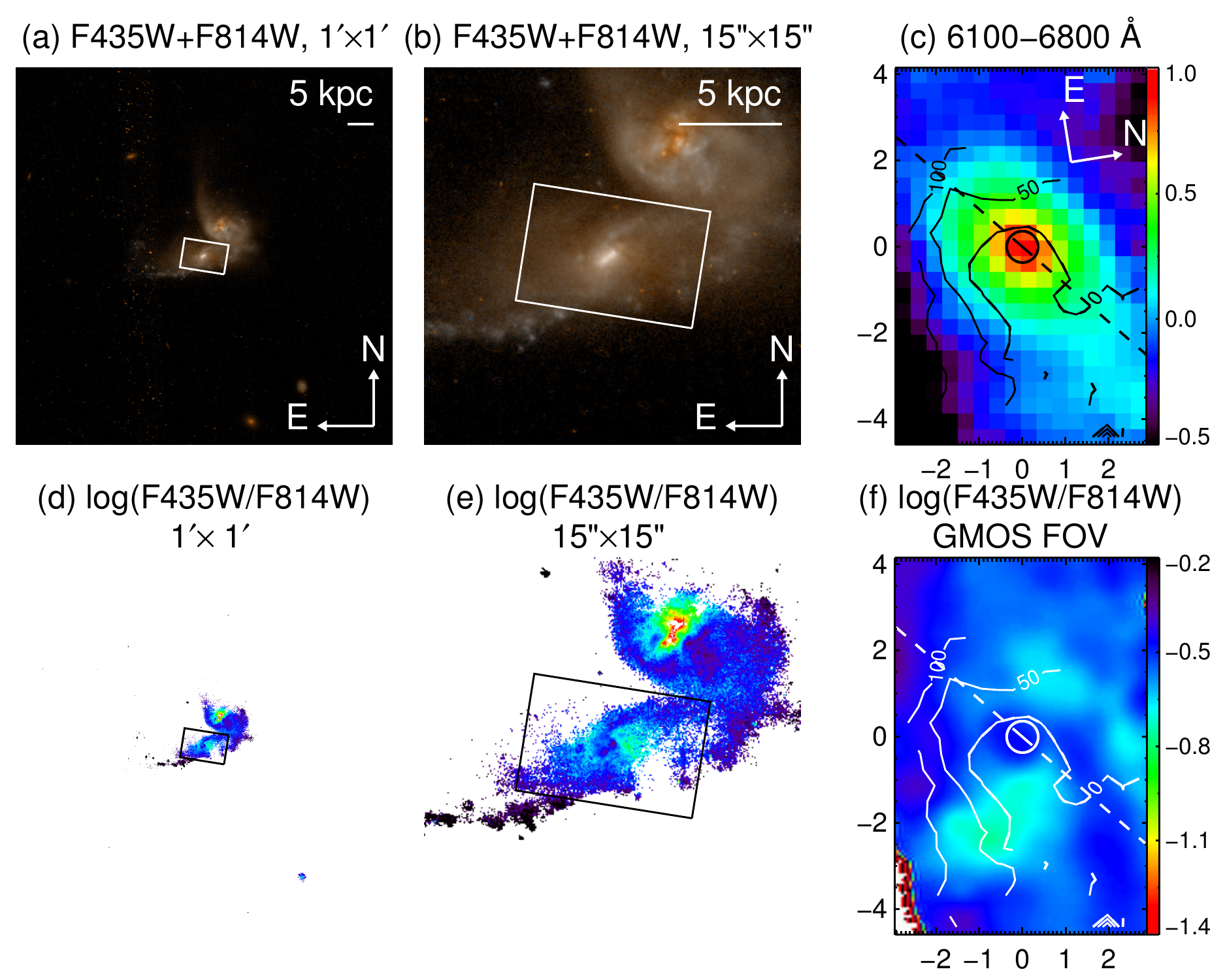}
  \caption{$(a)-(b)$: Two-color \hst\ image of F08572$+$3915:SE at
    sizes $1\arcmin\times1\arcmin$ and
    $15\arcsec\times15\arcsec$. (See
    Figure~\ref{fig:map_cont_f08572nw} for more details.) $(c)$
    Logarithm of total GMOS flux integrated from 6100 to 6800 \AA, in
    arbitrary units. See Figure~\ref{fig:map_cont_f08572nw} for more
    details. Velocities are relative to $z_{sys} = 0.0585$. The dashed
    line is along the major axis of the galaxy bulge, with PA $=$
    140\arcdeg\ E of N. $(d)-(e)$ Maps of the \hst\ continuum flux
    ratio, log(F435W/F814W), with sizes and orientations identical to
    $(a)$ and $(b)$, respectively. $(f)$ Map of \hst\ color within the
    GMOS FOV, rotated and smoothed with a Gaussian kernel of
    $\mathrm{FWHM}=1\arcsec$.}
  \label{fig:map_cont_f08572se}
\end{figure}

\begin{figure}
%  \plotone{f07.eps}
  \centering \includegraphics[width=6.5in]{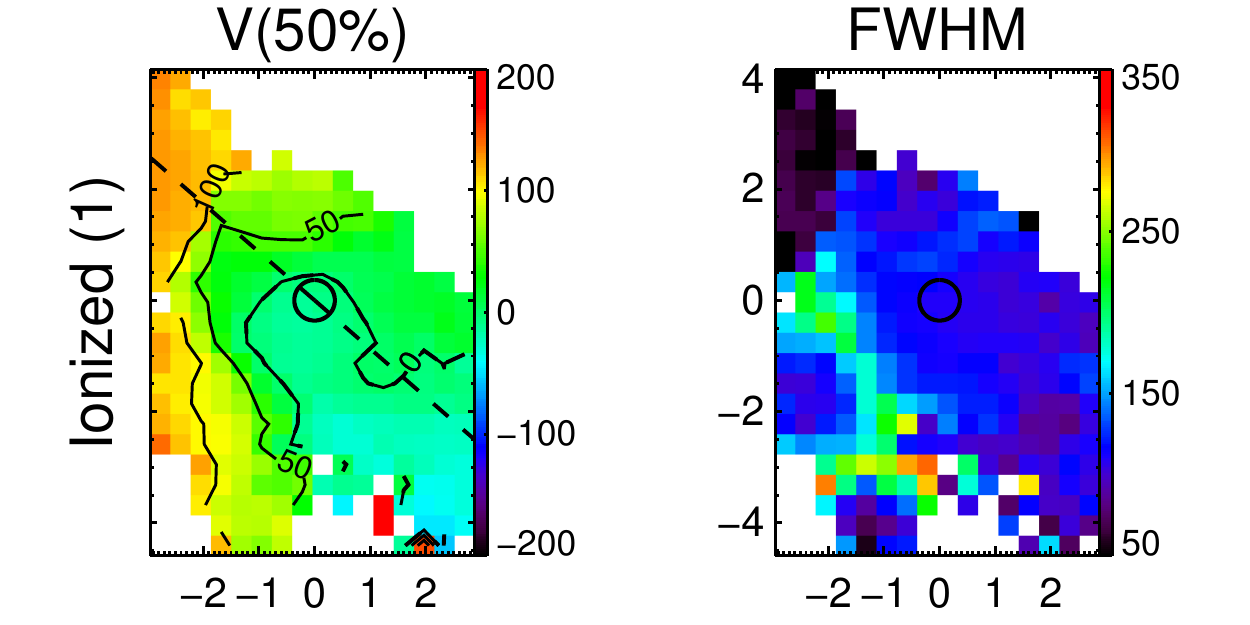}
  \caption{The same as Figure \ref{fig:map_vel_f08572nw}, but for
    F08572$+$3915:SE.}
  \label{fig:map_vel_f08572se}
\end{figure}

\begin{figure}
%  \plotone{f08.eps}
  \centering \includegraphics[width=6.5in]{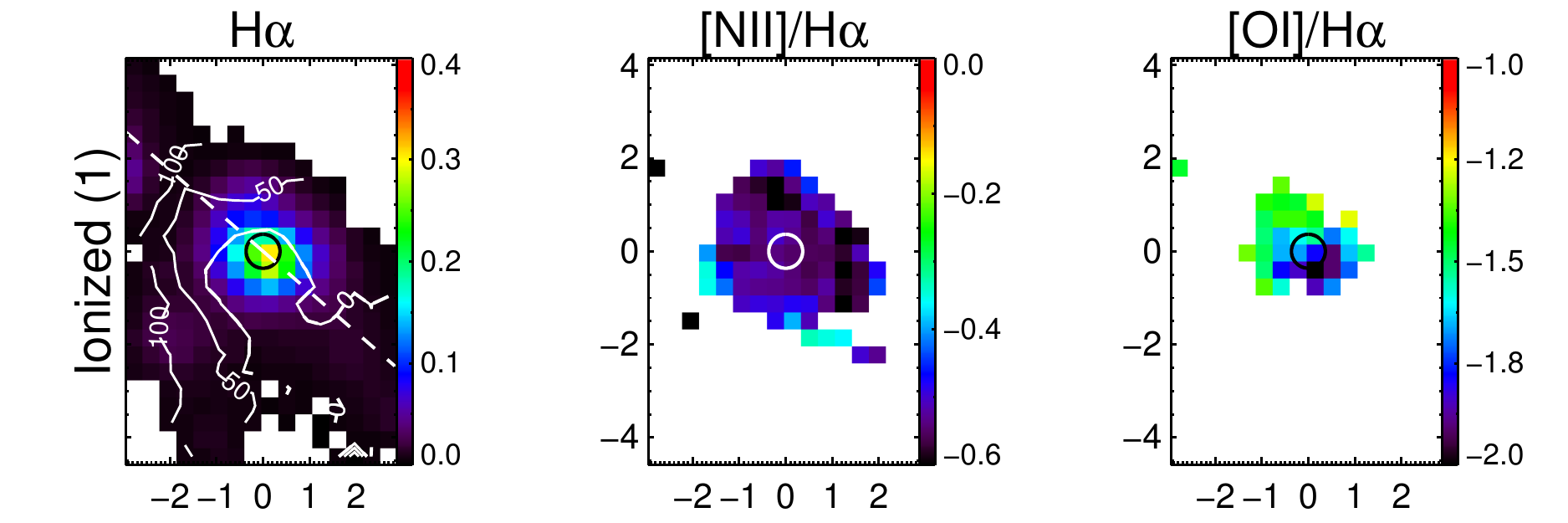}
  \caption{The same as Figure \ref{fig:map_lrat_f08572nw}, but for
    F08572$+$3915:SE.}
  \label{fig:map_lrat_f08572se}
\end{figure}

\begin{figure}
%  \plotone{f09.eps}
  \centering \includegraphics[width=6.5in]{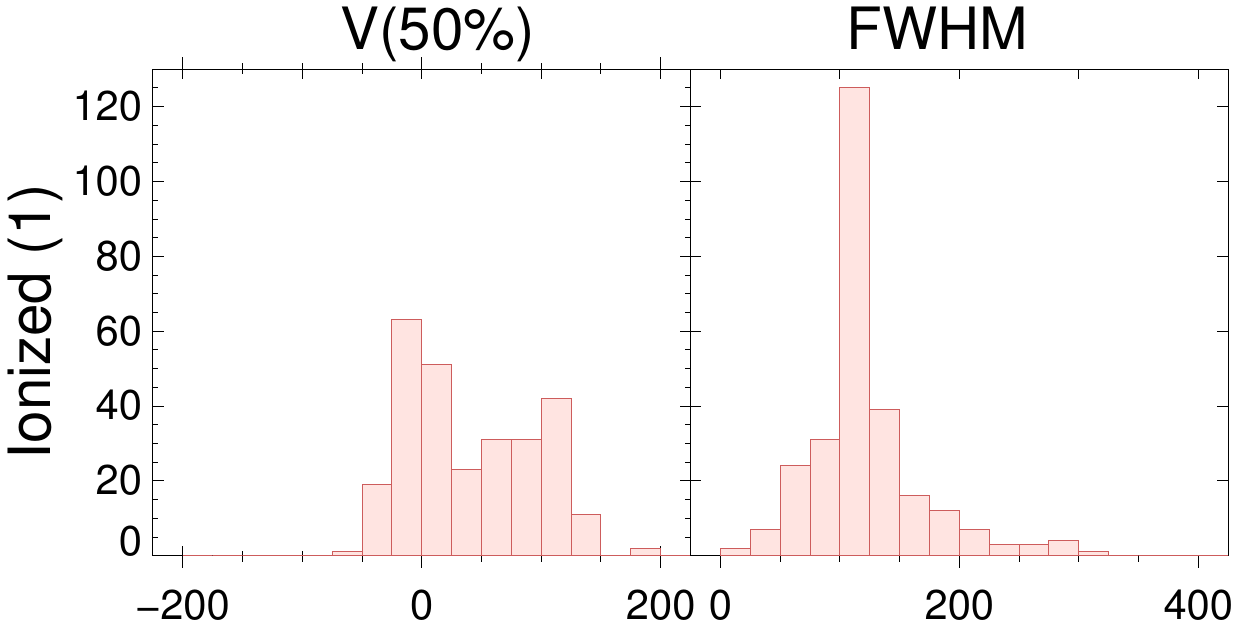}
  \caption{The same as Figure \ref{fig:veldist_f08572nw}, but for
    F08572$+$3915:SE.}
  \label{fig:veldist_f08572se}
\end{figure}

This system is a close pair that is likely near second pericenter. The
system's energetics are dominated by a buried AGN in the NW nucleus,
based on copious evidence. We review this evidence and then discuss
the outflow in this system, which shows the highest velocities in our
sample. Figures
\ref{fig:map_cont_f08572nw}$-$\ref{fig:veldist_f08572se} present maps
and velocity distributions for each nucleus.

%\clearpage

\subsubsection{Power Source} \label{sec:f08572_ps}

The NW nucleus is the source of the prodigious far-infrared (FIR)
luminosity, as it completely dominates the radio and mid-infrared
(MIR) luminosity of the system \citep{condon90a,soifer00a}. Inspection
of {\it Spitzer Space Telescope} archival images at 3.6\micron\ and
8.0\micron\ shows that the SE nucleus contributes less than 1\% of the
system luminosity at these wavelengths. The dominance of the NW source
is consistent with the $\sim$$2\times10^9$ \msun\ of molecular gas
detected in the NW nucleus \citep{evans02a}; only upper limits are
obtained in the SE.

Clear evidence for AGN in this system is absent in ground-based
optical spectra. Previous optical spectral types of LINER
\citep{veilleux99a} or Seyfert 2 \citep{yuan10a} in each nucleus were
based on shallow spectra that did not significantly detect \othl\ or
\hb\ \citep{kim98b}. The SDSS spectrum of the NW nucleus is much
deeper than these previous spectra, with a similar spatial
footprint. An analysis of the SDSS spectrum using our stellar
continuum and emission line fitting software (\S\,\ref{sec:obs})
yields $A_V = 2.14$, and a spectral type that is borderline between H
and C \citep{kewley06a}. This spectral type is not truly nuclear
however, since the line ratios vary significantly in the E-W direction
across the SDSS aperture. Some of these extended line ratios are
consistent with AGN-excited gas, though not uniquely so in the absence
of \oth\ data (Fig. \ref{fig:map_lrat_f08572nw}). The optical
appearance in \hst\ images also shows ``complex extinction and
scattering'' (\citealt{surace98a};
Fig. \ref{fig:map_cont_f08572nw}). In contrast with the NW nucleus,
the SE nucleus is clearly of spectral type H
(Fig. \ref{fig:map_lrat_f08572se}), even in the absence of \oth\ or
\hb\ data, consistent with its optical morphology (a ``typical galaxy
core''; \citealt{surace98a}).

In the NW nucleus, NIR recombination lines point to heavier
obscuration than in the optical: $A_V = 6.7$, based on Pa$\alpha$ and
Br$\gamma$ fluxes \citep{veilleux99b} and assuming Case B, $n_e =
10^4$ cm$^-2$, and $T = 10^4$~K \citep{osterbrock89a}. This nucleus is
unresolved in the NIR with \hst, and it becomes more prominent with
respect to the extended host between 1.1 and 2.2\micron\ (the
half-light radius decreases from 830~pc to 140~pc;
\citealt{scoville00a}). $L$-band and MIR spectra point to a very high
$A_V$, based on deep absorption features from carbonaceous and
silicate dust \citep{dudley97a,imanishi00a,soifer02a,armus07a}. The
MIR source radius is $<$130~pc at 12.5\micron\ \citep{soifer00a} and
$<$60~pc at 18\micron\ \citep{imanishi11a}. The increasing inferred
visual extinction and decreasing source size as tracers move to longer
wavelengths point to a highly obscured, compact source.

The most likely candidate to power the NW source is a heavily obscured
AGN, as confirmed by MIR spectroscopy. The infrared continuum
shape and lack of PAH emission provide evidence that a deeply buried
AGN powers $\sim$70\% of the total infrared luminosity
\citep{veilleux09a}. Hot dust emission also points to an AGN
\citep{armus07a}. This source is barely detected by the {\it Chandra
  X-ray Observatory}, though its hardness ratio is in principle
consistent with heavy obscuration of an AGN \citep{teng09a}.

\subsubsection{Kinematics} \label{sec:f08572_kin}

The SE galaxy reveals modest non-circular motions, indicative of tidal
dynamics, superimposed on ordinary rotation of an inclined disk. The
line of nodes is uncertain, but the major axis of the bulge is
130\arcdeg\ E of N. The projected peak-to-peak velocity amplitude
along this line is $\sim$150~\kms. The appearance of the spiral arms
in the \hst\ image and the redshifted emission in the SE point to the
near side of the disk lying in the NE. No absorption-line gas or
emission-line outflow is evident in this galaxy.

The narrow ionized gas component in the NW galaxy is undergoing
orderly rotation, again with modest non-circular motions mixed in. The
central isovelocity contours are not perpendicular to the line of
nodes, which we estimate to lie at 120\arcdeg\ E of N if it is to pass
between the projected rotational maxima ($\pm$120~\kms). If the
maximum possible disk velocity were 150~\kms\ (200~\kms), then the
disk inclination would be 53\arcdeg\ (37\arcdeg). The midpoint of the
rotational maxima is offset $\sim$0.5\arcsec NE of the nucleus. If the
tidal features near the nucleus are the elongations of spiral arms,
then the near side of the galaxy is to the SW of the nucleus.

\begin{figure}
%  \plotone{f10.eps}
  \centering \includegraphics[width=6.5in]{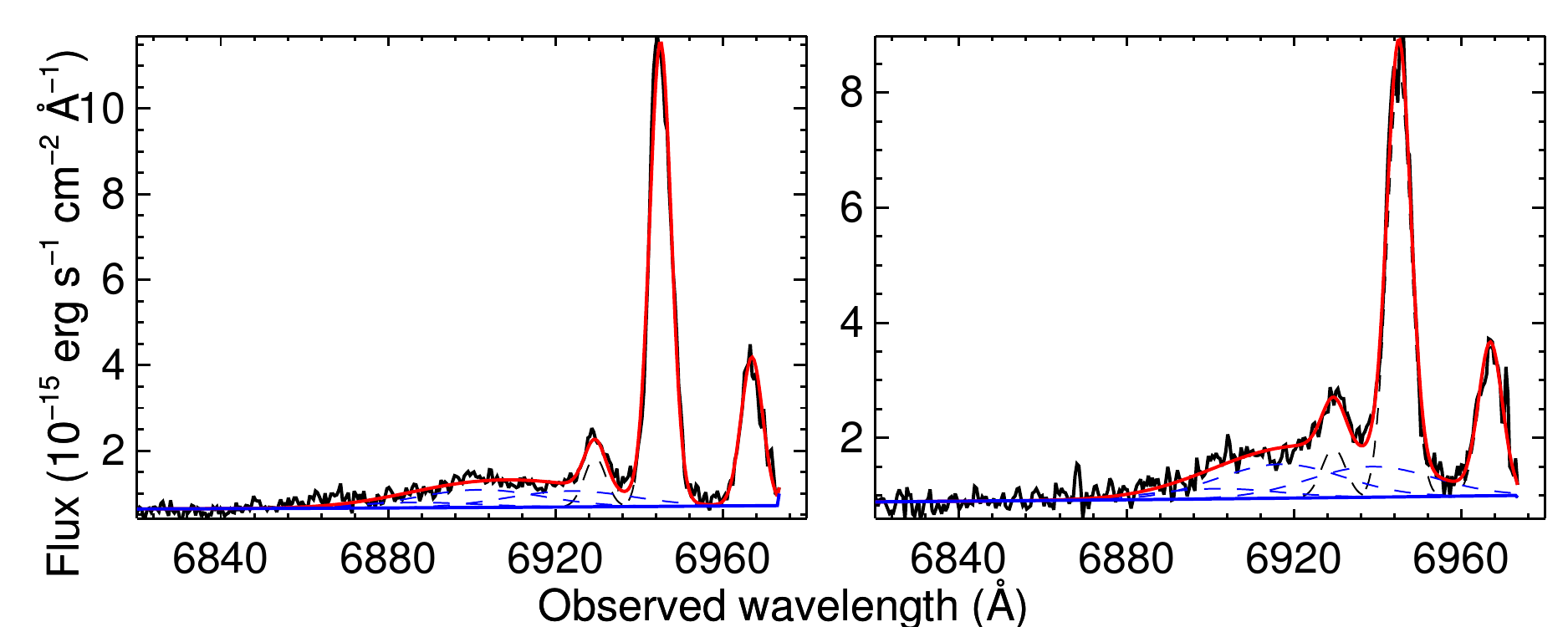}
  \caption{\ha$+$\ntll\ spectra of two spaxels in the emission-line
    outflow of F08572$+$3915:NW. The solid black line is the data, the
    solid red line the total fit, the solid blue line the continuum
    fit, and the dashed lines fits to individual lines. The different
    colors of the dashed lines represent different velocity
    components.}
  \label{fig:spectra_f08572nw}
\end{figure}

The broad emission line component in the NW nucleus of F08572$+$3915
required special treatment. We were unable to independently constrain
the line fluxes and width of the broad, outflowing gas component in
each spaxel because of the large width of this component, which caused
severe blending between \ha\ and \nt. The outflowing component of the
isolated line \oo\ was too weak to provide an independent constraint
on the line width. We instead used the single aperture Sloan Digital
Sky Survey (SDSS) spectrum of the NW nucleus to disentangle the line
widths and fluxes of the outflowing component. While our spectrum does
not have the wavelength range to include \sutl, the SDSS spectrum
does, and the outflowing component appears in this line. In both the
narrow and broad emission components of the SDSS spectrum, we fit
simultaneously all emission lines using a common galactocentric
velocity and line width in each component. We constrained the \sut\
line ratio to lie within its expected values based on atomic
physics. The \ha, \nt, and \sut\ lines together provided enough
leverage to simultaneously fix the line width and line fluxes of the
outflowing component. We then used these values to fix the \nt/\ha\
ratio and line width across the outflow in our data, and we fit only
the central velocity and absolute flux in the broad outflowing
component in each spaxel. Fig. \ref{fig:spectra_f08572nw} shows
example fits to the broad line in our data.

The broad ionized gas component reveals the highest velocity outflow
in our sample, with $\vtsigave= -2800$~\kms\ and a maximum value for
\vtsig\ of -3350~\kms. It peaks in velocity directly between the peaks
in projected rotational velocity, and shows lower velocities 2~kpc to
the S/SW (though still in excess of $\vtsig=-2000$~\kms). The
blueshifted component is very broad (FWHM $=$ 1500~\kms), but it is
{\it not} an AGN broad line region because it appears in other
forbidden lines (\sut) and is spatially resolved.

The neutral outflow in the NW galaxy is located $1-2$ kpc SE of the
nucleus, and reaches velocities up to $-1150$~\kms\ (with $\vtsigave =
-900$~\kms).

The outflow in F08572$+$3915 is not well enough resolved for a
structural model, but the data offer some clues. The emission-line
flow extends along the galaxy minor axis, though in a direction behind
the near side of the disk. However, at this point in the disk the
reddening is low. The decrease of outflow velocity in this direction
is consistent with the far-side of a minor-axis outflow. The other
direction along the minor axis shows heavy reddening; the outflow is
perhaps obscured along this direction. The obscuration along this
direction may also indicate a dusty flow. However, no neutral gas is
detected here; the neutral gas outflow instead appears SE of the
nucleus along the line of nodes. The neutral gas partly overlaps the
emission-line outflow, but its overall morphology is unresolved, and
it is unclear how the two are related.

\subsection{F10565$+$2448} \label{sec:f10565}

\begin{figure}
%  \plotone{f11.eps}
  \centering \includegraphics[width=6.5in]{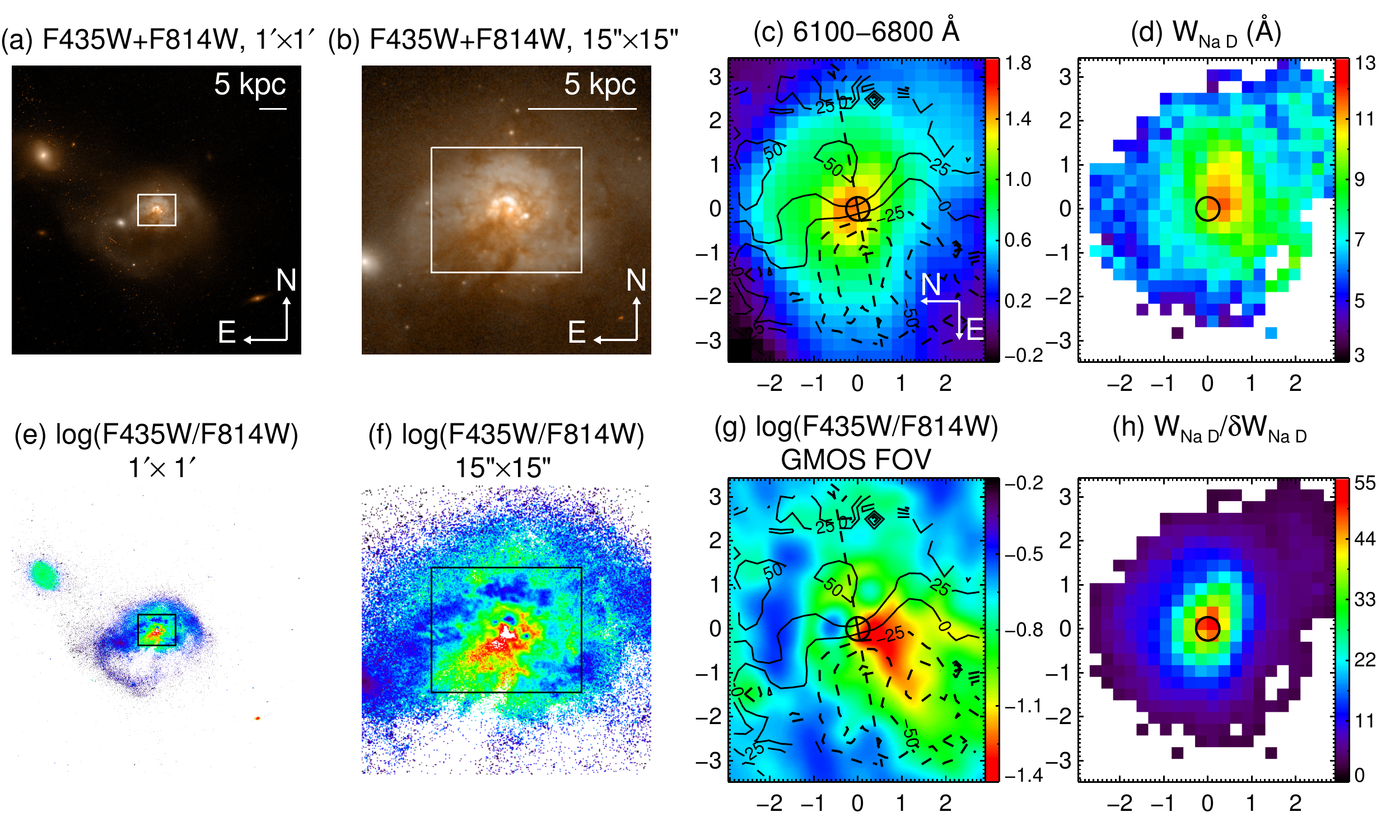}
  \caption{The same as Figure \ref{fig:map_cont_f08572nw}, but for
    F10565$+$2448. Velocities are relative to $z_{sys} = 0.0431$. The
    dashed line is along the CO line of nodes (\citealt{downes98a}; PA
    $=$ 100\arcdeg\ E of N). }
  \label{fig:map_cont_f10565}
\end{figure}

\begin{figure}  
%  \plotone{f12.eps}
  \centering \includegraphics[width=6.5in]{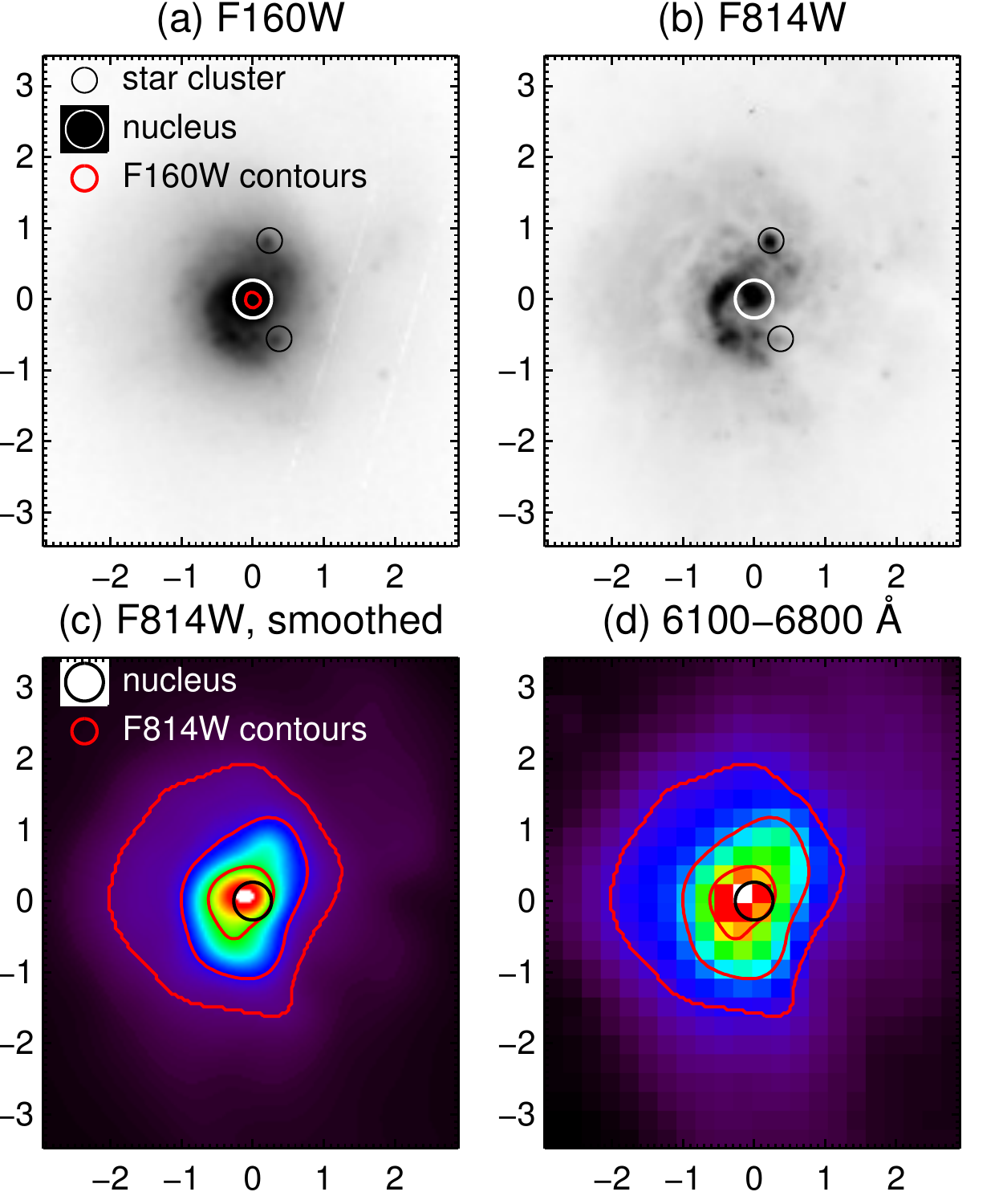}
  \caption{The same as Figure \ref{fig:register_f08572nw}, but for
    F10565$+$2448.}
  \label{fig:register_f10565}
\end{figure}

\begin{figure}
  % \plotone{f13.eps}
  \centering \includegraphics[width=6.5in]{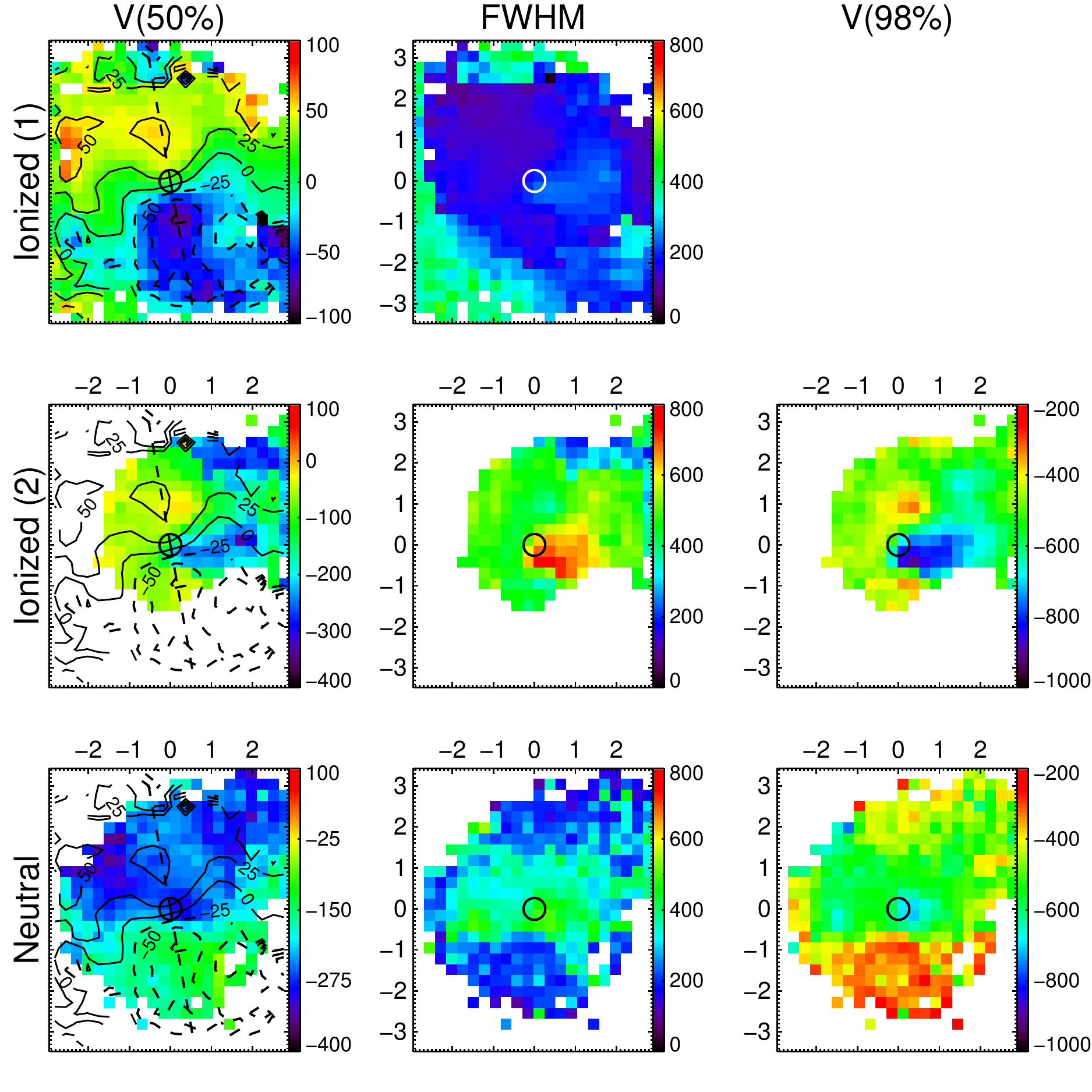}
  \caption{The same as Figure \ref{fig:map_vel_f08572nw}, but for
    F10565$+$2448.}
  \label{fig:map_vel_f10565}
\end{figure}

\begin{figure}
%  \plotone{f14.eps}
  \centering \includegraphics[width=6.5in]{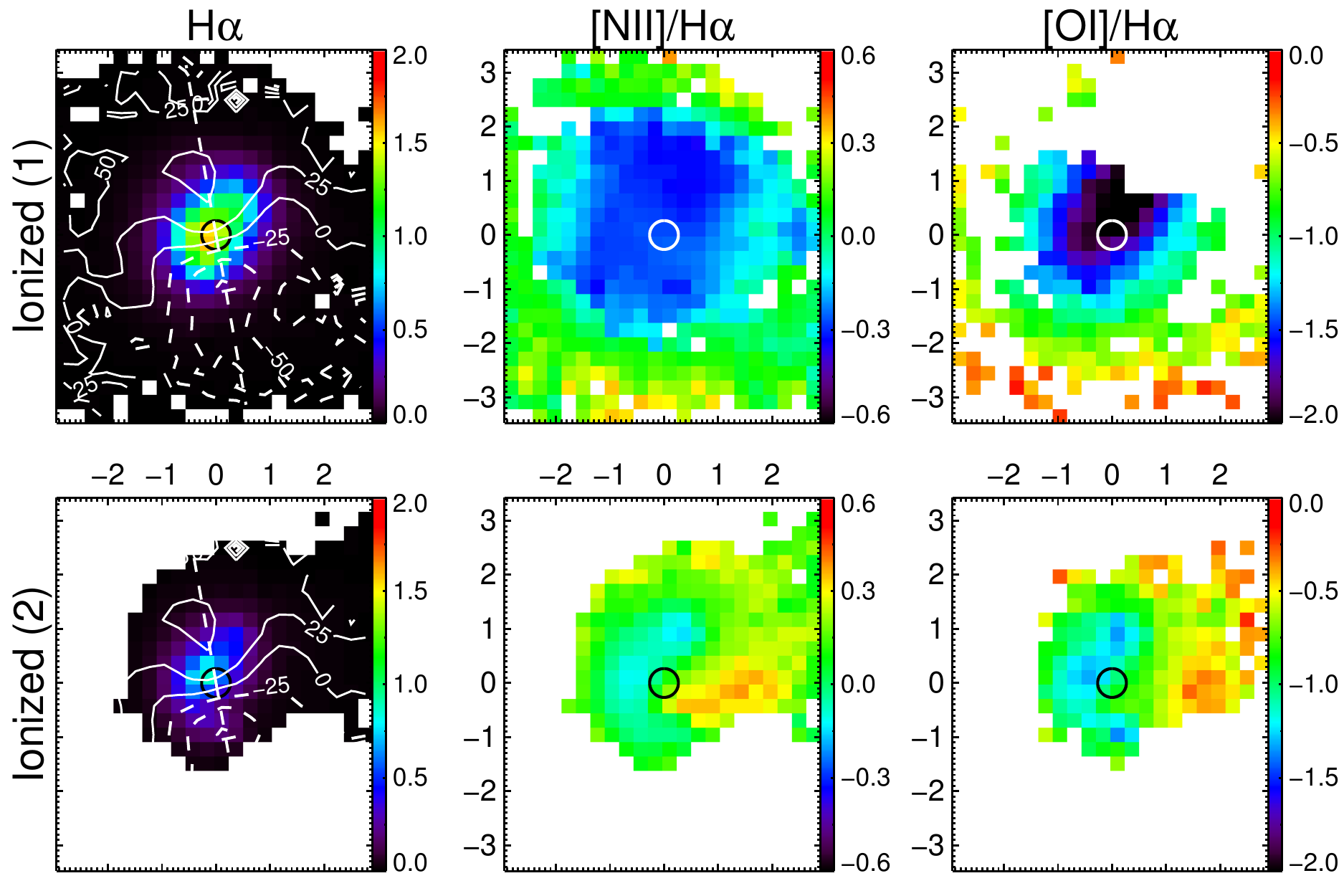}
  \caption{The same as Figure \ref{fig:map_lrat_f08572nw}, but for
    F10565$+$2448.}
  \label{fig:map_lrat_f10565}
\end{figure}

\begin{figure}
%  \plotone{f15.eps}
  \centering \includegraphics[width=6.5in]{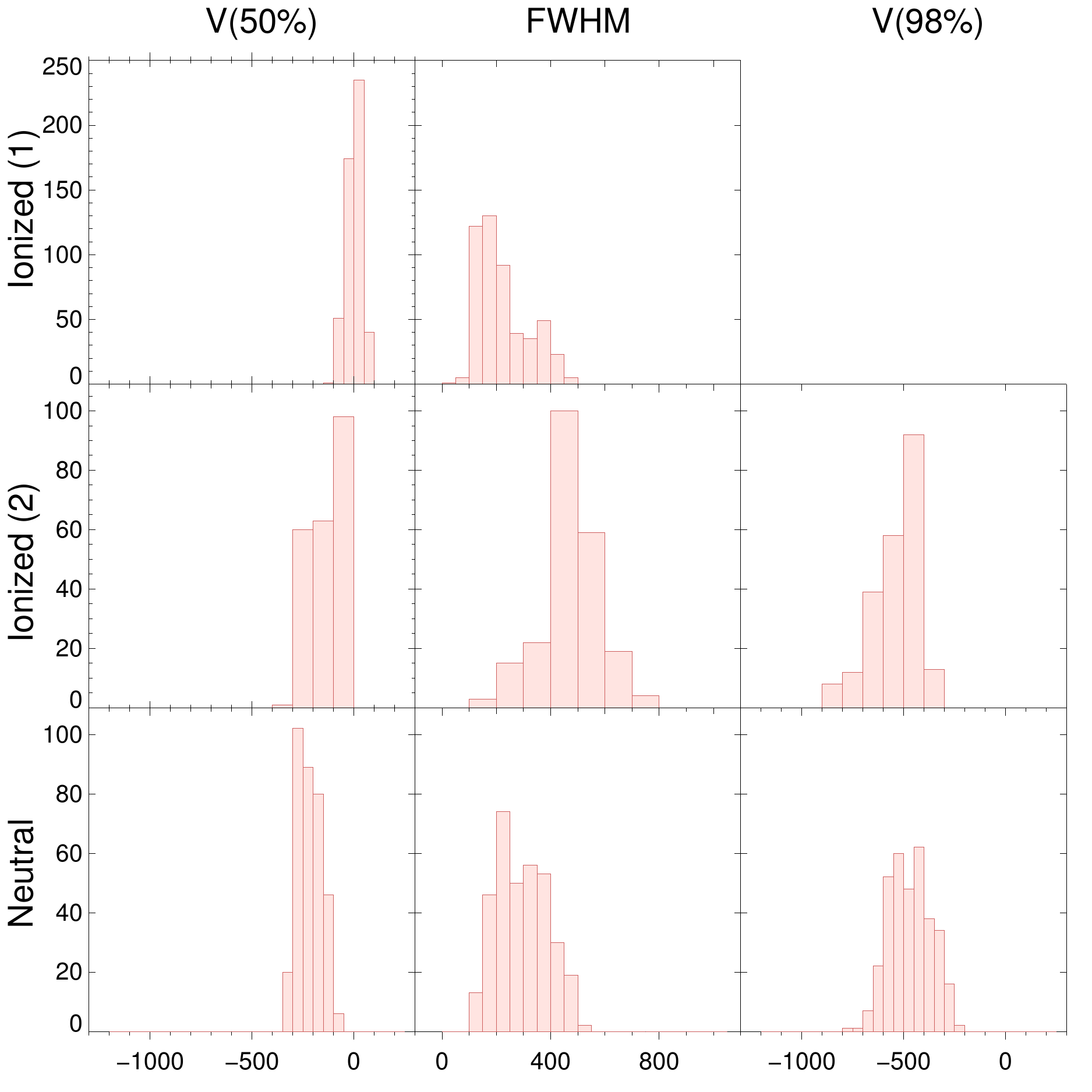}
  \caption{The same as Figure \ref{fig:veldist_f08572nw}, but for
    F10565$+$2448.}
  \label{fig:veldist_f10565}
\end{figure}

The IFS data for this galaxy, a triple system dominated in luminosity
by one member, was first presented in \citet{shih10a}. We review the
nature of the power source in this system and revisit the
interpretation of the IFS data based on a new reduction and analysis
with the current data pipeline and a finer binning (0\farcs3 spaxels,
vs. 0\farcs6 in \citealt{shih10a}). Figures
\ref{fig:map_cont_f10565}$-$\ref{fig:veldist_f10565} show maps and
velocity distributions for this system.

\subsubsection{Power Source}

Most of the luminosity in this triple, at all wavelengths, arises in
the westernmost nucleus. A second and third nucleus are located
8\farcs0 SE \citep{scoville00a} and 25\farcs8 E \citep{murphy96a} of
the W nucleus. The E and SE nuclei are not detected in 1.49~GHz VLA
images \citep{condon90a} or Chandra X-ray images
\citep{iwasawa11a}. All three nuclei have optical peaks in $R$, with
the W nucleus being $2\times$ as bright as the E nucleus in 2~kpc
square apertures \citep{armus90a}, and the sum of the W$+$SE nuclei
being $4\times$ as bright overall as the E galaxy
\citep{murphy96a}. The W nucleus also contains 95\%\ of the system's
total \ha\ flux (uncorrected for extinction; \citealt{armus90a}). In
the $K$ band, the W$+$SE nuclei are together $5.6\times$ as bright as
the E nucleus (based on 2MASS $K_{20}$ magnitudes retrieved via NED),
and the W nucleus is in turn significantly brighter than the SE
nucleus \citep{murphy96a}.

Judging from {\it Infrared Space Observatory} images, the E nucleus is
very faint compared to the W$+$SE complex at 7 and 15\micron\
\citep{dale00a}. Most of this MIR flux in turn arises from the W
nucleus, as seen in {\it Spitzer} archival images at
$4-24$~\micron. The W nucleus is the source of the prodigious FIR
emission from this system, which is obvious from archival {\it
  Spitzer} 70 and 160\micron\ images.

The E nucleus is a red spheroidal system with no star formation, as
evident from \hst\ optical images and the archival SDSS spectrum,
which show a smooth light distribution and an absorption-line
spectrum. The exact nature of the SE nucleus is unknown, but its blue
color is apparent in Figure \ref{fig:map_cont_f10565}. The W nucleus
is classified optically as \ion{H}{2}/composite
\citep{veilleux95a,yuan10a}, but it shows no high-excitation MIR
emission lines \citep{farrah07a} or X-ray signatures of an AGN
\citep{teng10a,iwasawa11a}. Based on PAH strength and MIR spectral
shape, \citet{veilleux09a} determine an AGN fraction of 0.17.

We conclude that a starburst in the W nucleus of F10565$+$2448 is
producing most of the energy in the system and that an AGN, if any is
present, contributes very little to this energy output.

\subsubsection{Kinematics} \label{sec:f10565_kin}

As discussed in \citet{shih10a}, the ionized gas rotates in the same
sense as the molecular gas, in a disk of PA $\sim$ 100\arcdeg\ and $i
\sim 20\arcdeg$ \citep{downes98a}. The near side of the disk is not
obvious because of the irregular morphology of the galaxy, but, as we
discuss below, it is likely to be the N side.

A neutral outflow was discovered in this system via nuclear spectra
\citep{heckman00a,martin05a,rupke05b}, and it is visible in molecular
gas, as well \citep{sturm11a}. \citet{martin06a} showed that this
neutral outflow extends eastward at least 8~kpc from the nucleus,
while \citet{shih10a} showed that its projected extent is at least
4~kpc from the nucleus in most other directions, as
well. \citet{shih10a} found that the neutral gas outflow is dusty and
has a velocity profile inconsistent with a symmetric minor-axis
outflow. They concluded that the outflow may be decelerated in the E
due to ambient interstellar gas, since significantly higher velocities
are observed W of the nucleus.

\begin{figure}
  % \plotone{f16.eps}
  \centering \includegraphics[width=6.5in]{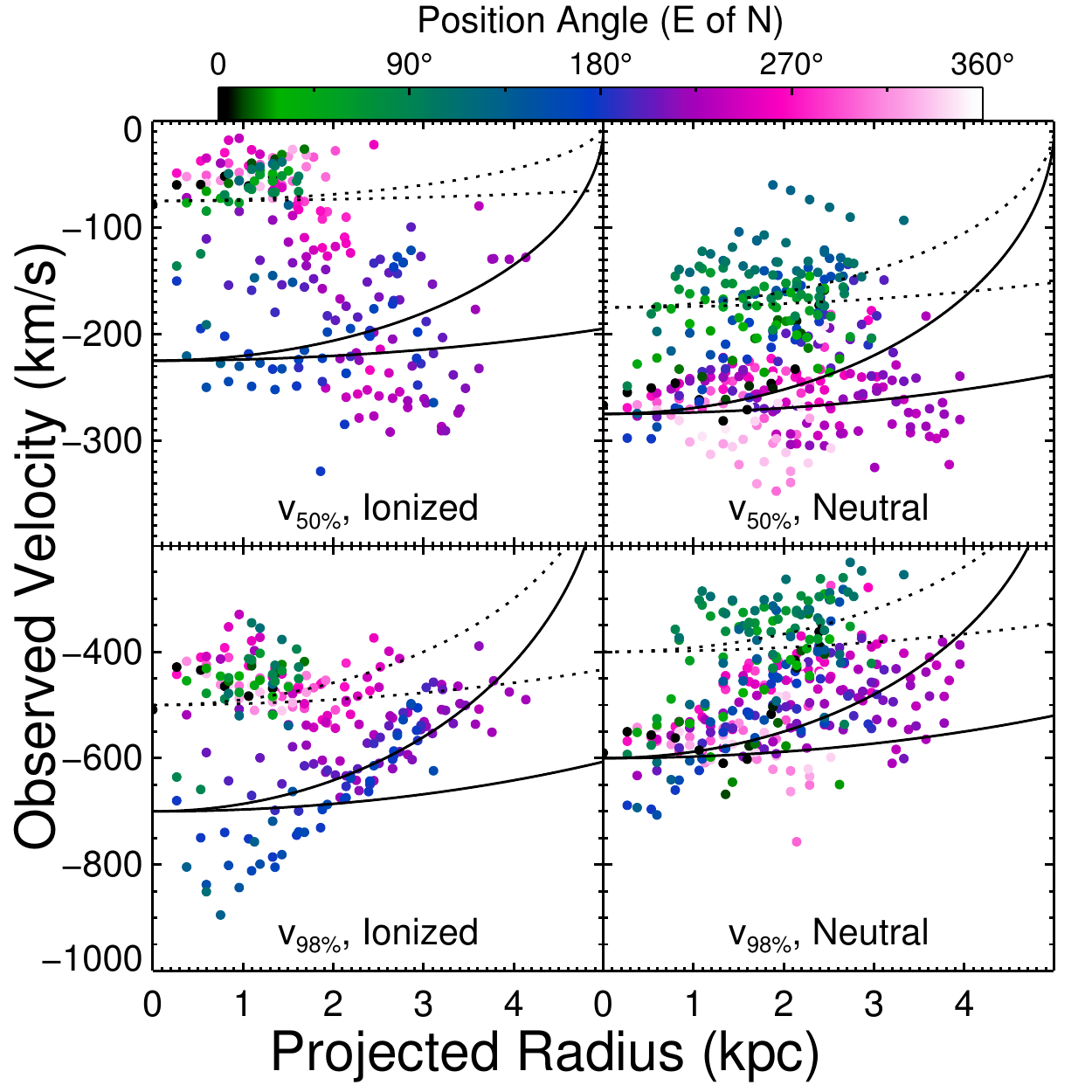}
  \caption{Observed velocity (\vfifty\ and \vtsig) vs. projected
    galactocentric radius in the outflowing ionized (broad component
    only) and neutral gas of F10565$+$2448. Each point represents one
    spaxel, and the position angle (E of N) of the spaxel with respect
    to the outflow center is represented by its color. The lines
    represent single radius free wind models (\S\,\ref{sec:models}) at
    constant velocity for $R = 5$~kpc and 10~kpc (the velocity of each
    model is given by the value at $R_{proj} = 0$~kpc). The flat
    profiles of \vfifty\ imply large wind radii ($\ga$5~kpc), and the
    strong asymmetry E and W of the disk requires asymmetric
    acceleration. The declining \vtsig\ values either place an upper
    limit on the wind radius ($\la$10~kpc) or reflect the decrease in
    unresolved (turbulent) motions with radius. (See
    \S\,\ref{sec:f10565_kin} for more details.)}
  \label{fig:vel_v_rad_f10565}
\end{figure}

We verify the extent and strongly asymmetric velocity structure of the
neutral and ionized gas found by \citet{shih10a}. The neutral gas
velocity varies slowly with projected galactocentric radius, as we
show in Figure~\ref{fig:vel_v_rad_f10565}. At a given position angle
(E of N with respect to the galaxy center), these figures show either
flat velocity profiles or perhaps a modest decline in \vfifty\ with
projected radius. SRFW models are able to reproduce this behavior as
long as the velocity is quite different E and W of the nucleus and the
wind radius is large enough (at least 5~kpc). The radial profiles of
\vtsig\ do show a noticeable decline with radius. If both \vfifty\ and
\vtsig\ represent bulk flow of gas at similar locations in the wind,
then this suggests that the gas radius is also not too large
($\la$10~kpc). Alternatively, the changes in \vtsig\ may reflect the
decline in unresolved bulk or turbulent motions with radius.

Compared to the analysis of \citet{shih10a}, the current study
increases the spatial information available for this source by Nyquist
sampling the seeing disk. With 0\farcs3 spaxels, a minor axis outflow
on scales of $\sim$1~kpc is evident in the data
(Figure~\ref{fig:dustpeak_f10565}). The primary evidence for this is
in the morphology and extent of the largest FWHM, highest velocity
regions in the neutral and ionized gas. If we assume that the blue
clumps to the N of the nucleus represent unobscured star forming
regions in the near side of a galaxy disk, then the high velocity
regions extend directly S along the minor axis of the CO nuclear
disk. We would expect blueshifted emission on the S side of the disk,
as is observed. The angle that the wind makes with respect to the line
of sight should be near that of the disk's inclination
($\sim$20\arcdeg; \citealt{downes98a}). The minor-axis flow must then
be relatively collimated (total opening angle, from one edge of the
wind to the other, of $\la2\times20\arcdeg$), since we observe it
primarily in projection away from the line of sight.

The regions of highest velocity are strikingly similar in morphology
to the dustiest regions of the disk, if we assume that the \hst\ color
just S of the nucleus is due to variable obscuration (consistent with
the patchy and filamentary structures of the colors in this
region). The largest values of FWHM in the gas (up to 800~\kms) are
coincident with the dustiest region of the galaxy (and region of
highest \nad\ equivalent width). Dust filaments extend S along the
minor axis, coincident with regions of high velocity, suggesting that
dust is being entrained in the wind along this direction.

\subsection{Mrk 231}

Mrk~231 is a late stage merger in which the nuclei have already
coalesced \citep{surace98a,veilleux02a}. It is the nearest QSO
\citep{boksenberg77a}, as well as the nearest broad absorption line
(BAL) QSO \citep{adams72a}. The kpc-scale neutral outflow in this
system was discovered by \citet{rupke05c}, and was more recently found
to extend in all directions \citep{rupke11a} and have a significant
molecular component \citep{fischer10a,feruglio10a,aalto12a,cicone12a}.

Mrk~231 is dominated by a central point source (3$\times$ brighter
than the host at $H$; \citealt{veilleux06a}) that has a QSO spectrum
with broad Balmer and Fe lines \citep{boksenberg77a}. It also hosts a
massive CO disk \citep{downes98a} with a high star formation rate,
based on the presence of PAH emission, \netl\ emission beyond that
required for an AGN, and MIR colors
\citep{genzel98a,veilleux09a}. The AGN produces $\sim$70\%\ of the
galaxy's luminosity \citep{veilleux09a}.

\begin{figure}
  % \plotone{f18.eps}
  \centering \includegraphics[width=6.5in]{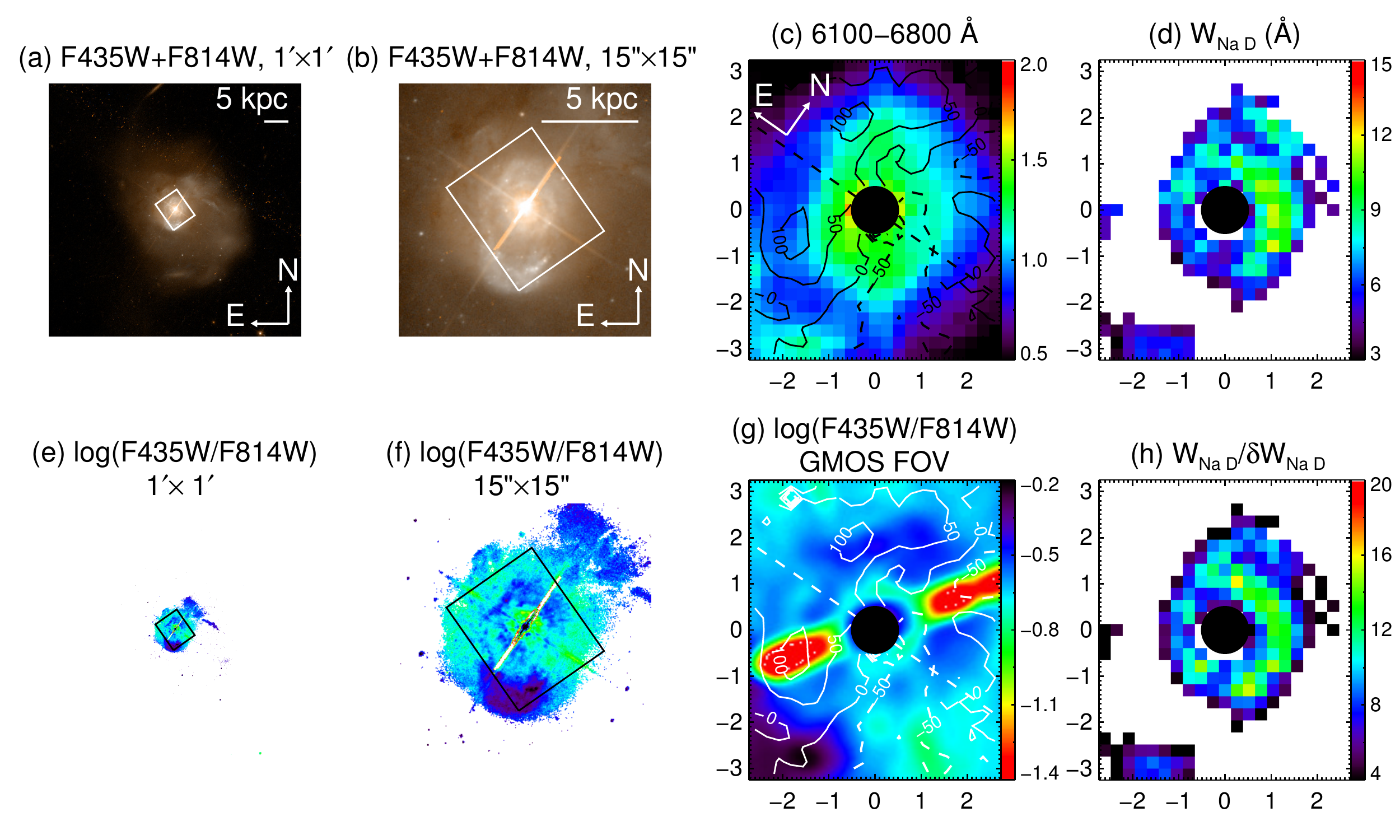}
  \caption{The same as Figure \ref{fig:map_cont_f08572nw}, but for
    Mrk~231. Velocities are relative to $z_{sys} = 0.0422$. The dashed
    line is along the CO line of nodes (\citealt{downes98a}; PA $=$
    90\arcdeg\ E of N).}
  \label{fig:map_cont_mrk231}
\end{figure}

\begin{figure}
%  \plotone{f19.eps}
  \centering \includegraphics[width=6.5in]{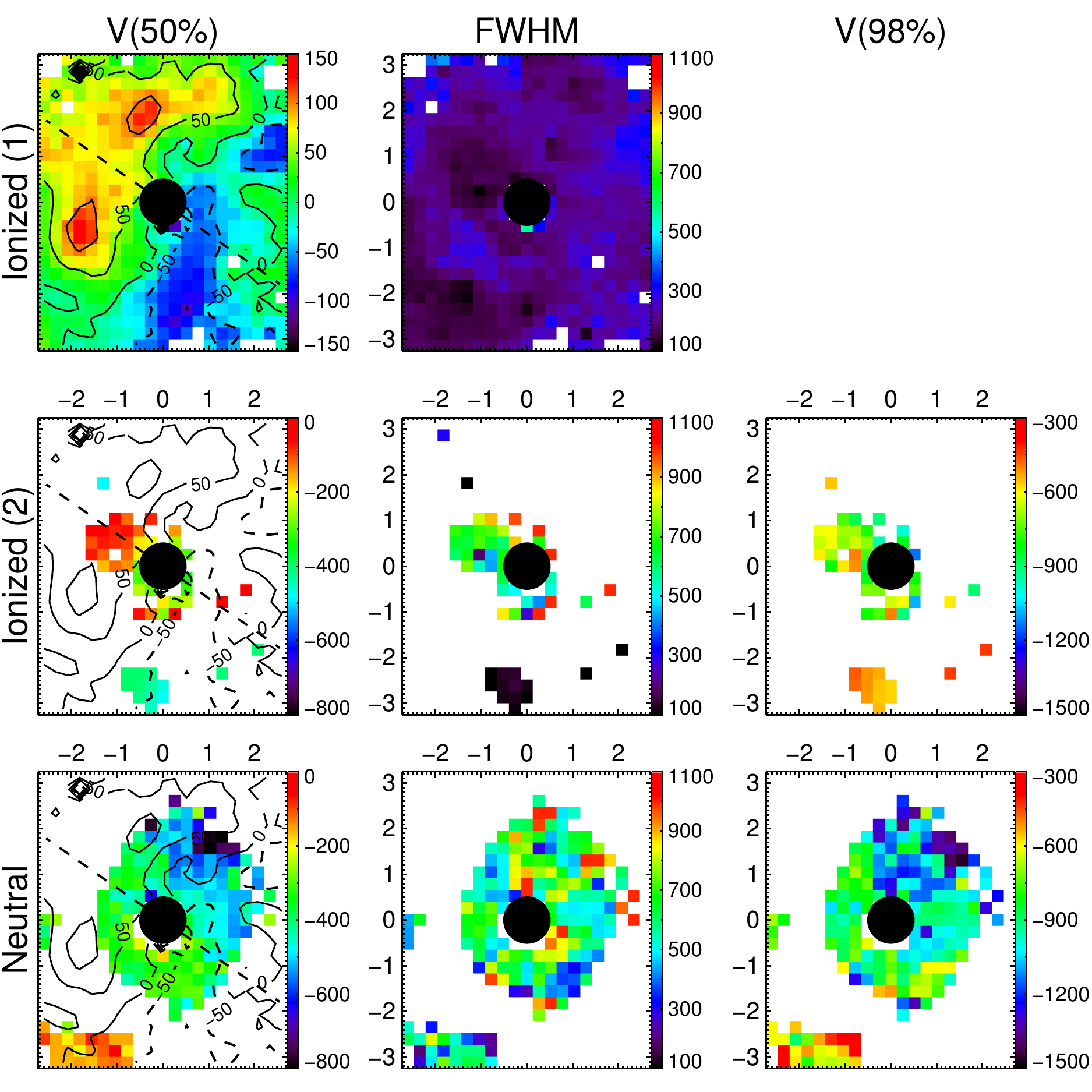}
  \caption{The same as Figure \ref{fig:map_vel_f08572nw}, but for
    Mrk~231.}
  \label{fig:map_vel_mrk231}
\end{figure}

\begin{figure}
%  \plotone{f20.eps}
  \centering \includegraphics[width=6.5in]{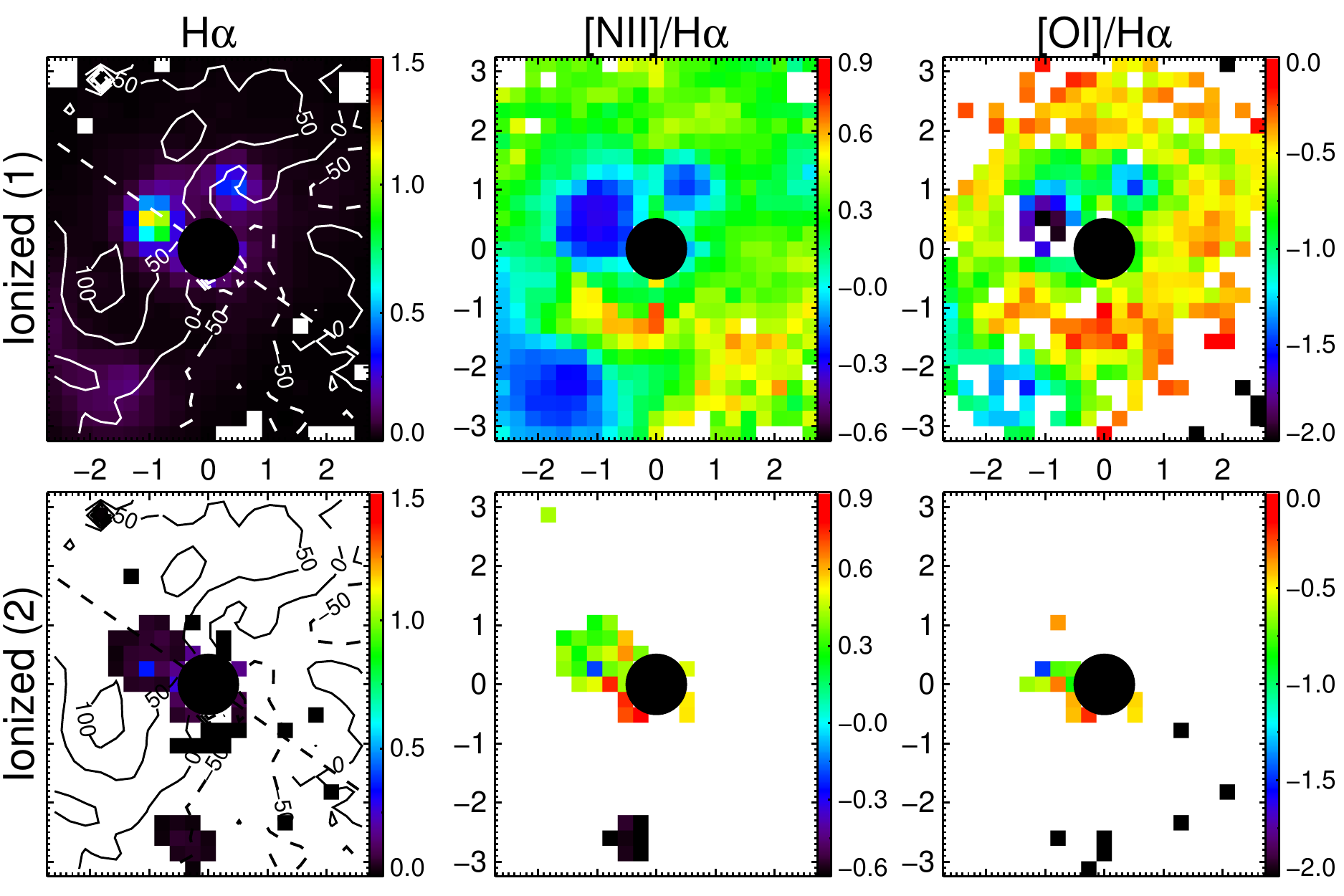}
  \caption{The same as Figure \ref{fig:map_lrat_f08572nw}, but for
    Mrk~231.}
  \label{fig:map_lrat_mrk231}
\end{figure}

\begin{figure}
%  \plotone{f21.eps}
  \centering \includegraphics[width=6.5in]{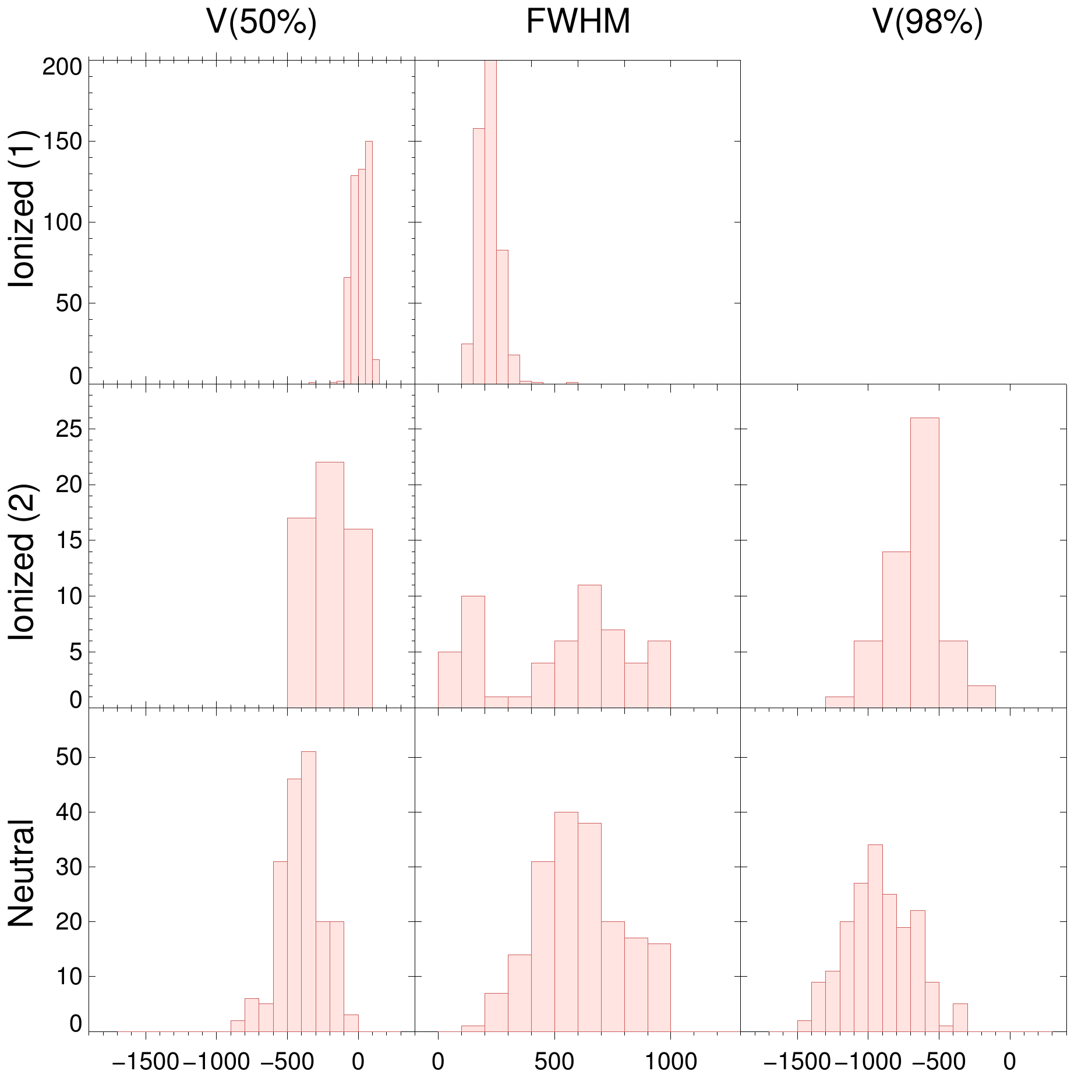}
  \caption{The same as Figure \ref{fig:veldist_f08572nw}, but for
    Mrk~231.}
  \label{fig:veldist_mrk231}
\end{figure}

The kinematics of the disk rotation and outflow in Mrk~231 are
discussed in \citet{rupke11a}. We present the entirety of the Mrk~231
data set here for easy comparison with the rest of our sample, but do
not discuss it further. Figures
\ref{fig:map_cont_mrk231}$-$\ref{fig:veldist_mrk231} show maps and
velocity distributions for this system.

\subsection{Mrk 273} \label{sec:mrk273}

\begin{figure}
  % \plotone{f22.eps}
  \centering \includegraphics[width=6.5in]{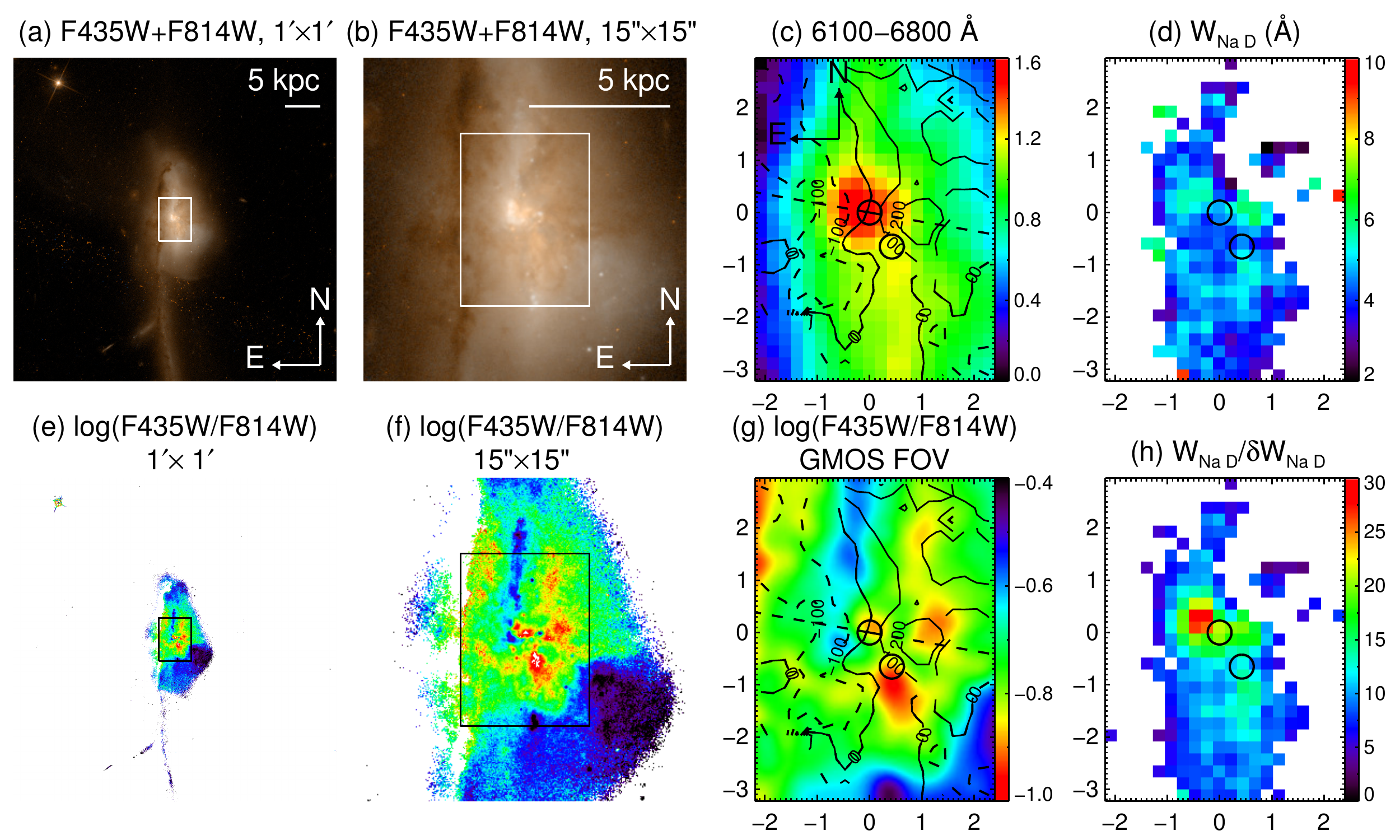}
  \caption{The same as Figure \ref{fig:map_cont_f08572nw}, but for
    Mrk~273. The open circles are the locations of the NIR
    nuclei. Velocities are relative to $z_{sys} = 0.0373$. The dashed
    line is along the approximate line of nodes of the gas disk around
    the N nucleus (\citealt{downes98a,cole99a,wilson08a}; PA $\sim$
    80\arcdeg\ E of N).}
  \label{fig:map_cont_mrk273}
\end{figure}

\begin{figure}  
%  \plotone{f23.eps}
  \centering \includegraphics[width=6.5in]{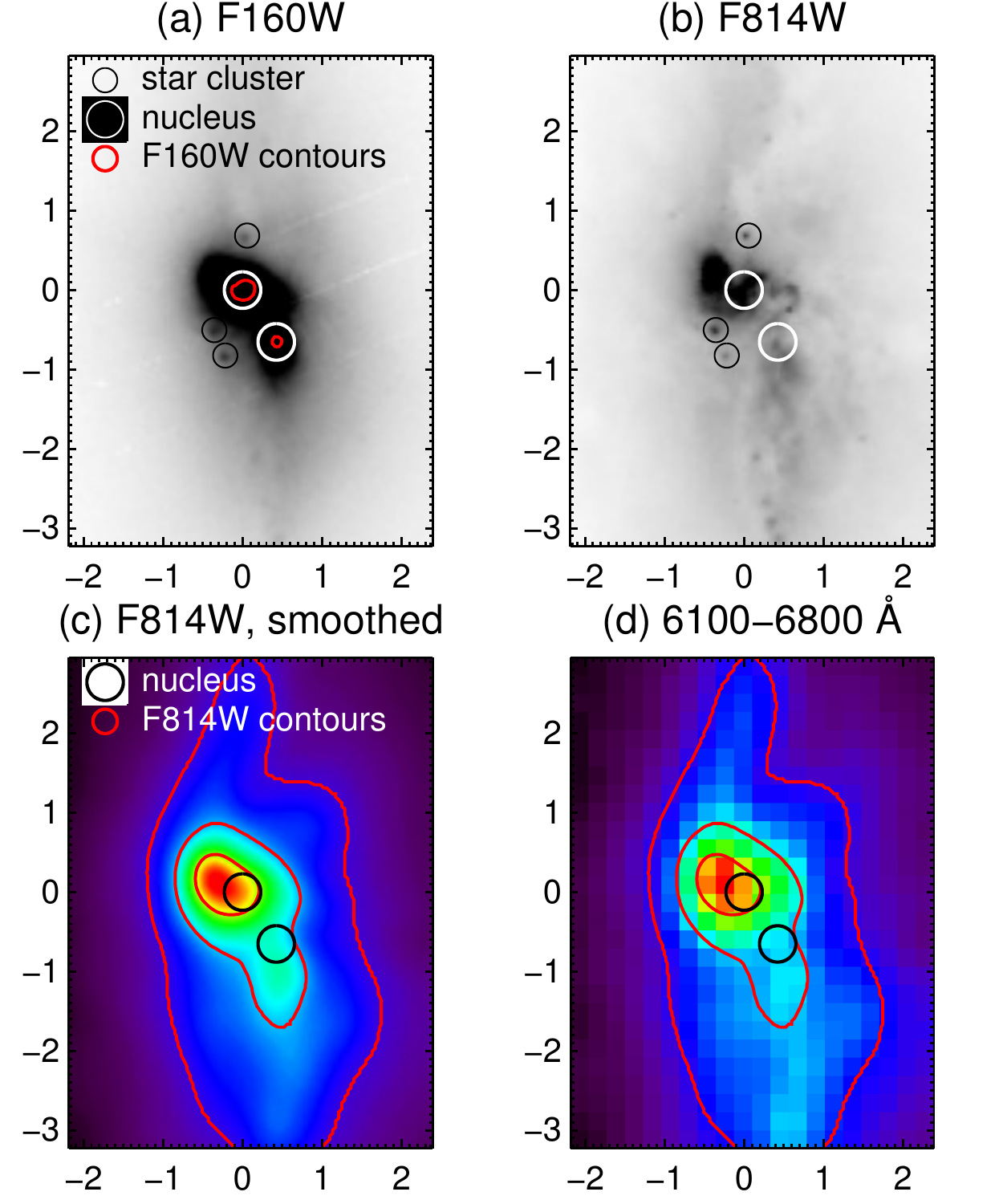}
  \caption{The same as Figure \ref{fig:register_f08572nw}, but for
    Mrk~273.}
  \label{fig:register_mrk273}
\end{figure}

\begin{figure}
%  \epsscale{0.8}
%  \plotone{f24.eps}
  \centering \includegraphics[width=5in]{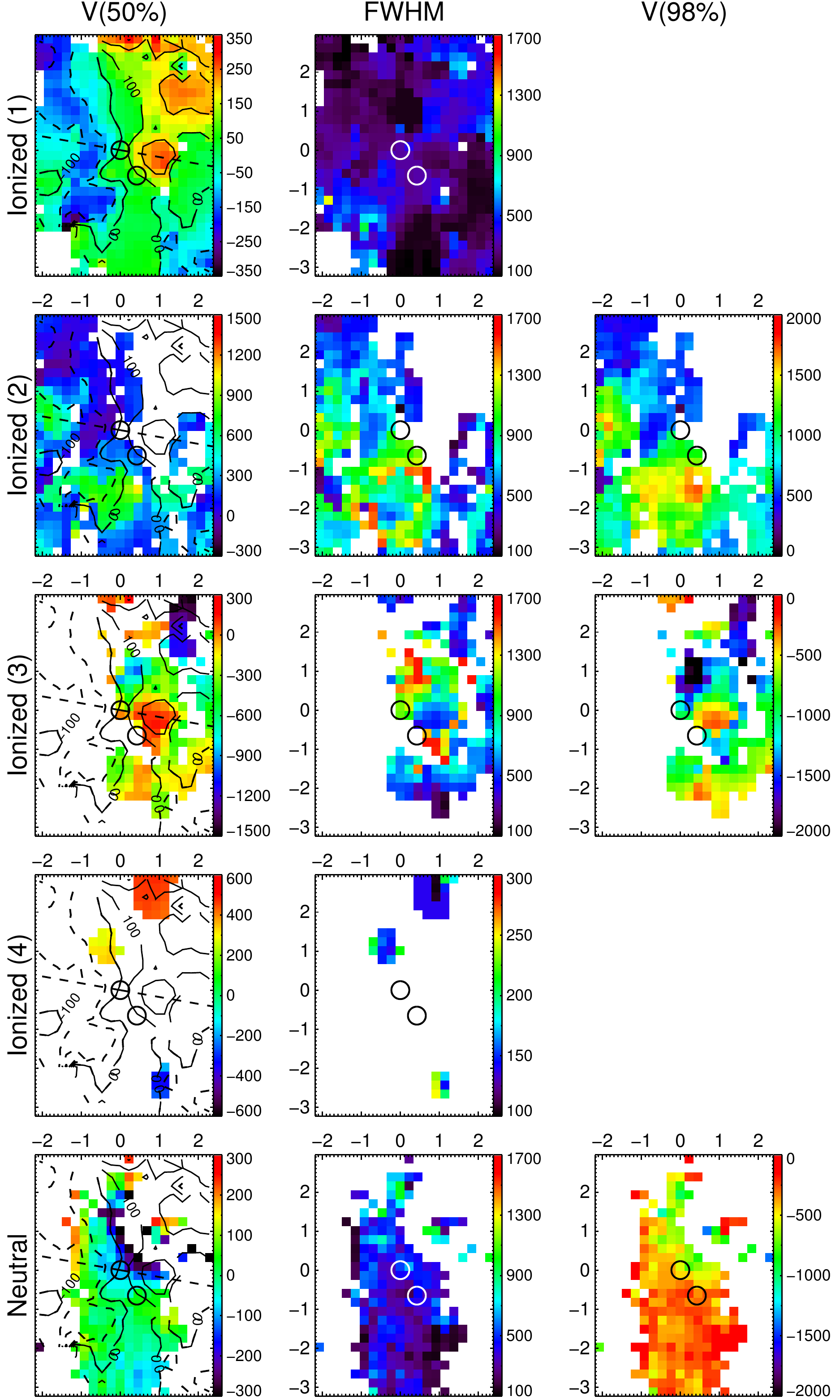}
  \caption{\scriptsize The same as Figure \ref{fig:map_vel_f08572nw},
    but for Mrk~273. For this sytem, the four ionized gas components
    are: (1) rotating, (2) broad and redshifted, (3) broad and
    blueshifted, and (4) isolated clouds with narrow profiles
    (\S\,\ref{sec:mrk273-kinematics}). For component 2, \vtsig\ is
    computed by adding, rather than subtracting, 2$\sigma$
    (\S\,\ref{sec:maps}).}
  \label{fig:map_vel_mrk273}
\end{figure}

\begin{figure}
%  \epsscale{0.8}
%  \plotone{f25.eps}
  \centering \includegraphics[width=6.5in]{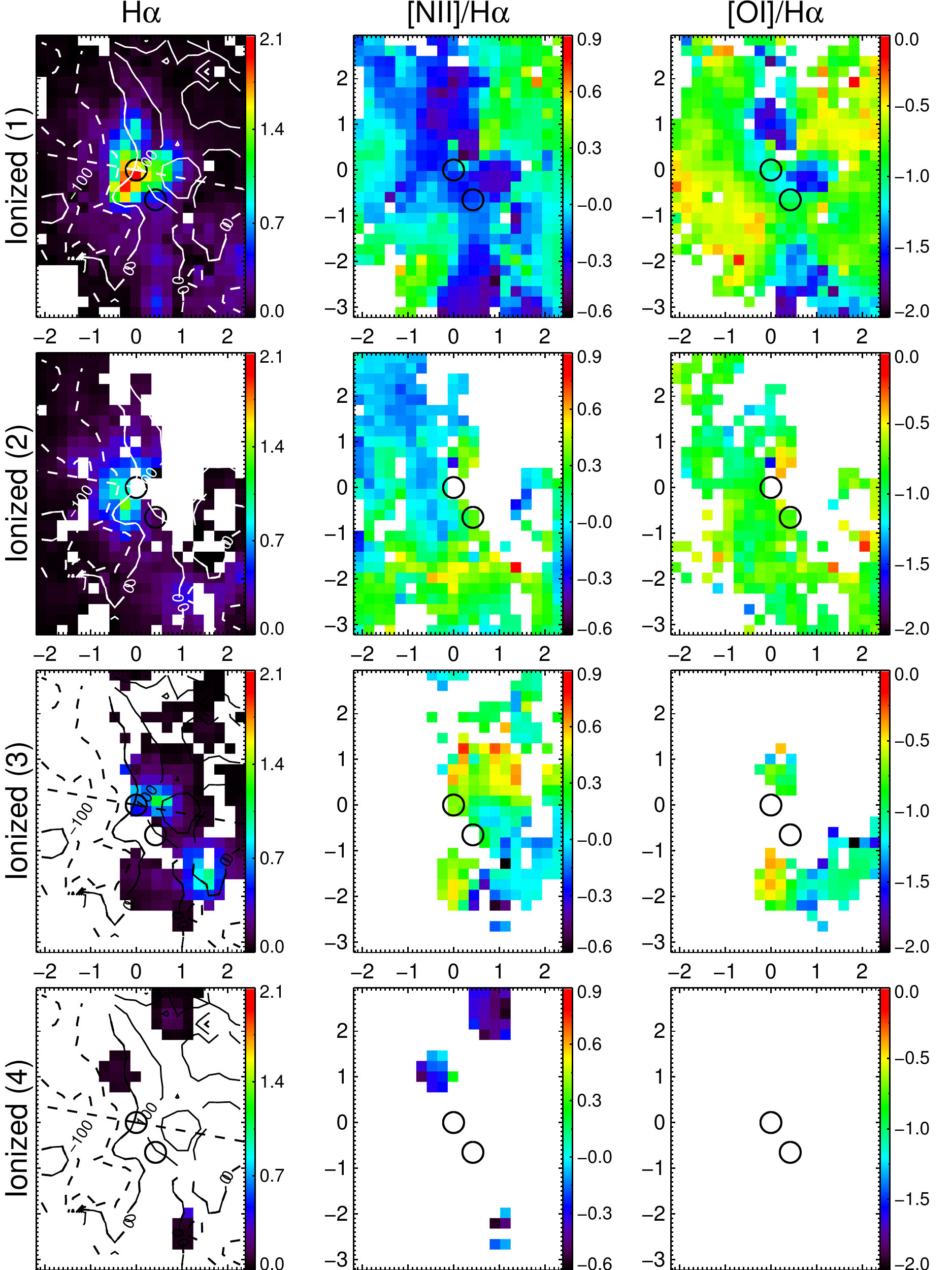}
  \caption{The same as Figure \ref{fig:map_lrat_f08572nw}, but for
    Mrk~273.}
  \label{fig:map_lrat_mrk273}
\end{figure}

\begin{figure}
%  \epsscale{0.8}
%  \plotone{f26.eps}
  \centering \includegraphics[width=5in]{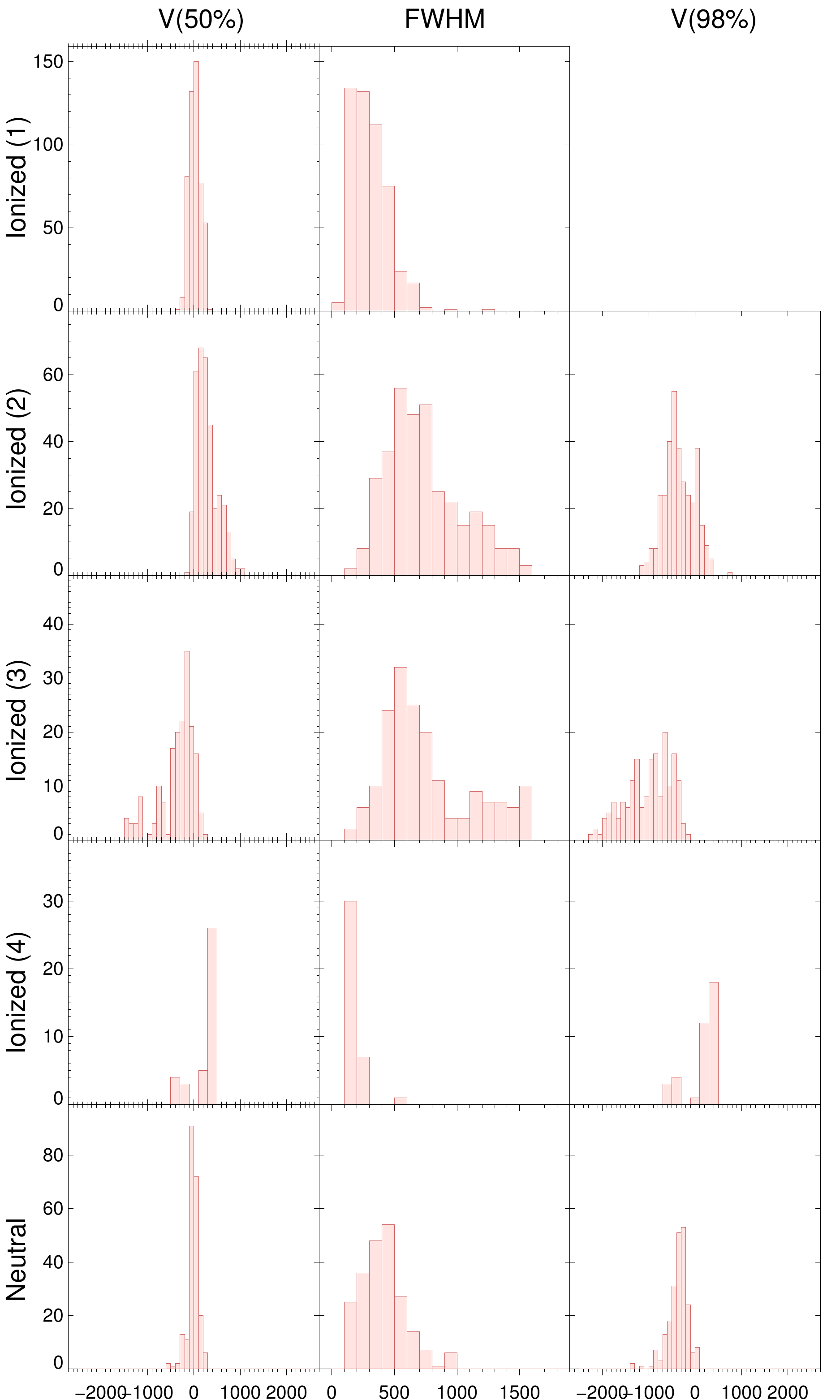}
  \caption{The same as Figure \ref{fig:veldist_f08572nw}, but for
    Mrk~273.}
  \label{fig:veldist_mrk273}
\end{figure}

Mrk~273 is a late merger with a binary nucleus and an AGN. The
bolometric luminosity of the galaxy is probably dominated by a
starburst, but the AGN contribution is still significant. The
kinematics in this system are complicated, but clearly show a
high-velocity outflow.  Maps of derived properties and velocity
distributions can be found in Figures
\ref{fig:map_cont_mrk273}$-$\ref{fig:veldist_mrk273}.

\subsubsection{Power Source}

The nuclear properties of Mrk~273 are enigmatic. Two nuclei (typically
labeled N and SW) are evident in the NIR and MIR
(\citealt{majewski93a,knapen97a,scoville00a,soifer00a}; see
Figure~\ref{fig:register_mrk273}). The N nucleus contains most of the
molecular gas in the system within an extremely compact core (radius
$\la120$~pc; \citealt{downes98a}), and is the source of most of the
MIR luminosity \citep{soifer00a}. In the radio, Mrk~273 resolves into
a diffuse radio source with two primary peaks \citep{carilli00a}. The
N radio peak (coincident with the N nucleus) is further resolved as
multiple sources at the highest resolutions, and the SE radio peak is
inconsistent with an AGN based on spectral slope \citep{bondi05a}. The
optical line emission from the center of Mrk~273 classifies it as a
Seyfert 2 \citep{koski78a,veilleux99a}. Using IFS, \citet{colina99a}
locate the Seyfert 2 emission near, but to the SW of, the two nuclei.

Recently, the SW nucleus has been identified as an AGN through
localization of the hard X-ray point source \citep{iwasawa11a}, which
is Compton thick based on spectral modeling from $2-40$~keV
\citep{teng09a}. The AGN shows strong high-ionization narrow-line
emission in the NIR and MIR: possible [\ion{Si}{6}]
\citep{veilleux99b} and strong [\ion{Ne}{5}] and [\ion{O}{4}]
\citep{genzel98a,armus07a,veilleux09a}. The ratios of these
high-ionization lines to a low-ionization line, [\ion{Ne}{2}], point
to a dominant contribution of the AGN to the FIR luminosity
\citep{veilleux09a}. Combined PAH and emission-line diagnostics
suggest that the starburst and AGN contribute similarly
\citep{genzel98a,veilleux09a}, while MIR continuum diagnostics point
to a dominant starburst \citep{veilleux09a}.

NIR IFS of the inner kiloparsec of Mrk~273 localizes high-ionization
coronal line emission ([\ion{Si}{6}]) to the SW nucleus and the SE
radio source \citep{u12a}. These new data suggest that the N nucleus
may harbor a heavily-obscured AGN (see also \citealt{iwasawa11a}) that
photoionizes the gas in the SE radio source. Alternatively,
high-velocity shocks may produce this emission. Regardless, the
existence of an obscured AGN in the N nucleus remains to be
investigated.

The concentration of CO and MIR continuum emission in the northern
nucleus would seem to imply that the SW AGN contributes little to the
bolometric luminosity of this dusty system. However, MIR dianostics
suggest otherwise, with an average AGN fraction of 0.46
\citep{veilleux09a}. This conclusion relies on the combined MIR
spectrum of both nuclei, and rests primarily on the high emission-line
ratios. The bolometric AGN luminosity inferred this way is entirely
consistent with the X-ray emission of the SW nucleus if we compare it
to optically-selected QSOs. If the AGN luminosity were lower by more
than a factor of $2-3$, then $L_{2-10~keV}/L_{AGN}$ would be too high
\citep{teng10a}.

\subsubsection{Basic Kinematics} \label{sec:mrk273-kinematics}

The gas kinematics in Mrk~273 are the most complex in our sample. The
atomic and molecular gas kinematics centered around the N nucleus are
characterized at 100~pc to kpc scales by a rotating disk of gas
\citep{downes98a,cole99a,carilli00a,yates00a}. The position angle of
the line of nodes of this disk is $70-90\arcdeg$, and it is inclined
by $45-55$\arcdeg\ with respect to the line of sight
\citep{downes98a,cole99a,u12a}. The disk is rotating rapidly, with a
projected velocity gradient of $400-500$~\kms\ over a few hundred pc
(velocity increasing from east to west;
\citealt{downes98a,cole99a,carilli00a,yates00a}). On larger scales,
there is a molecular feature that extends 5~kpc N and one that extends
5~kpc S of the N nucleus; the N feature is redshifted by
$\sim$300~\kms, and the S feature is near systemic \citep{downes98a}.

There are significant departures from simple E-W rotation, from
sub-kpc to kpc scales. The spatially resolved CO lines are broad, and
the integrated profile spans 1000~\kms\ \citep{downes98a}. Two
components emerge in the \othl\ spectra of \citet{colina99a}, and they
decompose these into a systemic component and a component that is
redshifted in the E and blueshifted in the W, with a peak-to-peak
amplitude of 2400~\kms. They suggest that the latter component
represents a starburst-driven superwind.

The present IFS data represent a significant improvement in
sensitivity, spatial resolution, and spectral resolution over the
\citet{colina99a} dataset, as well adding information on the neutral
gas phase.

It should be noted that the CO/\ion{H}{1} gas disk is tiny compared to
the FOV of our maps, subtending only $1-2$ GMOS spaxels
\citep{downes98a,carilli00a}.

\begin{figure}
  %\plotone{f27.eps}
  \centering \includegraphics[width=6.5in]{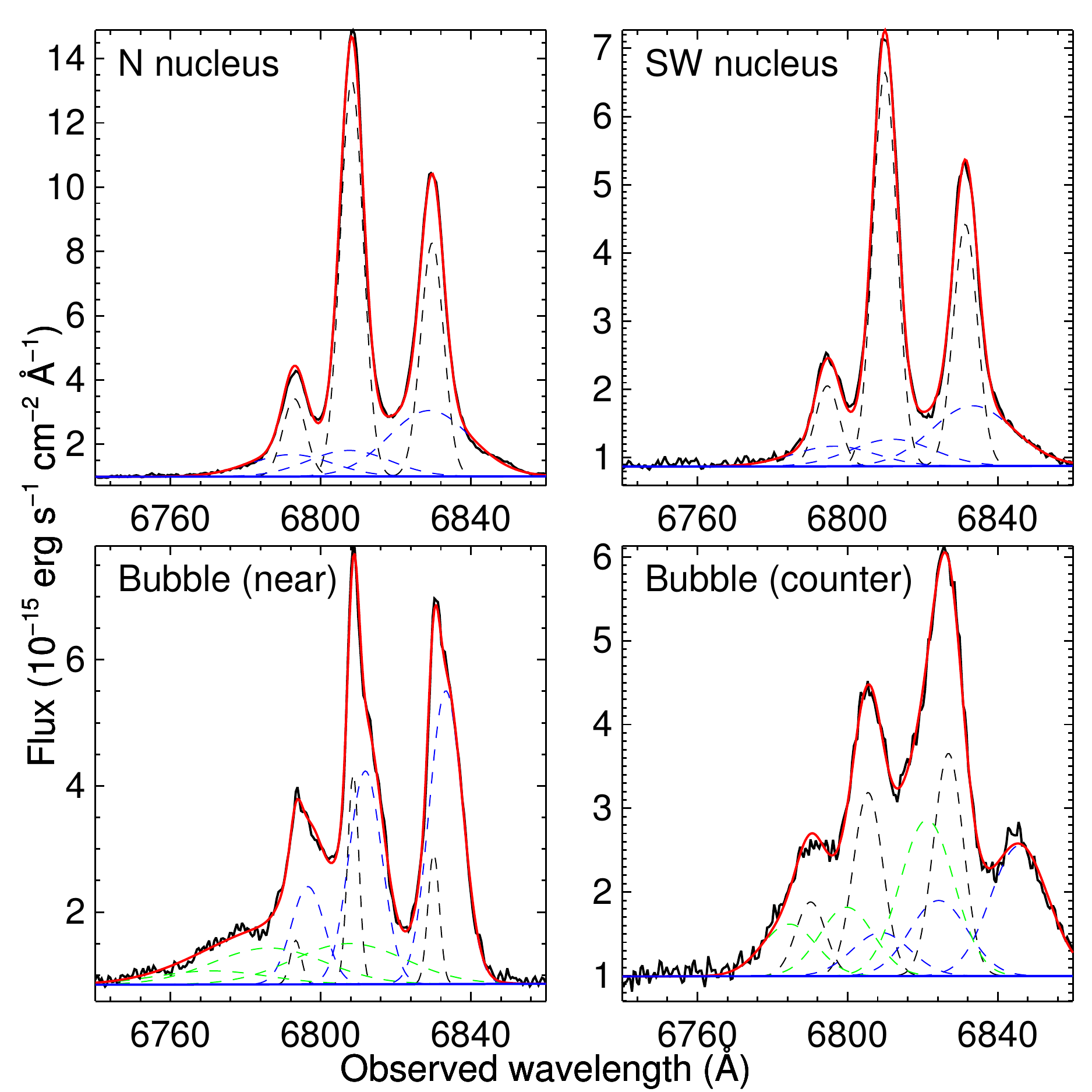}
  \caption{\ha$+$\ntll\ spectra of four spaxels in Mrk~273: nearest
    the N and SW nuclei, in the near-side superbubble, and in the
    counter-propagating superbubble. The solid black line is the data,
    the solid red line the total fit, the solid blue line the
    continuum fit, and the dashed lines fits to individual lines. The
    different colors of the dashed lines represent different velocity
    components. Note the three components in each superbubble: the
    near- (blueshifted) and far- (redshifted) sides of the bubble, as
    well as the rotating component.}
  \label{fig:spectra_mrk273}
\end{figure}

\begin{figure}
  % \plotone{f28.eps}
  \centering \includegraphics[width=6.5in]{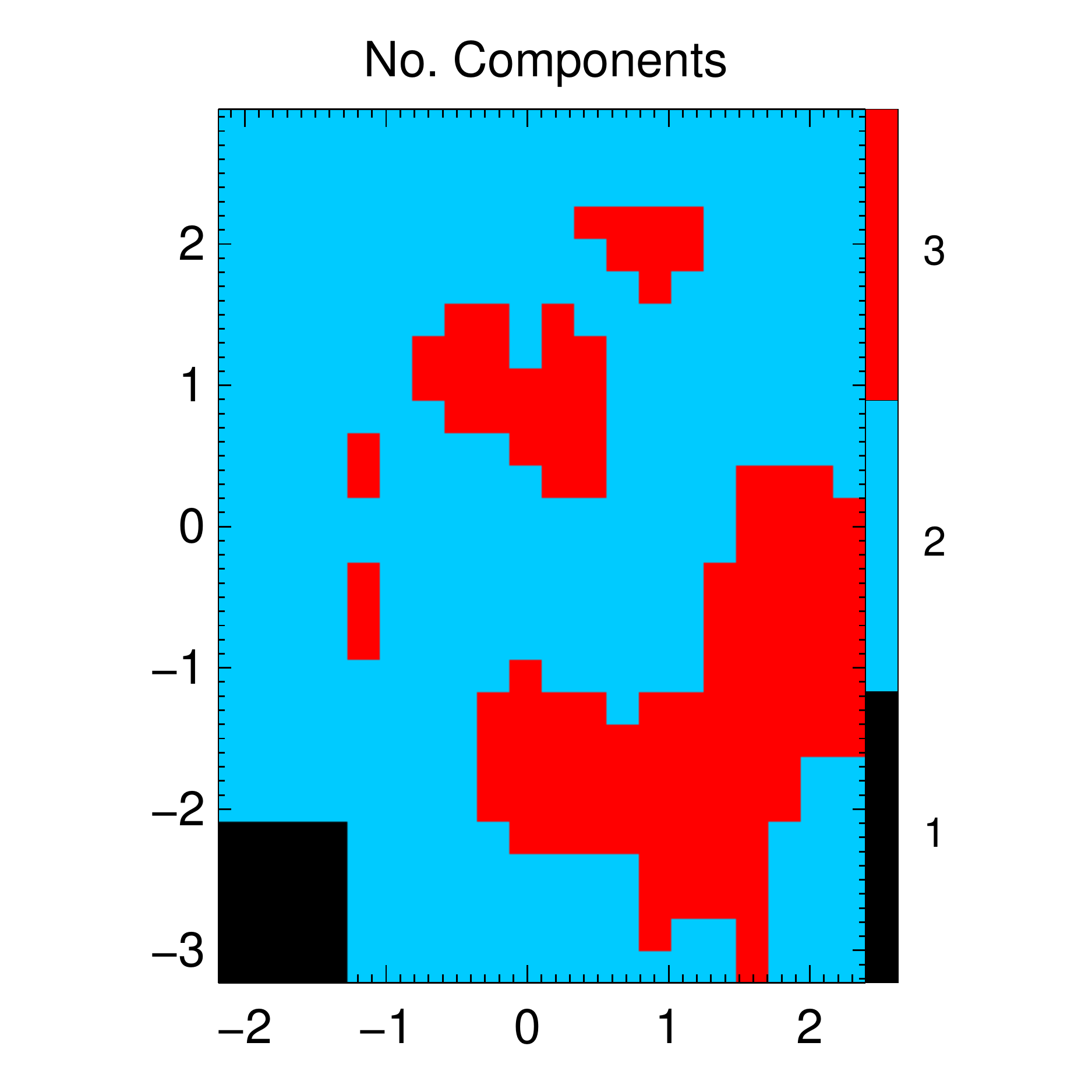}
  \caption{A color map of the number of emission-line components fit
    to each spaxel in Mrk~273. The axis labels are in kpc.}
  \label{fig:ncomp_mrk273}
\end{figure}

The ionized gas profiles are extremely broad in many spaxels, and are
in most cases best fit by two or three
components. Fig. \ref{fig:spectra_mrk273} shows several examples of
these fits, and Fig. \ref{fig:ncomp_mrk273} shows the number of
components in each spaxel. We order these components into one of four
categories as follows: (1) rotating, (2) broad and redshifted, (3)
broad and blueshifted, and (4) isolated clouds with narrow
profiles. In a two-component spaxel, we assume the narrower component
is rotating, though we tweak some spaxels by hand because the FWHM of
the two components are similar. The second component is then assigned
to the broad and redshifted bin or broad and blueshifted bin depending
on its velocity with respect to the rotating component. The isolated
clouds with narrow line profiles are easily distinguished from the
rotating narrow component because of the significant red- or
blueshifts of the isolated clouds, and from broad components by
linewidth and spatial footprint. If there are three components in a
spaxel, we generally choose the central component to be rotating,
though there are some exceptions.

The rotating ionized component follows the CO(1-0) rotation curve of
\citet{downes98a} very closely, including the twisting of the systemic
isovelocity contour, the peak redshifted velocity just W of the
nuclei, and the redshifted gas to the N. The rotating component is on
average narrow compared to the other components ($\langle$FWHM$\rangle
= 300$~\kms). The line ratios are low along a NNE/SSW axis, with
log(\nt/\ha) $\sim-0.6 - 0$ and log(\oo/\ha) in the range -1.8 to
-0.8. These ratios increase to 0.3 and -0.5 to the ESE and WNW.

The redshifted and blueshifted broad, ionized components are broadly
consistent with \citet{colina99a}: the blueshifted components lie
primarily W of the nuclei, while the redshifted components lie to the
E. However, our higher spatial and spectral resolution shows
significant substructure. Projected velocities rise away from the
nucleus, peaking at a remarkable $\vfifty = -1500$~\kms, $4-5$~kpc NW
of the nuclei. However, the broadest components (FWHM up to 1500~\kms)
with the highest \nt/\ha\ and \oo/\ha\ line ratios (indicative of
shock ionization) arise directly N and S of the nuclei. (These
locations also display modest peaks in $|\vfifty|$ of
$\sim1000$~\kms.)

The neutral gas in Mrk~273, as probed by \nad, is backlit by the
stellar continuum across most of the N-S extent of the GMOS FOV in a
band that is $1-2$~kpc wide. There are few departures from velocities
near systemic.  However, there is blueshifted gas north of the
nuclei. In the SRFW model calculations for this system, we include
only spaxels in which the neutral gas is blueshifted from the ionized
gas narrow component in that spaxel by more than 50~\kms.

The blueshifted ionized and neutral gas in the N/NW of the system is
clearly outflowing, as is the redshifted ionized gas in the S/SE. The
velocities (reaching $1500$~\kms\ in projection) are too high to be
explained simply by tidal motions, and are much larger than the
projected velocities in the narrow component (between $-350$ and
350~\kms). However, the redshifted and blueshifted ionized and neutral
components in other locations have more modest velocities (\vfifty\
and FWHM of a few hundred \kms) that could in principle be explained
by tidal motions.

The evidence for a bipolar outflow aligned E-W, as proposed by
\citet{colina99a}, is not the most likely explanation based on the
current data, though a variation on this theme emerges from our
analysis. The data conclusively point to a bipolar nuclear superbubble
in the center of the system, aligned N-S. Gas further from the
nucleus, if it is indeed outflowing, then traces gas blown out from
this bubble or earlier ejections from the nuclei.

\subsubsection{Nuclear Superbubble} \label{sec:mrk273-bsb}

\begin{figure}
%  \plotone{f30.eps}
  \centering \includegraphics[width=5in]{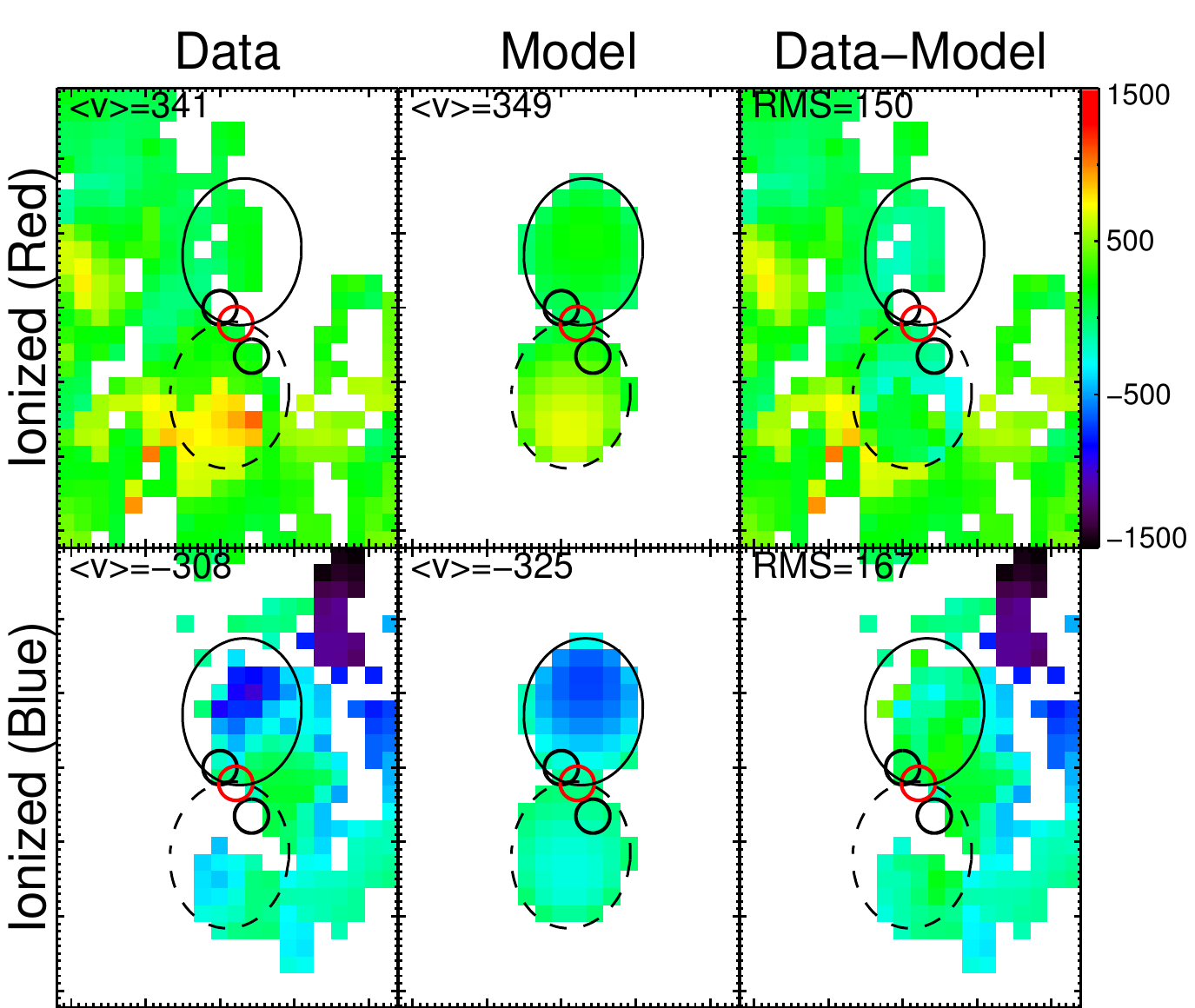}
  \caption{Color maps of superbubble model fits to the emission line
    velocity fields in Mrk~273. The fit shown is for $n = 2$. The
    three columns show the data (left), model (middle), and their
    difference (right). In each panel, the solid line outlines the
    near-side bubble, and the dashed line the counter-bubble. Small
    black circles locate the two nuclei, while the red circle is the
    best-fit outflow center, assuming symmetry along the outflow
    axis. The numbers in the left and middle columns give the average
    velocity of that map within the model confines, while the numbers
    in the right panel are the RMS velocity differences within the
    same area. Model parameters are given in
    Table~\ref{tab:modbbl}. (See \S\S\,\ref{sec:bsb} and
    \ref{sec:mrk273-bsb} for more details.)}
  \label{fig:modbbl_mrk273}
\end{figure}

Position-velocity (PV) diagrams of the line emission in the center of
Mrk~273 reveal the classic signature of a bipolar superbubble (Figure
\ref{fig:pv_e_mrk273}). Applying our BSB model (\S\,\ref{sec:bsb}) and
optimizing its parameters yields the values given in Table
\ref{tab:modbbl}. The model is overplotted in the PV diagrams, and the
two-dimensional projected velocity fields and bubble shapes are
compared to the data in Figure \ref{fig:modbbl_mrk273}. The bubble is
fairly symmetric in emission ($R_{xy}=0.8R_z$).

The absorption lines in Mrk~273 are not of high enough signal-to-noise
near the possible superbubble to accurately parameterize its neutral
gas content. There are higher absorption line velocities near the
edges of the near-side bubble in the N, suggesting that some neutral
material is entrained by an expanding bubble.

Overall, a bubble shape fits the inner radii of the wind well (at
projected radii $\la2$~kpc). However, the high velocities seen
immediately adjacent to it but at larger radii in the NW and SE
strongly suggest that the bubble is broken on its surface and gas is
escaping the bubble in some directions. This gas has as high or higher
velocities than in the bubble itself, perhaps indicating further
acceleration by a hot medium that could be blowing the bubble.

There are also areas of high velocity directly E and W of the nucleus,
along a position angle perpendicular to the bubble axis. These may
represent earlier episodes of gas ejection, or ejection due to an
unrelated process.

The fitted outflow center is formally located between the two nuclei,
0\farcs4 ($\sim$300~pc) from the N nucleus and 0\farcs7 ($\sim$500~pc)
from the SW nucleus. However, the errors in the bubble center are 1
spaxel (0\farcs3), so an origin at one or the other of the two nuclei
is plausible. If in fact the outflow arises between the two nuclei,
then it could be powered by a combination of energy from the AGN and
starburst. However, the outflow may also be asymmetric (extending
further on the near or far side due to, e.g., asymmetric
acceleration), lending further uncertainty to the true outflow center.

If the bubbles were centered on the N nucleus, their position angle
($-5\arcdeg$) would be consistent with collimation by the tilted
CO/\ion{H}{1} disk at that location (PA $70-90\arcdeg$;
\citealt{downes98a,cole99a}). The disk's inclination is $i=45\arcdeg$
from CO modeling \citep{downes98a}, which is inconsistent with the
superbubble inclination of 75$\pm$5\arcdeg. However, a higher disk
inclination is not out of the question, given the large velocity
gradients in the nucleus \citep{cole99a,carilli00a,yates00a}. A more
recent analysis of NIR IFS data, taken with adaptive optics, finds a
more inclined disk ($i\sim55\arcdeg$) based on ionized gas lines
\citep{u12a}.

The fitted outflow center is further from the SW nucleus. If the
bubbles were centered on the SW nucleus, it is unclear what would
collimate the outflow. Furthermore, the N disk has a large gas
reservoir to couple with the energy driving the outflow, while the SE
nucleus does not \citep{downes98a,cole99a,carilli00a}.

Our structural analysis is inconclusive on the power source of the
entire outflow (superbubbles $+$ extended flow), but more consistent
with the superbubbles being powered by the N nucleus than the SW
nucleus. Data of higher spatial resolution are needed, as well as
constraints on the presence of an obscured AGN in the N nucleus. As we
discuss below, however, the high velocities suggest an AGN-driven flow
(\S\,\ref{sec:comp-vel}).

\subsection{VV 705} \label{sec:vv705}

\begin{figure}
  % \plotone{f31.eps}
  \centering \includegraphics[width=6.5in]{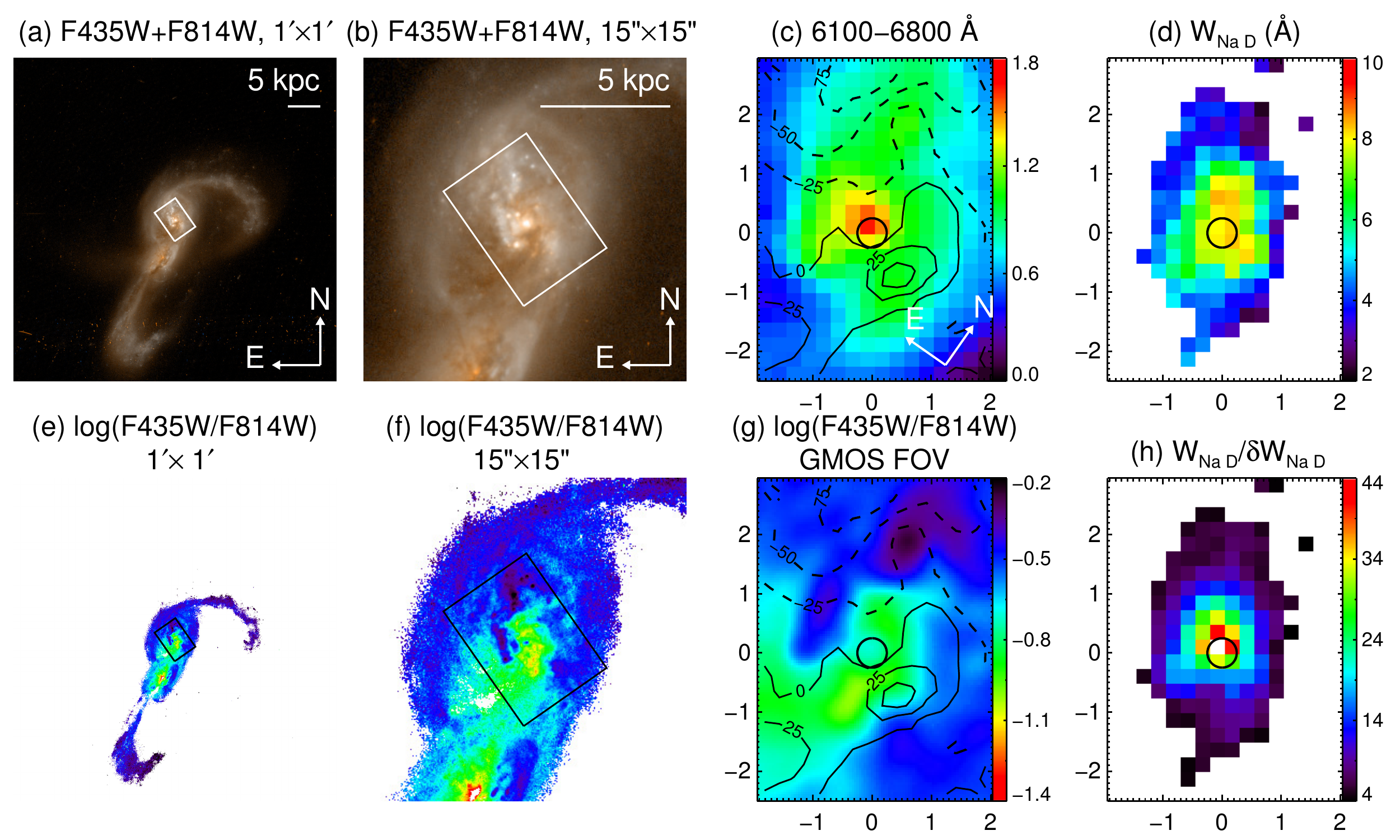}
  \caption{The same as Figure \ref{fig:map_cont_f08572nw}, but for
    VV~705:NW. Velocities are with respect to $z_{sys} = 0.04035$.}
  \label{fig:map_cont_vv705nw}
\end{figure}

\begin{figure}
  
  %\plotone{f32.eps}
  \centering \includegraphics[width=6.5in]{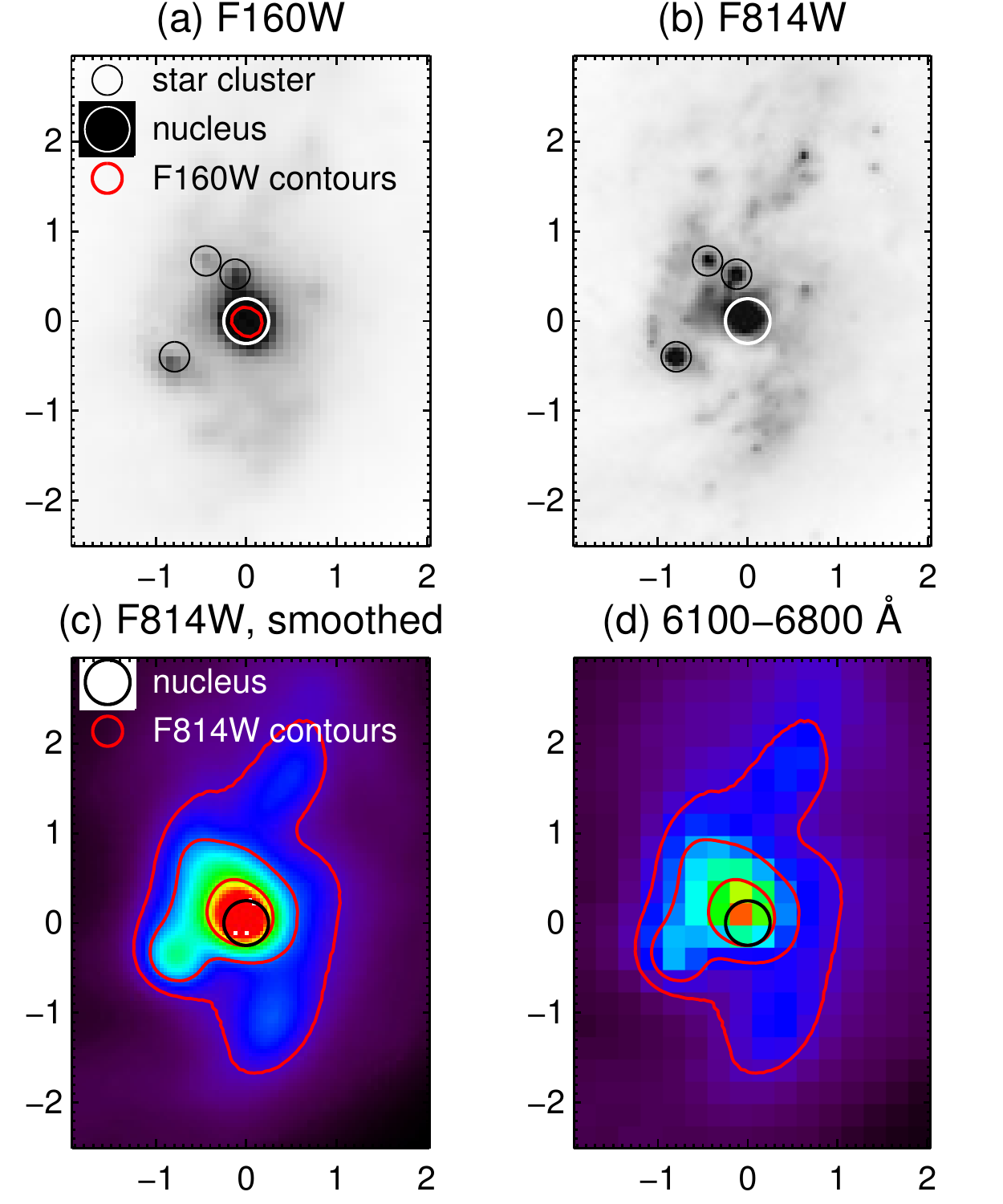}
  \caption{The same as Figure \ref{fig:register_f08572nw}, but for
    VV~705:NW.}
  \label{fig:register_vv705nw}
\end{figure}

\begin{figure}
  %\plotone{f33.eps}
  \centering \includegraphics[width=6.5in]{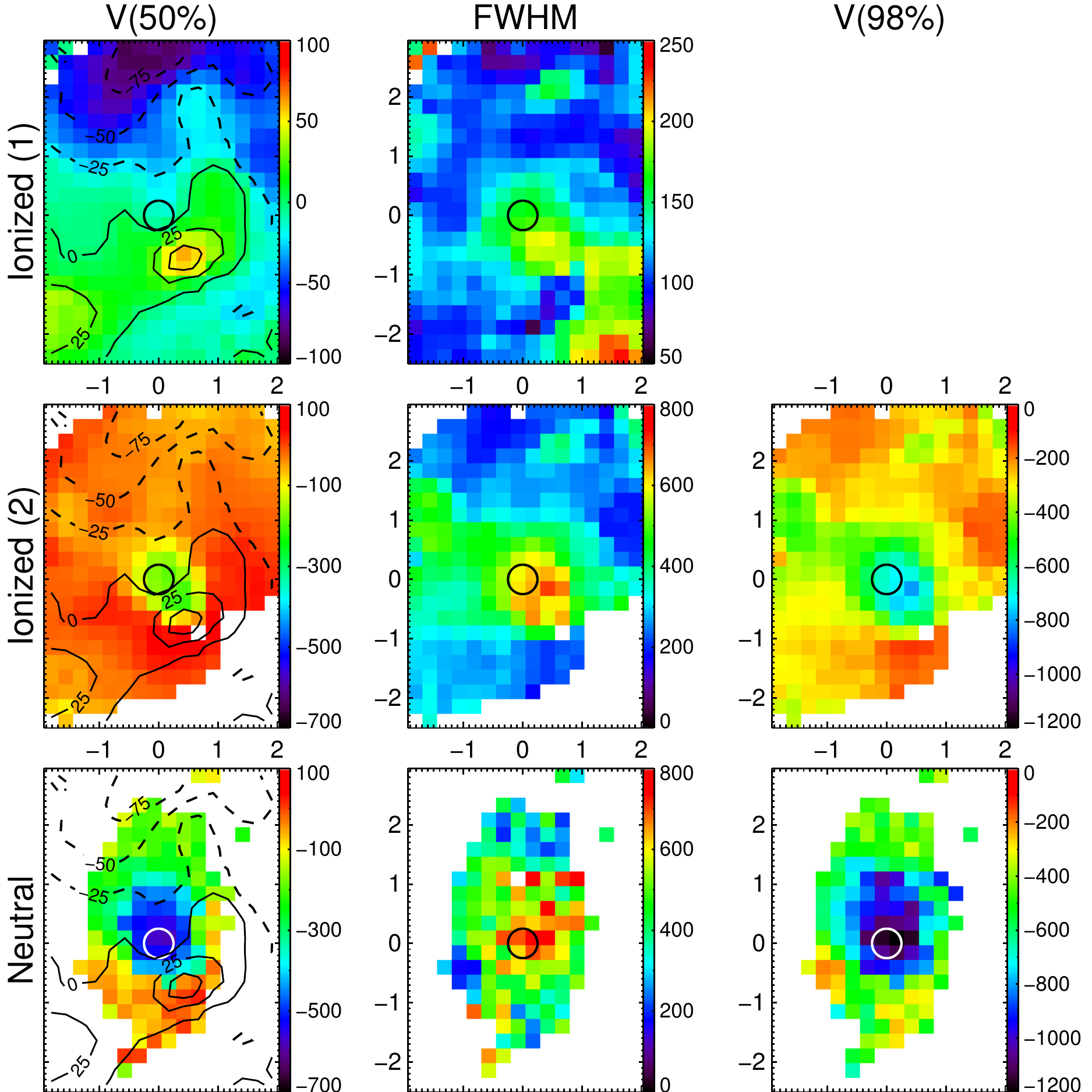}
  \caption{The same as Figure \ref{fig:map_vel_f08572nw}, but for
    VV~705:NW.}
  \label{fig:map_vel_vv705nw}
\end{figure}

\begin{figure}
  %\plotone{f34.eps}
  \centering \includegraphics[width=6.5in]{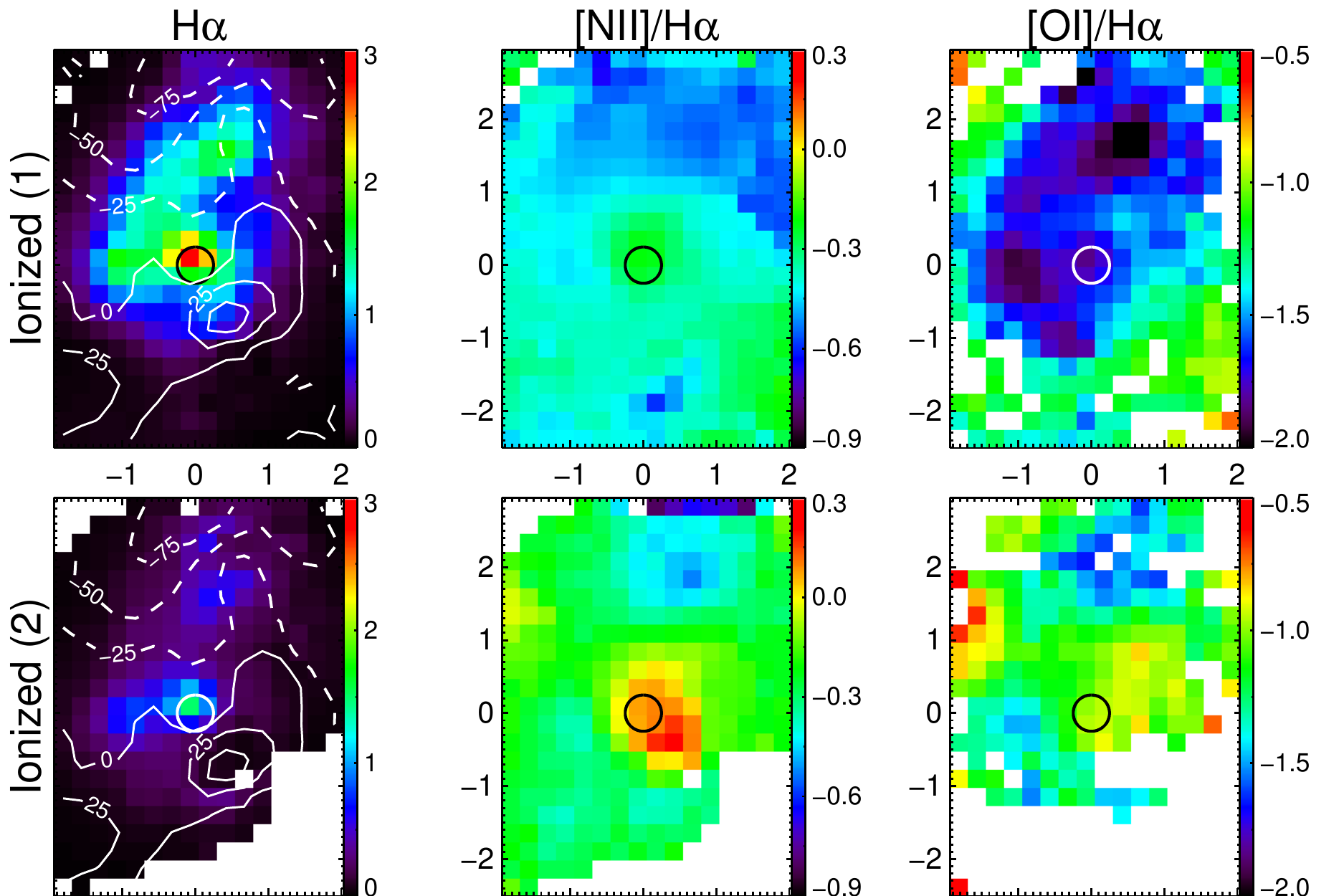}
  \caption{The same as Figure \ref{fig:map_lrat_f08572nw}, but for
    VV~705:NW.}
  \label{fig:map_lrat_vv705nw}
\end{figure}

\begin{figure}
  %\plotone{f35.eps}
  \centering \includegraphics[width=6.5in]{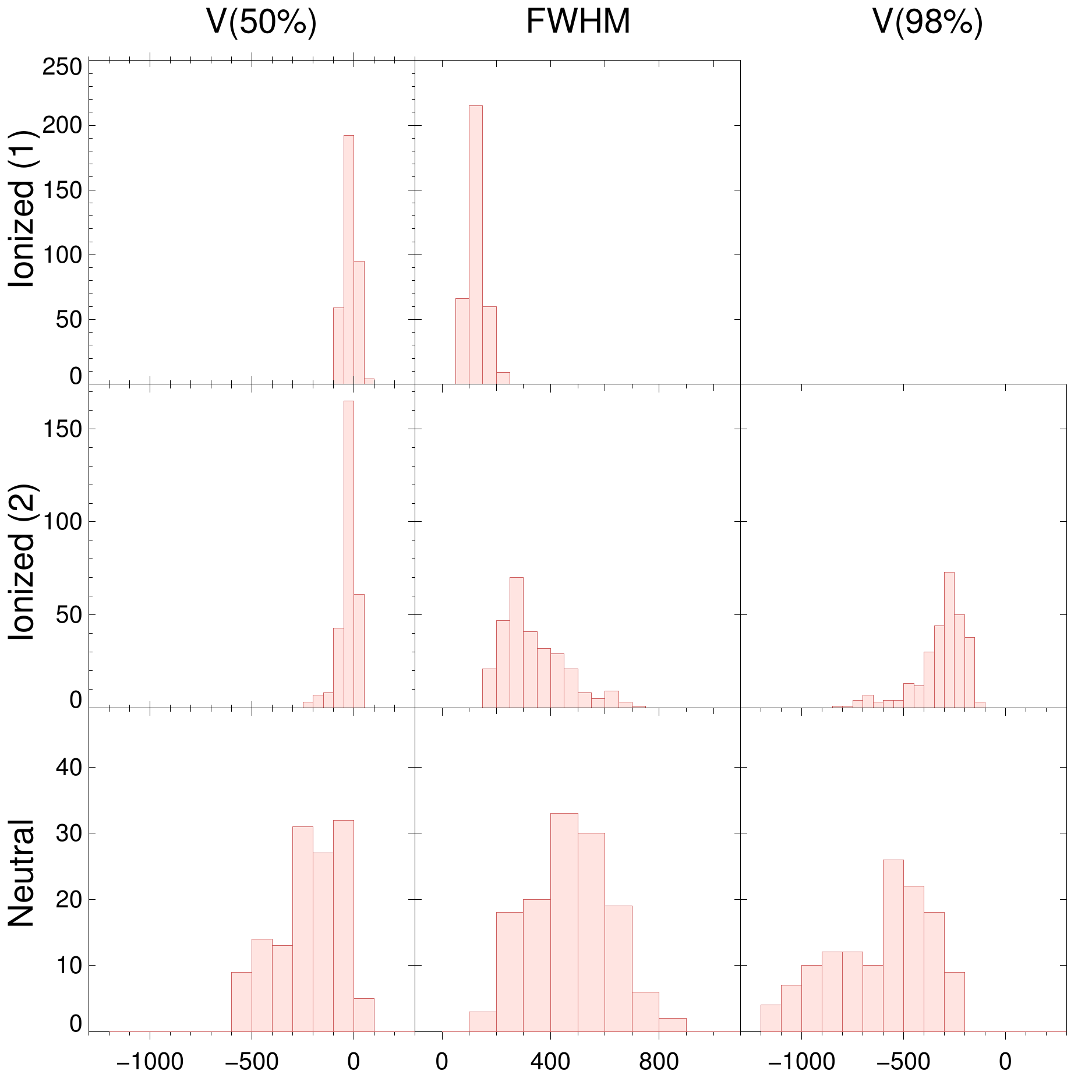}
  \caption{The same as Figure \ref{fig:veldist_f08572nw}, but for
    VV~705:NW.}
  \label{fig:veldist_vv705nw}
\end{figure}

\begin{figure}
  
  %\plotone{f36.eps}
  \centering \includegraphics[width=6.5in]{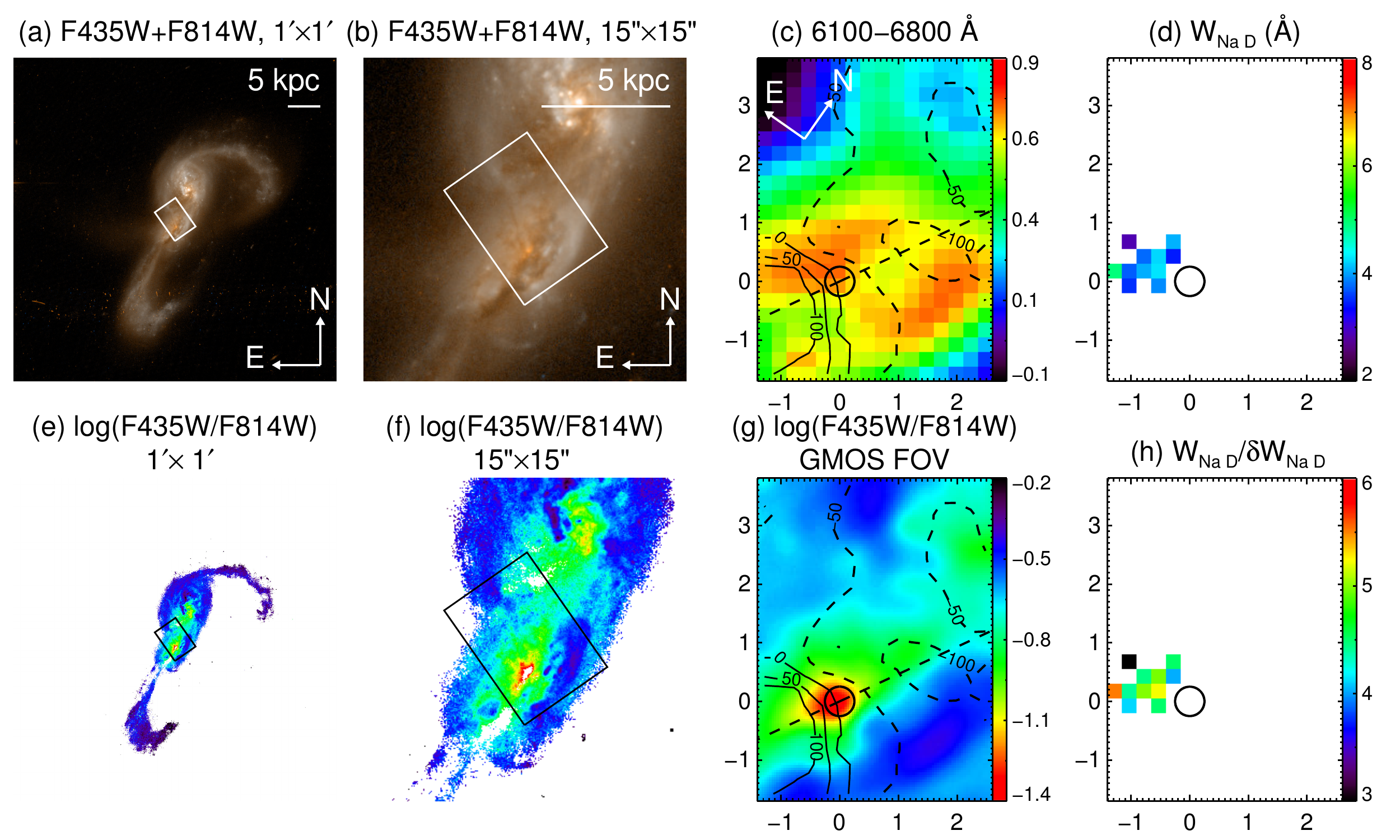}
  \caption{The same as Figure \ref{fig:map_cont_f08572nw}, but for
    VV~705:SE. Velocities are with respect to $z_{sys} = 0.0407$. The
    dashed line is the estimated line of nodes based on the current
    dataset (PA $= -30\arcdeg$). The nuclear position is based on the
    $I$-band peak.}
  \label{fig:map_cont_vv705se}
\end{figure}

\begin{figure}
  %\plotone{f37.eps}
  \centering \includegraphics[width=5in]{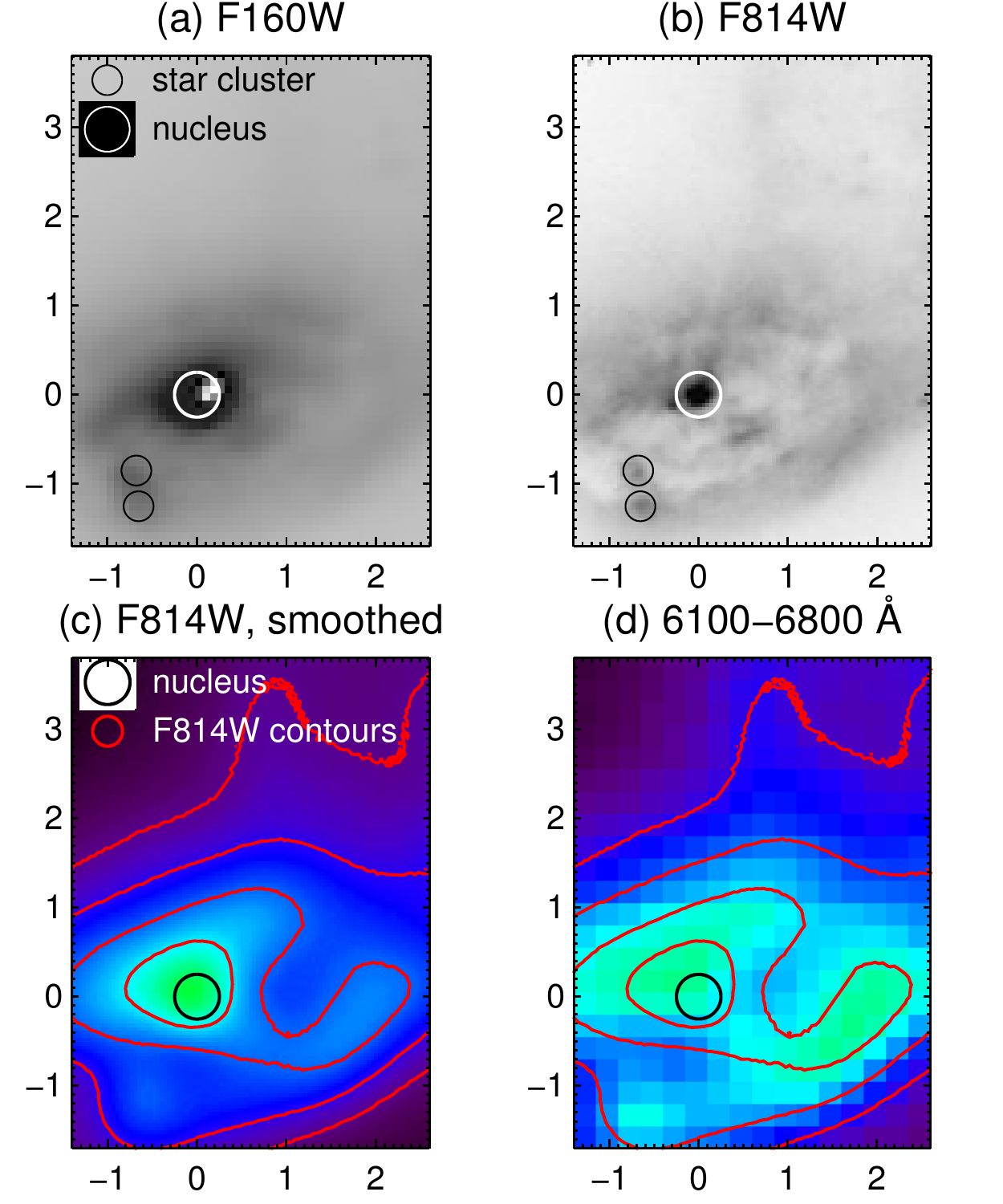}
  \caption{The same as Figure \ref{fig:register_f08572nw}, but for
    VV~705:SE. Note that saturated or bad pixels are present near the
    nucleus in the F160W image, and that the star clusters do not
    appear in the F160W image; as a result, the F814W image is used to
    locate the nucleus.}
  \label{fig:register_vv705se}
\end{figure}

\begin{figure}
  %\plotone{f38.eps}
  \centering \includegraphics[width=6.5in]{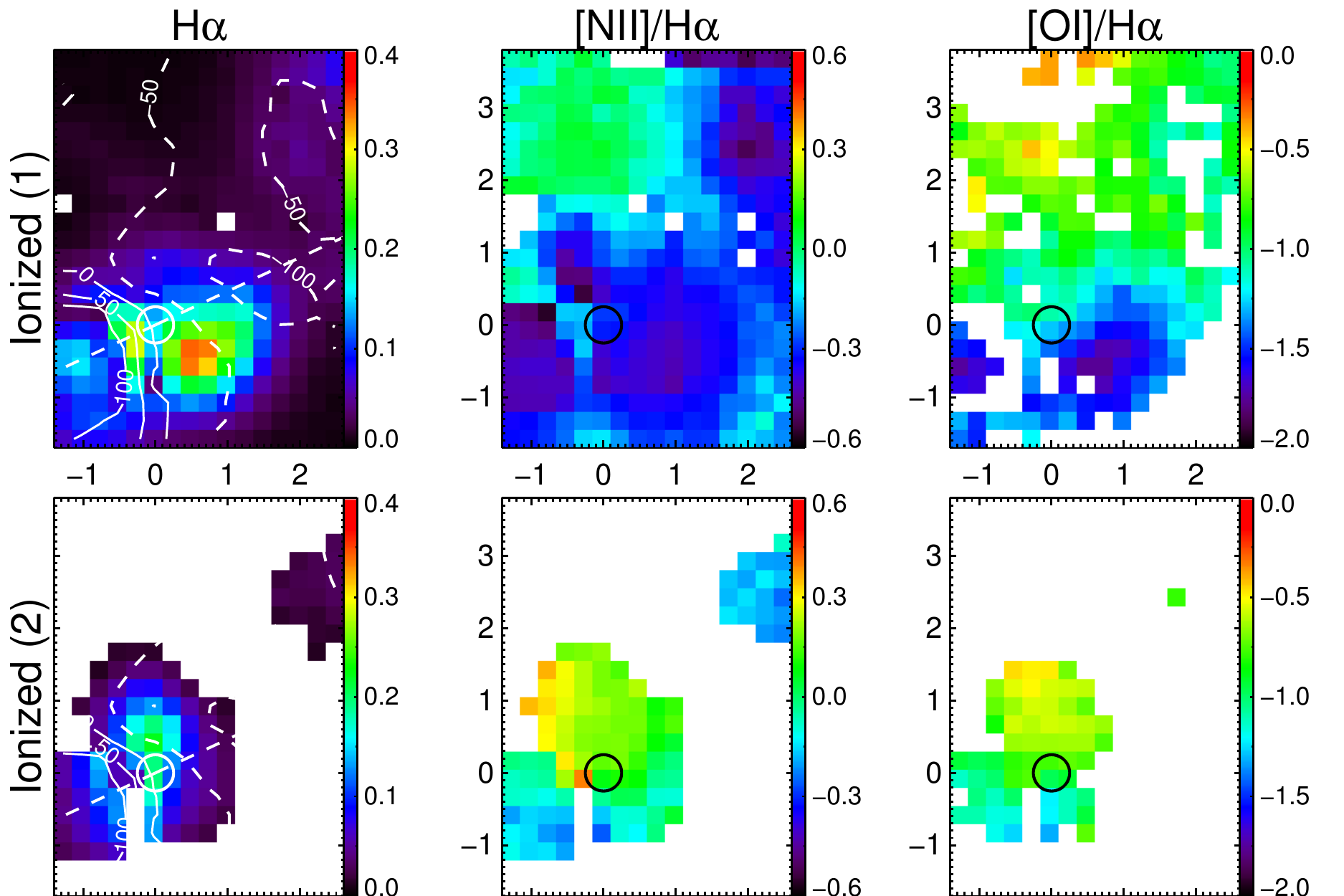}
  \caption{The same as Figure \ref{fig:map_vel_f08572nw}, but for
    VV~705:SE.}
  \label{fig:map_vel_vv705se}
\end{figure}

\begin{figure}
  %\plotone{f39.eps}
  \centering \includegraphics[width=6.5in]{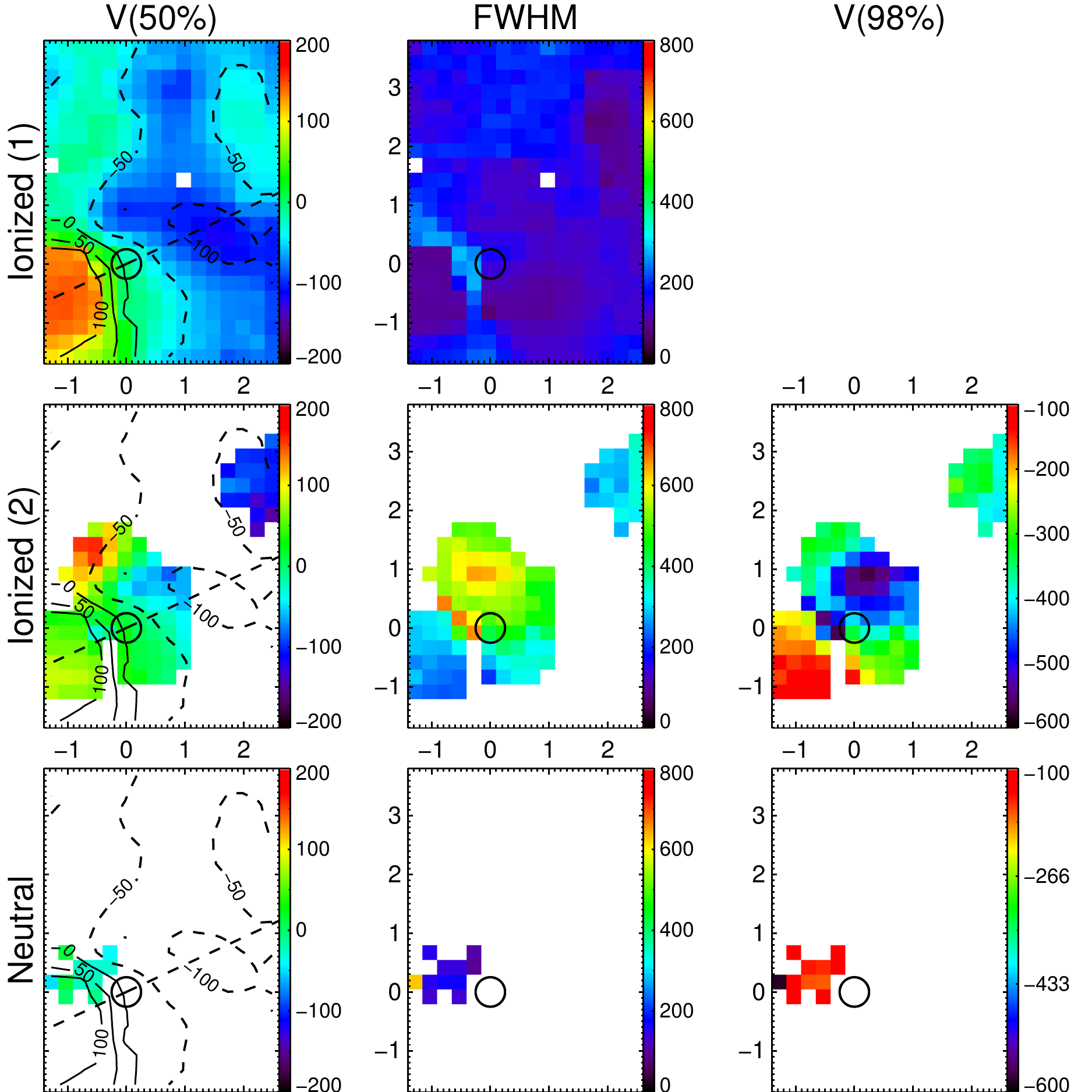}
  \caption{The same as Figure \ref{fig:map_lrat_f08572nw}, but for
    VV~705:SE.}
  \label{fig:map_lrat_vv705se}
\end{figure}

\begin{figure}
  %\plotone{f40.eps}
  \centering \includegraphics[width=6.5in]{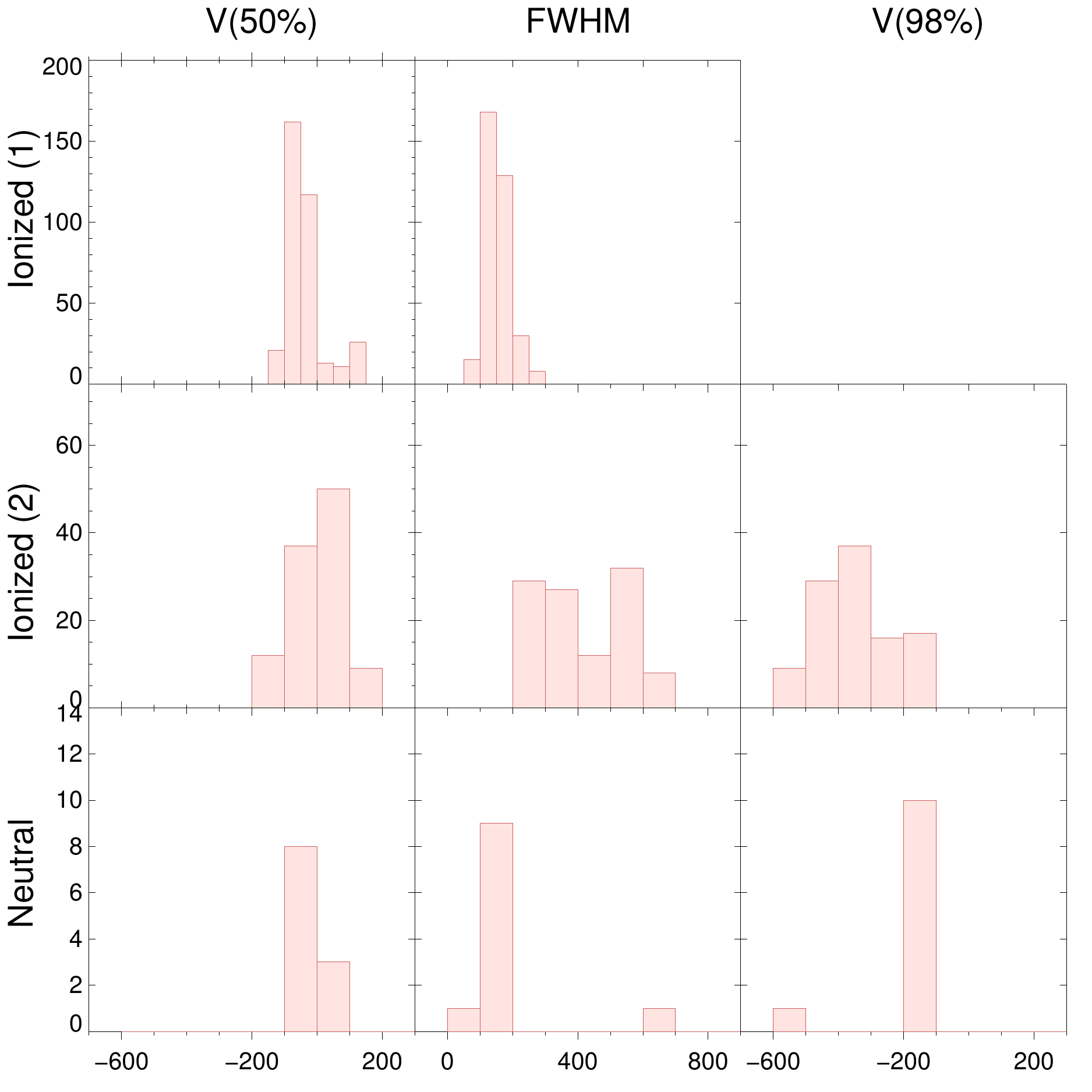}
  \caption{The same as Figure \ref{fig:veldist_f08572nw}, but for
    VV~705:SE.}
  \label{fig:veldist_vv705se}
\end{figure}

VV~705 is a binary merger with nuclear separation of 7.5~kpc, based on
inspection of an archival \hst\ Wide Field Camera 3 F160W image. Maps
of derived properties and velocity distributions for each nucleus are
shown in Figures
\ref{fig:map_cont_vv705nw}$-$\ref{fig:veldist_vv705se}. The
superbubble emerging from the NW nucleus is one of two in our sample.

\subsubsection{Power Source}

The NW nucleus contains most of the luminosity in this system at
multiple wavelengths: 8.44~GHz \citep[NW:SE luminosity ratio of
3:1;][]{condon91a}, $2-10$~keV \citep[8:1;][]{iwasawa11b}, and
$4-24$~\micron\ (archival {\it Spitzer} IRAC and MIPS 24\micron\
data). Though exact measurements are not available, at least 90\%\
(and perhaps closer to $\sim$99\%) of the 24~\micron\ emission emerges
from the NW nucleus, based on the strength of the 24~\micron\ PSF of
the NW nucleus. This strongly suggests that this nucleus is the source
of most of the FIR luminosity of the galaxy, as well. This would be
consistent with it hosting most of the molecular gas
($\sim$10$^{10}$~\msun; \citealt{sanders91a,chini92a}), based on the
narrow FWHM of CO (100~\kms; \citealt{chini92a}) and the measured CO
velocity near that of the NW nucleus ($cz = 12070-12090$~\kms;
\citealt{chini92a,yao03a}; and this work).

Both nuclei are classified as composite by \citet{yuan10a}, based on
spectra obtained by \citet{kim95a}. This is confirmed for the NW
nucleus by nuclear data from \citet{moustakas06a}. However, there is
no evidence for an AGN in either nucleus; X-ray imaging shows soft
spectra in both nuclei \citep{iwasawa11b}, MIR spectra reveal no
[\ion{Ne}{5}] or [\ion{O}{4}] fine structure lines \citep{dudik09a},
and compact radio emission is consistent with a nuclear starburst
\citep{smith98a}.

In the absence of evidence for an AGN, we conclude that any outflows
in this system will be powered by star formation. We assume an AGN
fraction of 10\%\ for consistency with the other two galaxies in our
sample without luminous AGN, but the actual fraction could be lower.

\subsubsection{Kinematics} \label{sec:vv705-kinematics}

The SE nucleus shows ordered rotation with non-circular motions
superimposed. We estimate a line of nodes PA of $-30\arcdeg$. The
peak-to-peak velocity amplitude along this PA reaches
250$-$300~\kms. The near side of the disk is likely to the W, given
that this side of the nucleus is less obscured (Figure
\ref{fig:map_cont_vv705se}). The line of nodes and rotation amplitude
of the NW nucleus are more difficult to quantify due to tidal motions
and a near face-on orientation. We do not attempt to estimate a PA or
inclination for this nucleus.

Non-circular motions and multiple narrow components in the SE nucleus
include a region of LOS overlap between the two gas disks, 3~kpc N of
the SE nucleus. Multiple components are also evident within 1~kpc of
the SE nucleus; some of these components have FWHM exceeding
600~\kms. These broad components have velocities within
$100-200$~\kms\ of the narrow component in the same spaxel. They may
arise in a weak outflow along the minor axis and/or with infalling
shocked tidal debris. Either of these interpretations are consistent
with the observed elevated FWHM and line ratios, but the data is not
sufficient to distinguish between these possibilities. Regardless, any
outflow in this nucleus is not powered by the primary source of
luminosity in the merger.

\begin{figure}
  %\plotone{f42.eps}
  \centering \includegraphics[width=6.5in]{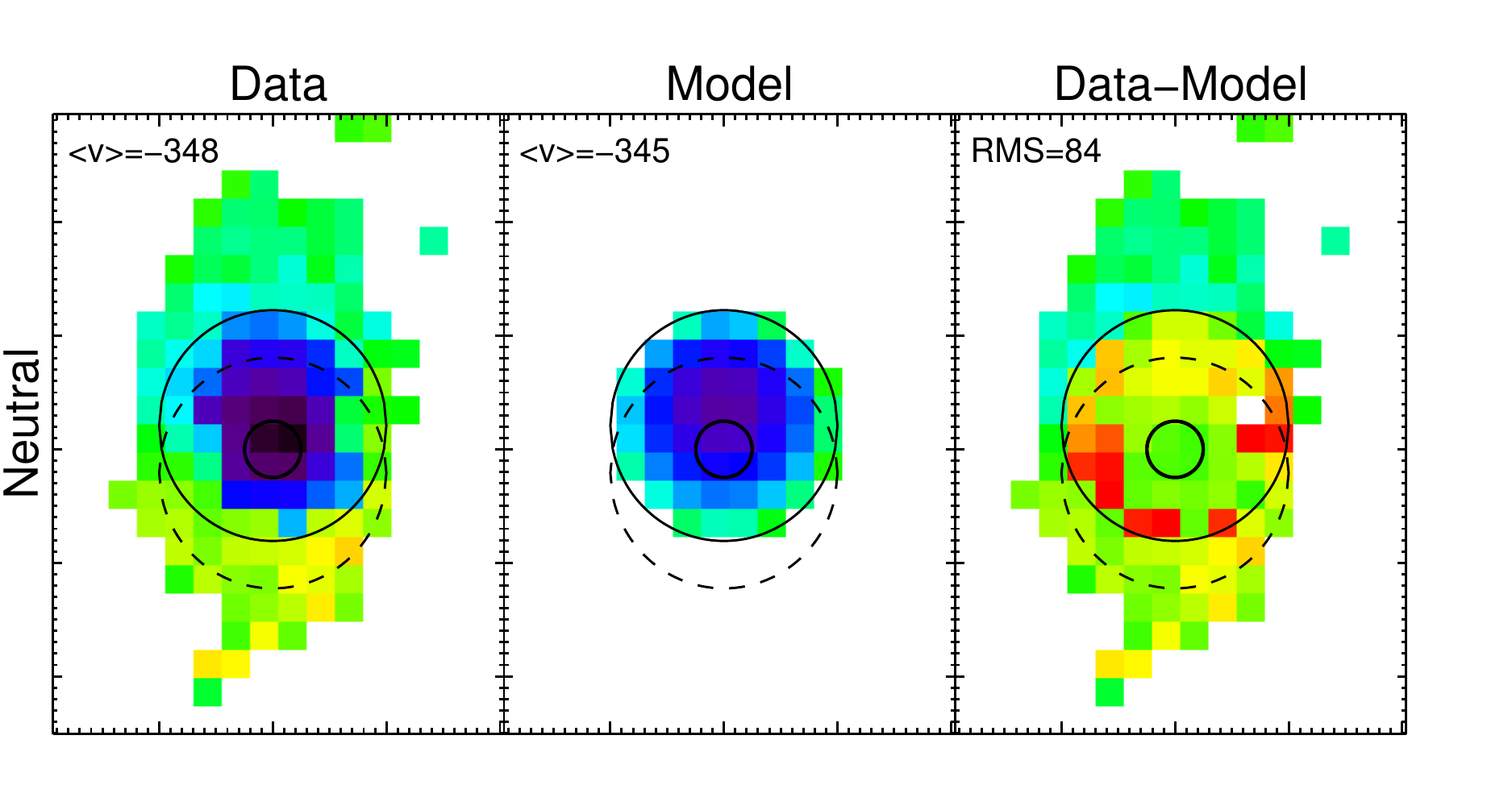}
  \caption{Color maps of superbubble model fits to the absorption line
    velocity field in VV~705:NW. The fit shown is for $n = 2$. The
    three columns show the data (left), model (middle), and their
    difference (right). In each panel, the solid line outlines the
    near-side bubble, the dashed lines the expected counter-bubble
    (the counter-bubble is hidden behind the disk). A small black
    circle locates the nucleus and outflow center. The numbers in the
    left and middle columns give the average velocity of that map
    within the model confines, while the number in the right panel is
    the RMS velocity difference within the same area. Model parameters
    are given in Table~\ref{tab:modbbl}. (See \S\S\,\ref{sec:bsb} and
    \ref{sec:vv705-kinematics} for more details.)}
  \label{fig:modbbl_vv705nw}
\end{figure}

The NW nucleus shows evidence for a superbubble emerging almost
directly along the line of sight. This bubble is most evident in
absorption, as seen in the PV diagram in Figure
\ref{fig:pv_a_vv705nw}. Applying our BSB model (\S\,\ref{sec:bsb}) and
optimizing its parameters yields the values given in Table
\ref{tab:modbbl}. The fits and residuals in two dimensions are shown
in Figure\,\ref{fig:modbbl_vv705nw}.

The superbubble model fits the inner 2~kpc well, but outflowing gas at
lower velocities is also visible NE of the bubble and, to a lesser
extent, to its SW. The best-fit parameters show that the outflow
emerges almost directly along the line of sight and reaches a velocity
of $\vfifty = -500$ to $-600$~\kms\ along the bubble axis. The
elongation of the bubble is fairly uncertain, and is consistent with
either a symmetric bubble ($\sim$2~kpc in diameter) or one that is
longer perpendicular to the galaxy disk by factors of $2-3$.

The outflow is also seen in emission
(Figure~\ref{fig:map_vel_vv705nw}). However, the ionized gas only
shows acceleration in a fraction of the outflow area that is seen in
absorption. The broad component in most spaxels has the same velocity
as the narrow, rotating component. As a result, in the SRFW model
calculations for this system, we include only spaxels in which the
broad component is blueshifted from the narrow component in that
spaxel by more than 50~\kms.

\subsection{F17207$-$0014}

\begin{figure}
  % \plotone{f43.eps}
  \centering \includegraphics[width=6.5in]{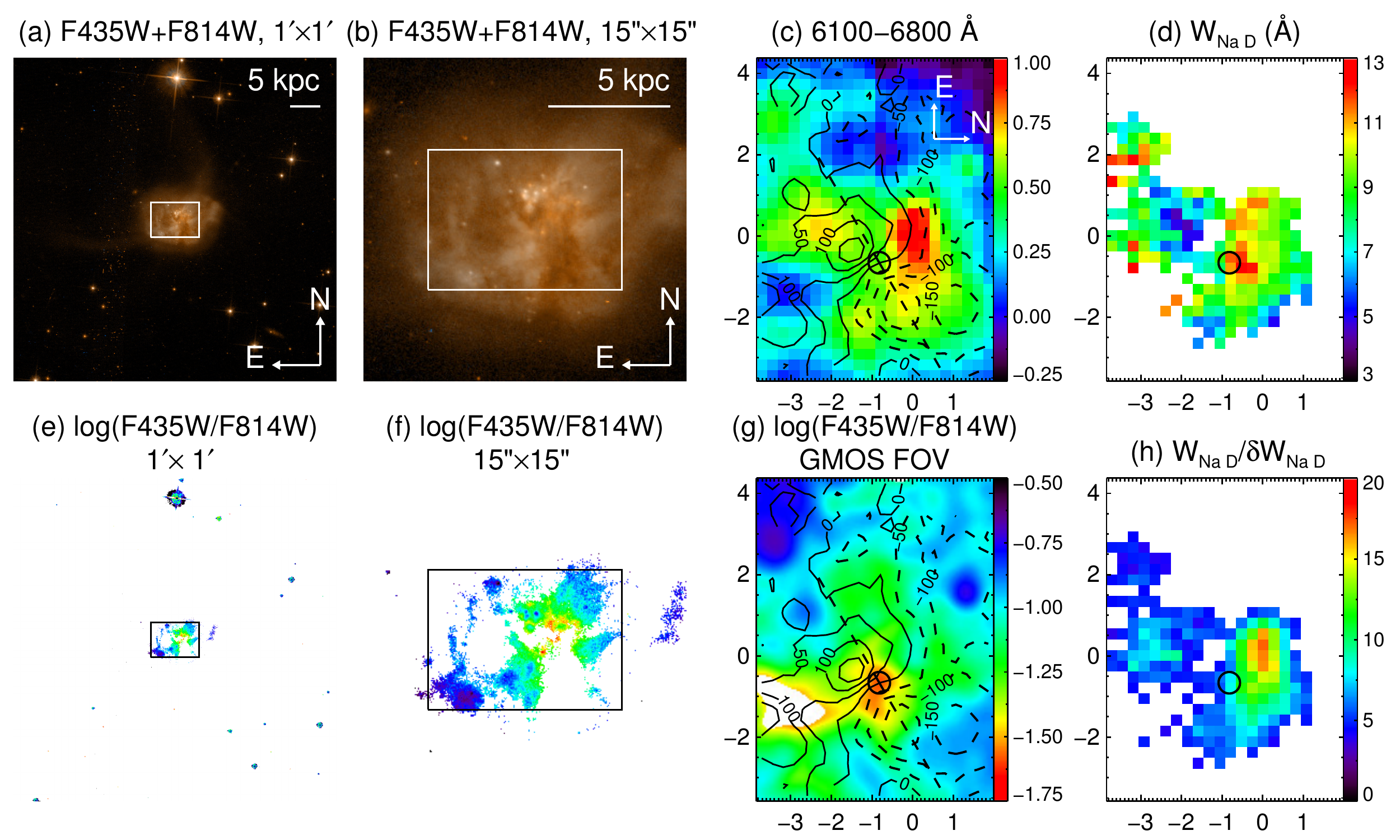}
  \caption{The same as Figure \ref{fig:map_cont_f08572nw}, but for
    F17207$-$0014. Velocities are with respect to $z_{sys} = 0.0430$,
    and the dashed line represents the position angle of the CO disk
    \citep{downes98a}.}
  \label{fig:map_cont_f17207}
\end{figure}

\begin{figure}
  %\plotone{f44.eps}
  \centering \includegraphics[width=5in]{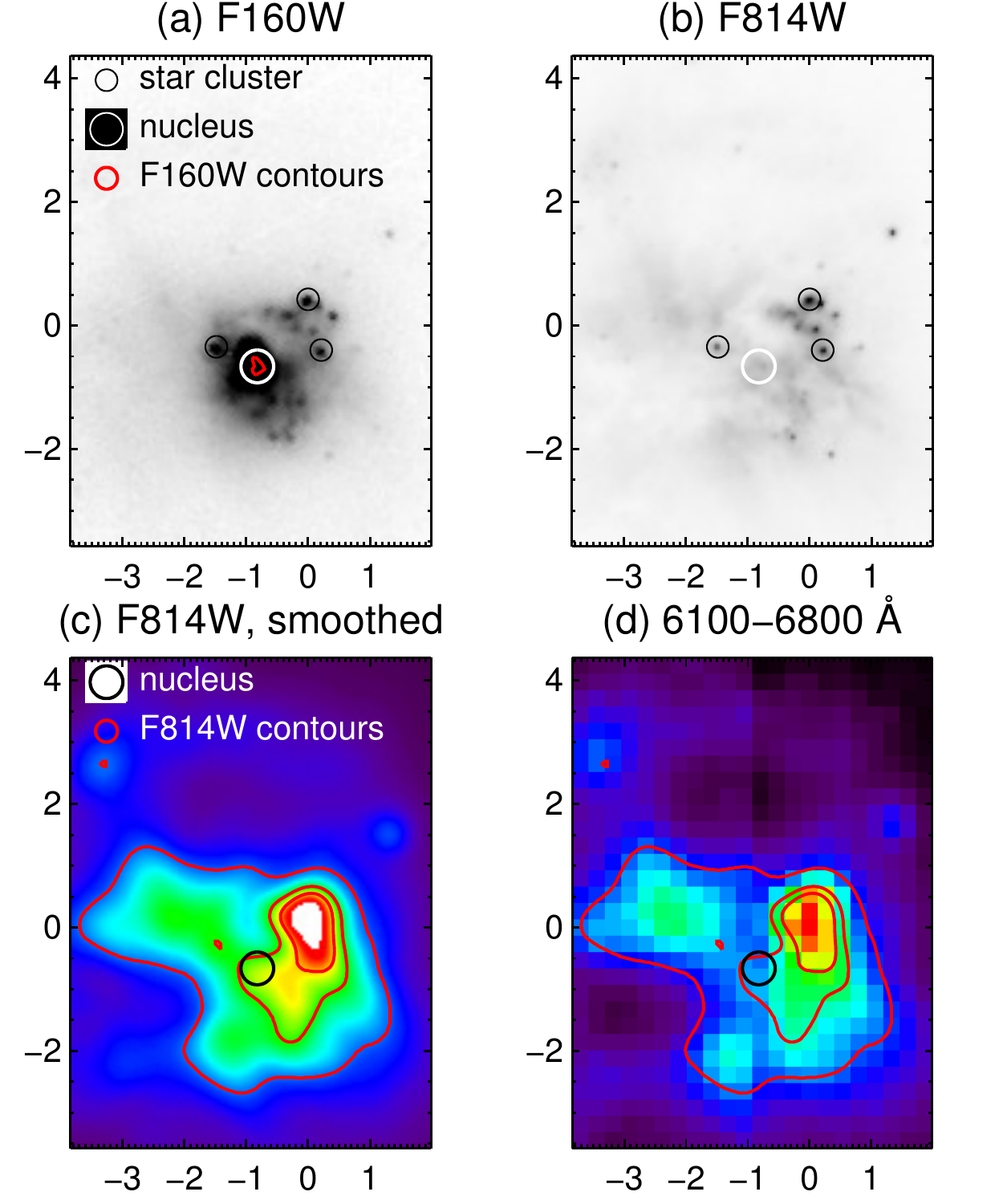}
  \caption{The same as Figure \ref{fig:register_f08572nw}, but for
    F17207$-$0014.}
  \label{fig:register_f17207}
\end{figure}

\begin{figure}
  %\plotone{f45.eps}
  \centering \includegraphics[width=6.5in]{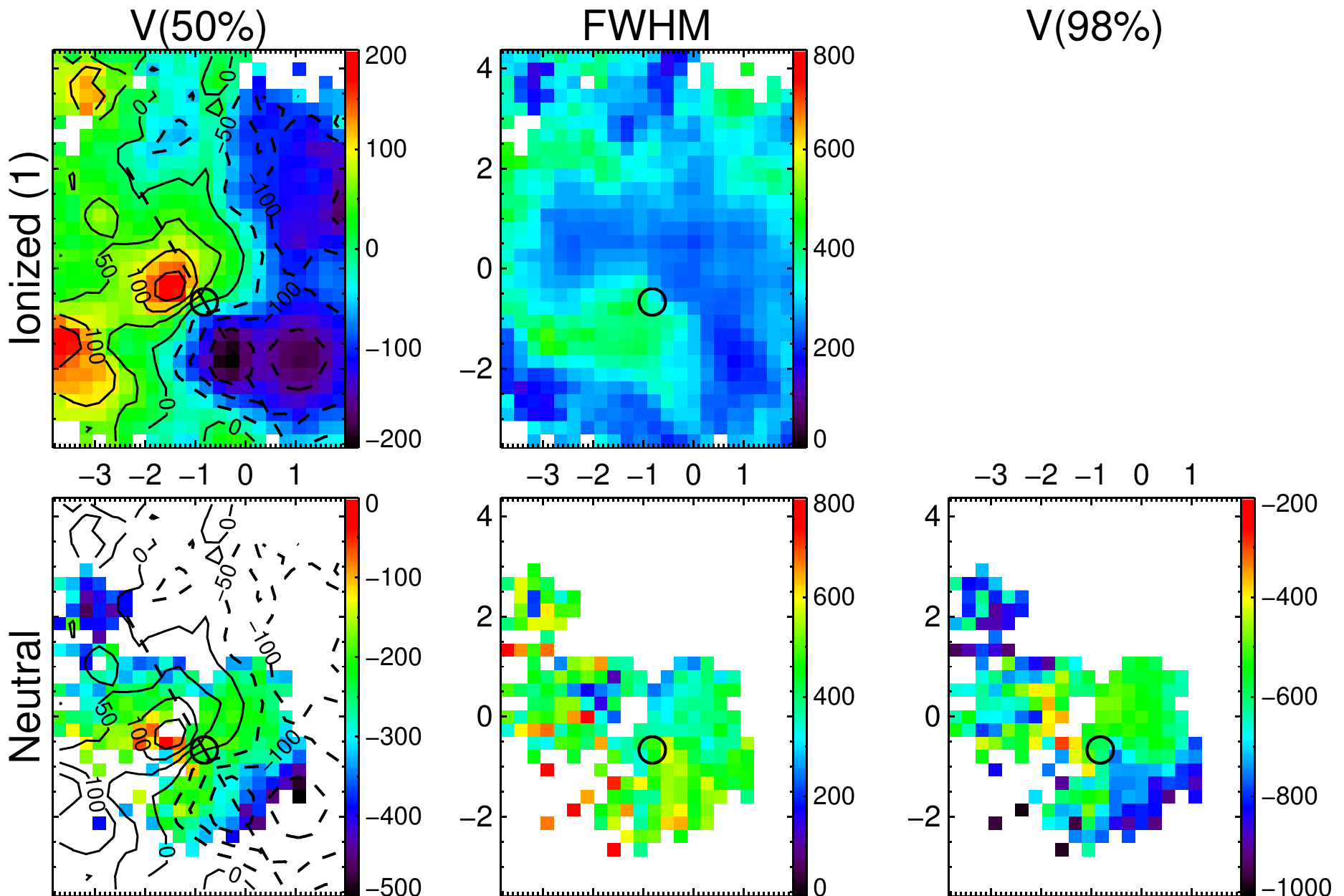}
  \caption{The same as Figure \ref{fig:map_vel_f08572nw}, but for
    F17207$-$0014.}
  \label{fig:map_vel_f17207}
\end{figure}

\begin{figure}
  %\plotone{f46.eps}
  \centering \includegraphics[width=6.5in]{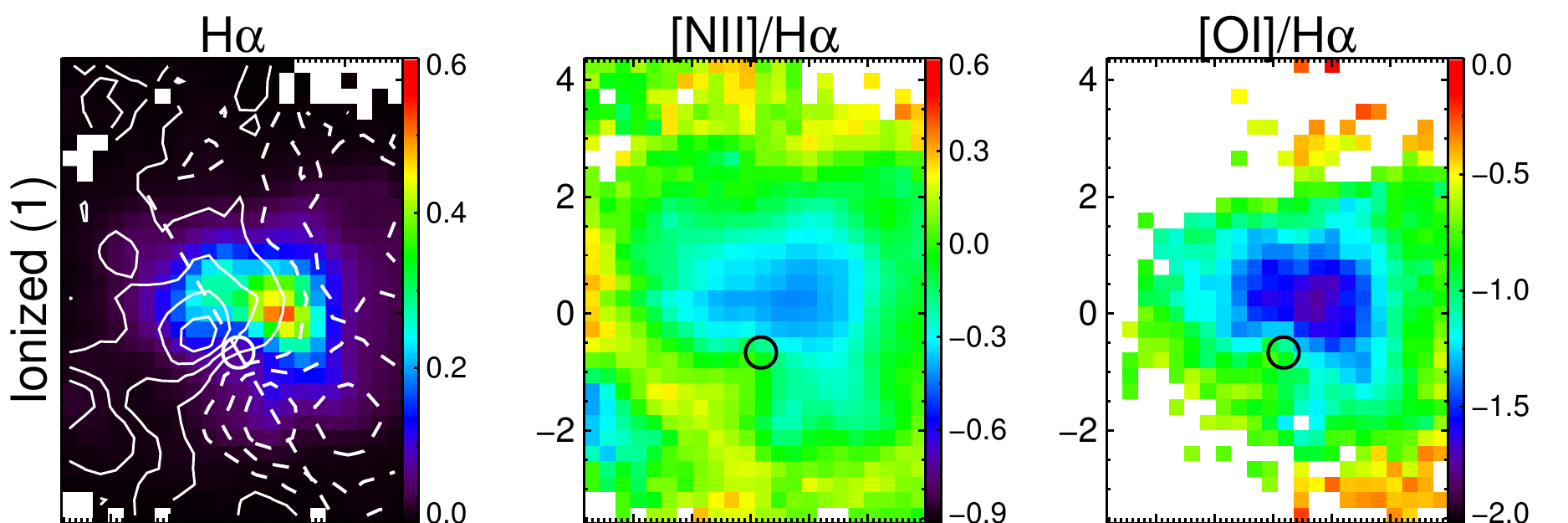}
  \caption{The same as Figure \ref{fig:map_lrat_f08572nw}, but for
    F17207$-$0014.}
  \label{fig:map_lrat_f17207}
\end{figure}

\begin{figure}
  %\plotone{f47.eps}
  \centering \includegraphics[width=6.5in]{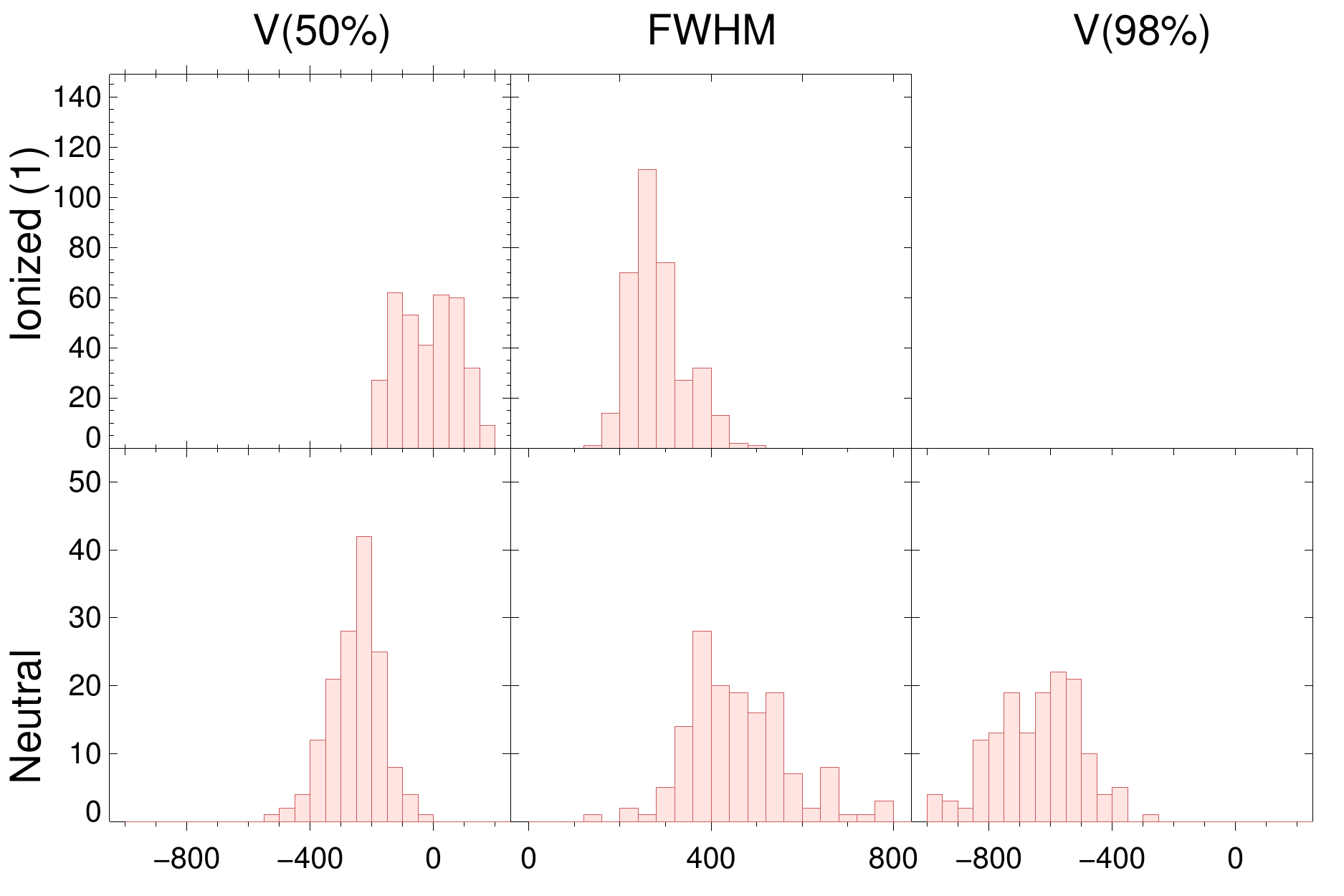}
  \caption{The same as Figure \ref{fig:veldist_f08572nw}, but for
    F17207$-$0014.}
  \label{fig:veldist_f17207}
\end{figure}

This system is a coalesced merger with a diffuse, patchy optical
morphology. We present maps and velocity distributions based on our
GMOS data in
Figures~\ref{fig:map_cont_f17207}$-$~\ref{fig:veldist_f17207}. F17207$-$0014
is the only system in our sample without a detected ionized outflow,
though it does reveal a large scale neutral gas outflow.

\subsubsection{Power Source}

All of the data on this system is consistent with an obscured
starburst. The nucleus is obscured in the optical but resolved in the
NIR (\citealt{murphy96a,scoville00a};
Figure~\ref{fig:register_f17207}). It is a diffuse source in the radio
and MIR \citep{soifer00a,momjian03a,baan06a}, no [\ion{Ne}{5}] or
[\ion{O}{4}] emission is detected in the MIR \citep{farrah07a}, the
AGN fraction based on MIR diagnostics is only 0.11
\citep{veilleux09a}, and the X-ray data do not show an obscured AGN
\citep[e.g.,][]{franceschini03a,teng10a,iwasawa11a}.

\subsubsection{Kinematics}

The CO and \ion{H}{1} disk in F17207$-$0014 subtends $1-2\arcsec$
along a line of nodes position angle of $\sim$120\arcdeg\
\citep{downes98a,momjian03a}. \citet{arribas03a} presented ionized gas
maps of this system and showed that the molecular and neutral
kinematics are seen in the ionized gas, as well. The rotation of the
nuclear ionized disk is roughly centered on the NIR peak, but at
larger scales the isovelocity contours twist away from this pattern
\citep{arribas03a}. The kinematics on galactocentric radii $\ga$1~kpc
probably reflect tidal motions \citep{arribas03a}.

A neutral outflow in this system was discovered by \citet{rupke05b}
and \citet{martin05a}, and shown to extend over $5-6$~kpc
\citep{rupke05b,martin06a}. \citet{martin06a} found no variation in
the outflow velocity (with respect to the rotating ionized gas) along
a position angle of 170\arcdeg. Recently, \citet{westmoquette12a}
presented ionized gas maps of F17207$-$0014 that showed two velocity
components in the ionized gas in the center of this system. However,
the two components have very similar velocity, and are not obviously
an ionized outflow.

Our new maps of the ionized and neutral gas in F17207$-$0014 improve
on the \citet{arribas03a} and \citet{westmoquette12a} studies by
adding the neutral gas kinematics and increasing the spatial
resolution. Our ionized gas maps indeed resemble higher resolution
versions of the \citet{arribas03a} maps. We do not detect two ionized
gas components, as in \citet{westmoquette12a}, but the much larger
spaxels in the \citet{westmoquette12a} data may increase their
sensitivity.

We detect extended neutral gas that is everywhere outflowing in this
system, consistent with previous nuclear and extended
detections. However, we show that this outflow extends in all
directions, and to galactocentric radii of at least 4~kpc (with a
maximum extent of 6~kpc, as in \citealt{martin06a}).

\begin{figure}
%  \plotone{f48.eps}
  \centering \includegraphics[width=6.5in]{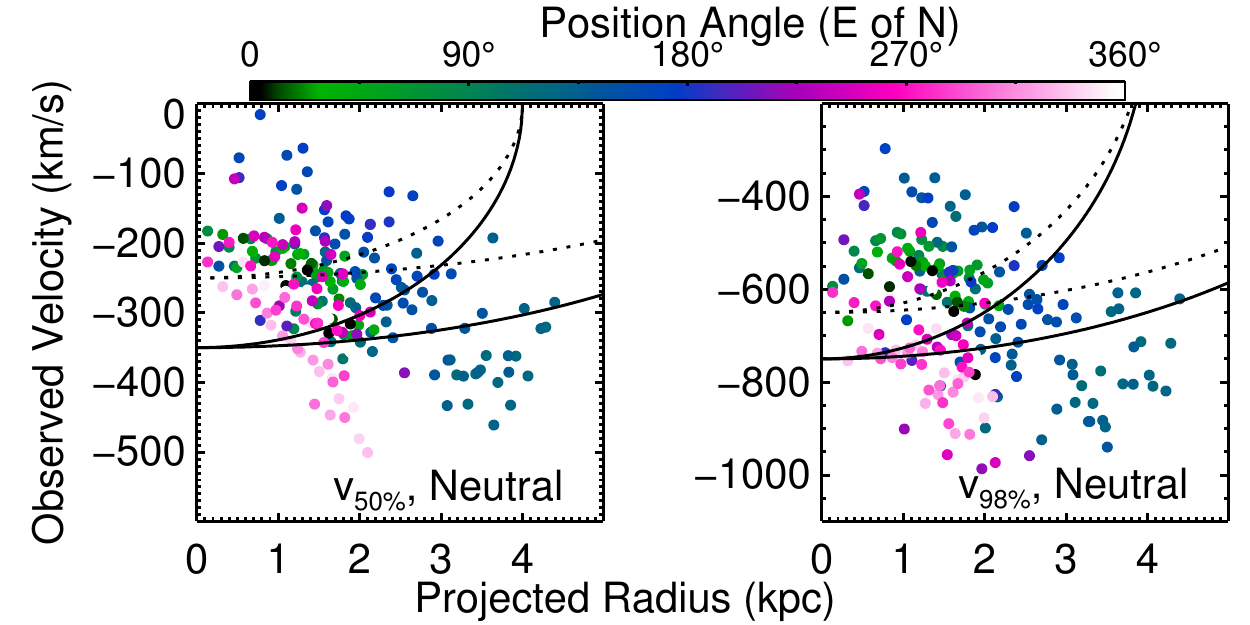}
  \caption{Observed velocity (\vfifty\ and \vtsig) vs. projected
    galactocentric radius in the outflowing neutral gas of
    F17207$-$0014. Each point represents one spaxel, and the position
    angle (E of N) of the spaxel with respect to the outflow center is
    represented by its color. The lines represent single radius free
    wind models (\S\,\ref{sec:analysis-modeling}) at constant velocity
    for $R = 4$~kpc and 8~kpc (the velocity for a given model is given
    by the value at $R_{proj} = 0$~kpc). The observed increase of
    velocity with projected galactocentric radius is inconsistent with
    the SRFW model.}
  \label{fig:vel_v_rad_f17207}
\end{figure}

Contrary to \citet{martin06a}, we find that the outflow velocity
increases with increasing projected radius
(Figure~\ref{fig:vel_v_rad_f17207}). This is inconsistent with our
fiducial SRFW model, which predicts decreasing velocity with
increasing projected radius. An alternative scenario is that the
projected galactocentric radii represent the true galactocentric
radii, and that the outflow speed is increasing with radius.

\section{SAMPLE TRENDS} \label{sec:sample}

Our sample of six merging systems reveals an extended outflow in every
case. In five systems there is an ionized outflow (F17207$-$0014 is
the odd galaxy out), and four of these are new ionized outflow
detections (the Mrk~273 outflow was previously known;
\citealt{colina99a}). In six systems there is a neutral outflow. Three
of these are new detections (F08572$+$3915, Mrk~273, and VV~705),
despite two of these three having been observed in neutral outflow
surveys using single-aperture spectroscopy.

In this section we combine the results from our six-galaxy sample to
draw broader conclusions about the multiphase structure and power
sources of galactic winds in ULIRGs. These conclusions are summarized
in \S\,\ref{sec:summary} and rely in part on the analysis in
Figures~\ref{fig:vel_v_fwhm}$-$~\ref{fig:vel_v_lir}.

\subsection{Wind Structure} \label{sec:wind-structure}

\subsubsection{Ionized Gas} \label{sec:wind-structure-ion}

We know a great deal about the structure of winds in local disk
galaxies hosting starburst-driven superwinds
\citep{veilleux05a}. These winds typically emerge along the minor axis
of the disk as either a bipolar superbubble or a bicone. The ionized
gas in such BSBs or bicones is often limb-brightened and filamentary,
suggesting it lies on the edges of the bubble or cone
\citep[e.g.,][among many
examples]{heckman87a,bland88a,veilleux94a,martin98a}. In the model of
a bipolar wind, these bubbles or cones have opening angles that
increase with radius, reaching angles (on one side of the disk) of
$45-100\arcdeg$ \citep{veilleux05a}.

The structure of ionized gas outflows in local mergers is less
understood, though the two nearest and most-studied examples have
possible minor-axis and/or bipolar flows: Arp~220 \citep{heckman87a}
and NGC~6240 \citep{veilleux03a,bush08a}. The kinematics of the
ionized gas in mergers are made complex by the prevalence of multiple
overlapping disks and tidal motions \citep{colina05a}, as well as
outflows. However, careful spectral decomposition of the emission
lines is able to separate gas rotation from other motions \citep[][and
this study]{shih10a,rupke11a,westmoquette12a}. The narrow component in
the ionized gas (always the strongest in flux overall) follows the
rotation of the resolved kpc-scale CO disk (when CO has been
previously observed via interferometry at arcsecond resolution, as in
4/6 systems in our sample). The good match between the ionized and
molecular gas in the inner $1-2$~kpc of these systems is somewhat
surprising, since these disks are dusty and probably optically thick
\citep{downes98a}; thus, the ionized gas is likely to trace the outer
``skin'' of the disk rather than its core, or emerges from a patchy
and inhomogeneous medium. Where the molecular gas is more extended
(e.g., Mrk~273), there continues to be a good correspondence between
the bulk motions of the molecular and ionized gas.

The linewidths in the narrow component of the ionized gas are
relatively large (the average FWHM in this component in each of the
eight nuclei in our sample is in the range $100-300$~\kms, and
typically near 200~\kms; see, e.g., Figure
\ref{fig:veldist_f08572nw}). Similar linewidths are seen in the
resolved molecular gas in these systems \citep{downes98a}. These large
linewidths may result from several effects. They do not likely result
from beam smearing, since the narrow component velocity fields
typically have smaller amplitudes than this. Sometimes overlapping
disks and/or regions where two disks are colliding (as in
F08572$+$3915 and VV~705:NW) amplify the linewidths in isolated
regions as a result of the combined motion of the two disks or the
tidal motions of collision. In other systems, the linewidth may be
inflated by either energy injected by star formation and/or AGN
activity in the inner disks, or on larger scales by tidal motions
resulting from the ongoing merger.

On scales larger than the central $1-2$~kpc, the ionized gas continues
the rotational pattern of the inner disks, but often with isovelocity
twists superimposed or changes in velocity amplitude. It is likely
that the motions of the ionized gas on large scales in most of these
systems better reflect the ongoing merger of the two initial systems
rather than the inner disk of either nuclei or the nuclear disk of the
coalesced system. The ionized gas motions at projected galactocentric
radii of $3-5$~kpc thus result from a combination of the internal
angular momenta of the original two systems with that of the merger
orbit.

The interpretation of the broader and blueshifted or redshifted
velocity components in the ionized gas is relatively
straightforward. In emission and without other information, outflow
and inflow are in principle indistinguishable. However, as we have
shown for Mrk~273 and VV~705, outflows can be distinguished from
inflow via structural modeling, since BSBs are obviously present. A
multiphase, multi-kpc minor axis outflow is also likely present in
F10565$+$2448, based on the morphological connections between the
ionized gas kinematics, neutral gas kinematics, and nuclear dust
filaments. In the cases of F08572$+$3915, the large-scale kinematics
in Mrk~273, and the inner kpc of Mrk~231, the observed velocities on
these scales are simply too high to be explained by tidally-induced
inflow.

\citet{soto12b} also argue that, in ULIRGs, the observed ionized gas
must be on the near side of the galaxy, and thus outflowing if
blueshifted, since far-side ionized gas would be obscured by
dust. There is some evidence from our data against this conclusion,
but there is also evidence to support it; the reality may be more
nuanced. First, the continuum obscuration by dust (as revealed by
multicolor \hst\ imaging; e.g., Figure \ref{fig:dustpeak_f10565}) is
patchy. Nonetheless, optically thick molecular disks (not traced by
the optical continuum) reside at the nuclei of these systems and
probably obscure any far side ionized gas very near the
nucleus. Second, we do observe the receding side of the superbubble in
Mrk~273. However, this bubble is highly inclined and extended; in most
other cases where we see evidence consistent with a near-side minor
axis ionized flow, evidence for the far-side flow is weaker or absent
(see discussion below and of individual objects above). The good
correspondence between ionized and neutral gas velocities
(\S\,\ref{sec:multiphase-wind}) also suggests that the ionized gas is
primariliy located on the near side of these systems, where the
absorbing gas must lie. Overall, it is clear that spatially resolved
observations of these systems, as we present here, are crucial to
resolve these structural degeneracies.

The ionized outflows we observe are in all cases consistent with being
collimated by a molecular disk within the galaxy, emerging along the
minor axis of that disk. In the three cases where CO disks and ionized
outflows are observed (F10565$+$2448, Mrk~231, and Mrk~273), there are
outflows reaching $1-2$~kpc away from the nucleus that have directions
and inclinations consistent with the CO minor axis. In the other two
cases of ionized outflows where CO disks are present (but have not
been observed via interferometry), the direction and inclination of
the outflows are consistent with the disk minor axis as inferred from
the rotation of the ionized gas.

In particular, in the two cases where we fit the 3D shape of the
outflow, the kinematics are consistent with 2~kpc diameter BSBs, which
would formally have (one-sided) opening angles of 180\arcdeg. In
reality, the footprint of the outflow is not a point, such that the
opening angle is smaller than this. For Mrk~273, let us assume the CO
core/disk edges from \citet{downes98a} as the extent of the launching
region (radii $100-400$~kpc, depending on the measure). If we draw a
line connecting the CO disk edge with the bubble edge at a disk height
of 1~kpc, then we infer an opening angle of $30-40\arcdeg$, similar to
local superbubbles.

Outflowing ionized gas is also observed away from the minor axis,
particularly in the cases of Mrk~273 and F10565$+$2448. The morphology
of these extended flows is more ambiguous, though in Mrk~273 it may in
part be breakout from the superbubble (\S\,\ref{sec:mrk273-bsb}).  It
is possible that these are traces of earlier bubbles or ejection
episodes that have moved to larger radii.
 
This idea is buttressed by the timescale for wind evolution over the
galactocentric radii probed by our observations, which is $t_{dyn}\sim
R_{wind}/\vfifty = 10$~Myr (for $R_{wind} = 3$~kpc and $\vfifty =
300$~\kms). The timescale for the orbital evolution of the galaxies in
our sample is of order $\ga$100~Myr
\citep[e.g.,][]{barnes96a,mihos96a}; i.e., an order of magnitude
larger. Thus, if we assume that we are observing the six galaxies in
our sample at a random time late in the merger, then the existence of
minor axis flows at roughly similar scales ($1-2$ kpc projected radii)
in each system suggests that they are a relatively continuous
phenomenon, either in the form of multiple ejection episodes or a
continuous flow. This in turn further suggests that the removal of gas
from the nuclei of mergers is an efficient process. We return to this
subject in \S\,\ref{sec:feedback}.

\subsubsection{Neutral Gas}

The prevalence of minor-axis outflows collimated by a CO disk is seen
in neutral, as well as ionized, gas. F10565$+$2448, Mrk~231, Mrk~273,
and VV~705:NW all show evidence of expanding neutral flows along the
minor axis, roughly consistent with the morphology of the ionized
flows. In the neutral gas, however, there are exceptions:
F08572$+$3915 and F17207$-$0014 do not obviously show flows along the
CO minor axis, despite revealing outflowing neutral gas at other
locations. Furthermore, in F10565$+$2448 and VV~705:NW, there are
extensive areas of outflowing neutral gas away from the minor axis
flow.

In the case of neutral gas, there is structural ambiguity for two
reasons: (1) the neutral gas requires a strong continuum background
source, so the absorption-line signal-to-noise ratio decreases away
from the galaxy nucleus; and (2) the \nad\ resonant line may suffer
from emission-line filling near systemic \citep{prochaska11a}. We try
to minimize the former by deep exposures, but there is still a
limiting radius at which the signal drops below detectability. Though
we cannot provide quantitative constraints, we believe the latter to
be minimal, since the strong dust obscuration in these systems would
prevent outflowing emission from the far side of the outflow from
reaching us (see discussion in \S\,\ref{sec:wind-structure-ion} for
more on far-side line emission). This qualitative conclusion is
supported by the lack of any observed redshifted \nad\ emission in
these systems, even at large radii where the dust obscuration is
smaller than in the nuclei.

\subsubsection{Maximum Extent and Opening Angle}

The extended outflows in our dataset are consistently large,
stretching across much of the FOV. Five of the six systems in our
sample show outflowing gas up to the edge of the FOV, at scales
ranging from $3-5$~kpc from the nucleus. Thus, the sizes inferred from
our data are ultimately lower limits to the overall wind sizes in
these systems. As shown by \citet{martin06a}, outflows continue to
larger radii in long slit spectra. Using larger radii in our SRFW
model would only increase the measured masses, momenta, and energies.

The location of the neutral gas in nearby galaxies is not fully
constrained because it is seen in absorption. However, absorption-line
surveys can measure the average integrated covering factor of the
neutral gas ($\Omega/4\pi$; \S\,\ref{sec:SRFW}) by observing many
different systems. As seen through {\it nuclear} lines of sight, the
integrated covering factor increases with increasing star formation
rate. Systems with star formation rate (SFR) $\sim$ 1~\msun~yr$^{-1}$
show $\Omega/4\pi\sim0.1$ \citep{chen10a,bouche12a}, nearby LIRGs (SFR
$= 7-70$~\msun~yr$^{-1}$) have $\Omega/4\pi\sim0.4$ \citep{rupke05b},
and nearby ULIRGs (SFR $\ga$ 70~\msun~yr$^{-1}$) have
$\Omega/4\pi\sim0.8$ \citep{rupke02a,rupke05b,martin05a}. (In the
model of a bipolar wind, these correspond to opening angles on one
side of the disk of 60\arcdeg, 110\arcdeg, and 160\arcdeg,
respectively.)

The integrated covering factors of the ionized and neutral gas can
also be estimated from the SRFW, using the radii we assume
(\S\,\ref{sec:SRFW}). This way of estimating $\Omega/4\pi$ is much
less certain, especially since we assume a wind radius. A typical
value is $\Omega/4\pi \sim 0.1$, if we double the near-side covering
factor to account for a far-side outflow (Table~\ref{tab:modsrfw}).

For the neutral gas, values of $\Omega/4\pi$ determined from the SRFW
deprojection (0.1) are much lower than those based on observations of
many nuclear sightlines (0.8). The explanation we favor for this
discrepancy is the fact that the background continuum source declines
in brightness with increasing projected radius. Thus, the neutral gas
becomes increasingly difficult to detect with increasing projected
radius. Furthermore, these larger radius sightlines dominate the
measured $\Omega/4\pi$, once deprojected. Thus, the integrated
covering factors measured from our IFS observations are lower limits
to the true values, which are probably closer to $\sim$0.8.

\subsection{Ionized Gas Excitation} \label{sec:gas-excitation}

The excitation of the ionized gas in a wind serves as a signpost of
galaxy-scale processes. In disk galaxies in the local universe, gas
excitation reveals wind-driven shocks and ionization by the wind power
source. Outflowing warm ionized gas is typically shock-excited in
cases where the wind is driven by a starburst alone
\citep{veilleux02b,sharp10a,rich10a}, and photoionized by the AGN
radiation field in cases where an AGN is also present
\citep{veilleux03a,sharp10a}.

In local mergers, shock ionization is also prevalent
\citep{monrealibero06a,monrealibero10a,rich11a,soto12b,westmoquette12a}. Within
a given system, forbidden-to-recombination line ratios typically
increase with increasing radius and decreasing H$\alpha$ surface
brightness \citep{monrealibero10a,westmoquette12a}. This is suggestive
of a line flux mixing model, with a varying contribution of star
formation and shocks determining the line ratios
\citep{rich11a,westmoquette12a}.

Merger shocks could be driven by tidally-induced motions or gas
outflows. \citet{monrealibero06a,monrealibero10a} found a correlation
between line ratios and gas velocity dispersion and argued for a tidal
origin of merger shocks. High spectral resolution, multiple profile
emission line fitting, and comparison to shock models confirmed that
more shock-like line ratios were associated with higher velocity
dispersion \citep{rich11a}. \citet{soto12b} identified two subclasses
of shock- (or possibly AGN-excited) gas -- a subset with lower
velocity dispersions ($\sigma \la 150$~\kms) which is consistent with
a tidal origin and a subset with higher velocity dispersions ($\sigma
\ga 150$~\kms) that is probably outflowing.

From our data, the \nt/\ha\ and \oo/\ha\ line ratio maps for the
galaxies in our sample (e.g., Figure~\ref{fig:map_lrat_f10565}) are
consistent with these previously determined relationships. In
individual objects, line ratios in the rotating gas component
anticorrelate with \ha\ flux, correlate with FWHM, and increase with
increasing projected galactocentric radius. Within each system, small
regions can deviate from these trends; careful inspection of the maps
of each system show that the overall correlations do not tell the
entire story. The exact nature of the spatial variation of gas
excitation in each source is outside the scope of this work,
particularly since we probe a limited number of strong emission lines.

However, we are able to improve on the analysis of \citet{soto12b} by
studying individual systems, and on the analysis of \citet{rich11a} by
being able to identify outflows. Figure~\ref{fig:vel_v_fwhm} shows the
ionized gas velocity (for outflowing components) as a function of
FWHM, color-coded by \nt/\ha\ emission-line ratio. In a simple mixing
model between stellar photoionization and shock excitation and/or AGN
photoionization, increasing the percentage of line flux from shock
excitation or AGN photoionization to a particular spaxel will increase
\nt/\ha.

We find that, within a given system, spaxels with a given FWHM show
increased line ratios with increasing velocity (\vfifty) with respect
to systemic. Correspondingly, spaxels with a given velocity show
increased line ratios as FWHM increases. It is clear that bulk motions
(as traced by \vfifty) and unresolved bulk or randomized motions (as
traced by FWHM) are both responsible for shock ionization of the broad
emission line components in ULIRGs. In some systems (F10565$+$2448 and
Mrk~273) these motions are due almost completely to powerful
outflows. In others (VV~705:NW), some of the motions ($\vfifty \la
-50$~\kms\ and FWHM $\ga$ 400~\kms) are clearly due to outflows and
some ($|\vfifty| \la 50$~\kms\ and FWHM $\la$ 400~\kms) may be tidally
induced motions.

In the three galaxies in our sample with luminous AGN, photoionization
by the AGN could in principle contribute to the observed line
ratios. In F08572$+$3915:NW, the SDSS spectrum shows no evidence of
NLR emission, though the AGN in this source is heavily obscured. In
Mrk~231, there is presently no published spatial information on
extended high-ionization emission lines.  Finally, in Mrk~273, there
is AGN-photoionized gas with low velocities near the SW nucleus
\citep{colina99a}. However, the contribution of the AGN to the
excitation of the outflowing gas in Mrk~273 is unclear
\citep{colina99a}. Future IFS observations of these sources in
high-ionization lines will help to clarify the origin of ionization.

\subsection{The Multiphase Wind} \label{sec:multiphase-wind}

The present dataset permits the first spatially-resolved comparison
between the neutral and ionized gas phases in merger winds. Previous
spatially-resolved surveys have focused on either the ionized gas
\citep{heckman90a,lehnert96a,veilleux03a,sharp10a,westmoquette12a,soto12a,soto12b}
or the neutral gas \citep{martin06a}. Single aperture surveys have
produced evidence that the velocities of the neutral and ionized gas
phases in merger winds are not identical or correlated on a
system-to-system basis, except perhaps in the case of AGN
\citep{heckman00a,rupke05c,rupke05b}. \citet{soto12b} argue that the
ionized gas in ULIRG winds is of comparable mass to the neutral gas.

In Figures~\ref{fig:vel_a_v_e} and \ref{fig:fwhm_a_v_e}, we show that
there is a significant correlation between the (projected) neutral and
ionized gas velocities within a given galaxy for three of the five
galaxies in our sample with both ionized and neutral outflows:
F10565$+$2448, Mrk~273, and VV~705:NW. This correlation is present
whether \vfifty, \vtsig, or FWHM is considered. Furthermore, in all of
the systems with both ionized and neutral outflows, the phase with
higher velocity varies from point to point and object to object. In
2/5 systems with both ionized and neutral outflows, the ionized gas is
typically higher in velocity; in 2/5 the situation is reversed; and in
1/5 the velocities are comparable, on average.

The correlations we observe point to a physical connection between the
ionized and neutral outflow phases within a given galaxy, even if this
connection is complex. This complexity relates to unresolved
substructure along the line of sight that may be different for each
phase, as well as the fact that the tracers of these phases reflect
the underlying gas properties in different ways (e.g., density:
emission lines scale as $n^2$ and optically thin absorption lines as
$n$).

Two systems, F08572$+$3915:NW and Mrk~273, show the highest velocities
in our survey, $2000-3000$~\kms. Velocities this high are seen only in
the ionized gas, and not the neutral gas (or molecular gas; see
below). These high velocities are also only seen in galaxies with an
AGN, a point to which we will return in the next section.

The ratio between the mass, momentum, and energy contained within the
ionized and neutral phases varies from object to object
(Table~\ref{tab:mperats_in}), but in all systems the neutral gas
dominates the mass. In two systems (F08572$+$3915:NW and Mrk~273) the
ionized gas contributes significantly to the momentum and energy. The
ionized gas, however, is clearly affected by extinction in these
systems. Correcting for extinction would only increase the
contribution of the ionized gas to a system's mass, momentum, and
energy. It is unclear how extinction corrections would affect the
typical velocity of the gas. IFS observations of other Balmer lines in
these systems (in order to correct for extinction), or of ionized gas
in the NIR and MIR, would provide better constraints on the ionized
gas properties of the wind.

The wind in F10565$+$2448 is also dusty, as shown by the correlation
between continuum color and \nad\ equivalent width on a
spaxel-by-spaxel basis (\citealt{shih10a};
Figure~\ref{fig:weq_v_hstcol}) and the dust filaments emerging along
the minor axis (\S\,\ref{sec:f10565_kin} and
Figure~\ref{fig:dustpeak_f10565}). Both lower covering factor and
lower optical depth reduce the absorption line equivalent width. If
the gas and dust are closely associated, then a lower covering factor
makes the continuum bluer by allowing more light to emerge dust-free,
and a lower optical depth makes it bluer by simply reddening the
continuum to a lesser degree. However, the underlying stellar
population can also vary with position, and will introduce scatter
and/or slope changes (within a given system) or systematic offsets (on
a system-to-system basis).

The other galaxies in our dataset do not show the large dynamic range
in \hst\ F435W/F814W color and \nad\ equivalent width that are seen in
F10565$+$2448. F17207$-$0014 comes the closest, and shows a less
significant correlation, with more scatter. The data for the other
galaxies in our sample (except for Mrk~231, whose continuum structure
is too strongly impacted by the central source for accurate colors) do
not show correlations, but do lie on or near the correlation defined
by F10565$+$2448.

If the neutral phase of these winds is indeed dusty, then this
outflowing dust is partly responsible for obscuring the visible galaxy
continuum. Most of the UV light in ULIRGs is heavily obscured by
concentrations of nuclear dust \citep{soifer00a,goldader02a}; however,
on larger scales outflowing dust may play an important role as a
foreground screen in extinguishing the visible continuum. This effect
is seen at high spatial resolution in the kpc-scale dust filaments in
F10565$+$2448 (Figure~\ref{fig:dustpeak_f10565}). Future detailed
analyses of the optical colors and other extinction probes (e.g., the
Balmer decrement) will better illuminate this connection in other
systems.

The winds in ULIRGs also have a molecular component \citep[][Veilleux
et al. 2013, in
prep.]{fischer10a,feruglio10a,sturm11a,aalto12a,cicone12a}. Except for
the case of Mrk~231, the only information at present on the molecular
phase in these systems comes from single-aperture {\it Herschel}
observations of OH transitions. Three galaxies in our sample
(F08572$+$3915:NW, Mrk~231, and F17207$-$0014) are in the 6-galaxy
sample of \citet{sturm11a}. The single-aperture integrated velocity
profiles are not directly comparable to our spatially-resolved
data. However, it is clear that the high neutral gas velocities in
F08572$+$3915:NW and Mrk~231 are also observed in the molecular
gas. \citet{sturm11a} measure \vfifty $= -700$ and $-600$~\kms\ and
$\vtsig = -1300$ and $-1200$~\kms, respectively, in these
galaxies. These \vfifty\ (\vtsig) values lie somewhere in between the
average and maximum \vfifty\ (\vtsig) values that we measure in the
neutral gas for these two galaxies, suggesting that the neutral and
molecular gas are connected. However, the ionized gas in F08572$+$3915
reaches much higher velocities. In F17207$-$0014 (\vfifty $= -100$ and
maximum velocity of $-$370~\kms\ in the molecular gas), the neutral
gas shows significantly higher velocities than the molecular gas,
though the highest velocity neutral gas is at large radius.

The presence of molecular outflows alongside minor-axis neutral and
ionized flows in our IFS sample suggests a helpful picture. On kpc
scales, the dusty, molecular, neutral, and ionized wind emerges along
the minor axis of a molecular disk. The molecular gas is entrained as
the wind emerges from this disk, and is roughly co-spatial with the
neutral and ionized gas at small scales.
 
\subsection{Wind Power Sources}

Perhaps the most important outstanding issue in our understanding of
ULIRG winds is the relative importance of a starburst and AGN in
accelerating the wind in a particular galaxy. This is an important
question because feedback from AGN and/or stars are frequently invoked
in models to explain the emergence of QSOs from dusty, major mergers
of disk galaxies \citep{sanders88a} and the subsequent truncation of
star formation and black hole activity
\citep{springel05a,hopkins05a}. Attention has recently turned to AGN
feedback in particular in this context, due to its higher expected
velocities compared to stellar feedback and its rapid emergence after
the onset of AGN activity.

Indirect arguments for the presence of QSO-mode feedback in mergers
abound, but until recently data have been slim. There is now evidence
from multiple gas probes that the large-scale outflow in Mrk~231 is an
example of such feedback \citep{rupke05c,fischer10a,feruglio10a,
  sturm11a,rupke11a,aalto12a,cicone12a}. The argument for this outflow
being AGN- rather than starburst-driven relies on the relative power
of the central engine compared to the starburst, and on the remarkably
high velocities observed, which are higher than those seen in ULIRGs
with lower luminosity AGN or no AGN \citep{sturm11a,rupke11a}.

Surveys of outflows in mergers and other systems based on nuclear
spectra suggested that starbursts are sufficient to power large-scale
outflows in systems where both are present, and found that there was
no strong evidence for the AGN dominating large-scale outflow dynamics
\citep{rupke05c,krug10a}. However, spatially-resolved and multiphase
data, as in the case of Mrk~231, and the advent of precise nuclear AGN
luminosities for individual major mergers \citep{veilleux09a} have the
potential to alter this conclusion. Nuclear {\it Herschel} spectra of
nearby ULIRGs (including Mrk~231) show a correlation of molecular
outflow velocity with AGN luminosity (\citealt{sturm11a}; Veilleux et
al. 2013, in prep.). This suggests that these (unresolved) molecular
outflows are powered primarily by an AGN at the highest velocities.

The current sample contains three systems whose infrared luminosity is
clearly dominated by star formation, and for which there is no clear
evidence of AGN activity: F10565$+$2448, VV~705:NW, and
F17207$-$0014. Our sample also has three systems that contain luminous
AGN: F08572$+$3915, Mrk~231, and Mrk~273. How do the outflows in the
systems with AGN compare to those without?

\subsubsection{Comparison of Wind Structures}

Structurally, the winds in galaxies with and without AGN are not
obviously different. The data are consistent with the presence of
flows along the minor axis of a nuclear disk in almost every system,
with or without an AGN. Furthermore, superbubbles (F08572$+$3915:NW
and VV~705:NW) and widespread extended outflows (F10565$+$2448,
Mrk~231, Mrk~273, and F17207$-$0014) are each present in nuclei with
and without AGN. Ionized gas excitation correlates with \vfifty\ and
FWHM in both types of systems, suggesting shock excitation (though AGN
photoionization may still be present; \S\,\ref{sec:gas-excitation}).

A wind perpendicular to a nuclear disk is consistent with the
structures of starburst-driven winds in the local universe. However,
it differs from those of the narrow line region (NLR) outflows in
nearby Seyfert galaxies, which have random orientations with respect
to the galaxy disk
\citep{crenshaw00a,crenshaw00b,fischert10a,fischert11a,mullersanchez11a}.
A collimated outflow perpendicular to a galaxy disk is thus, at face
value, more consistent with a starburst-driven wind than a canonical
Seyfert NLR outflow. However, this inference is based on the study of
local disk galaxies, rather than mergers with massive molecular disks
and luminous AGN. The AGN in our sample, with luminosities
$3-10\times10^{45}$ erg~s$^{-1}$ \citep{veilleux09a}, are much
brighter than Seyferts. They are in fact radio-quiet QSOs, whose
narrow line region sizes can reach several kpc or more
\citep{bennert02a,greene11a}. (NLR outflows in Seyferts are typically
confined to sub-kpc scales.) The structure of NLR outflows in
radio-quiet QSOs is not well understood. From our data, we cannot
determine the exact relationship between the outflowing ionized gas in
our sample and the classical NLR, since we do not observe
high-ionization emission lines and previous data is inconclusive
(\S\,\ref{sec:gas-excitation}).

It is entirely plausible that the nuclear molecular disk could, on
sub-kpc scales, collimate and redirect an outflow driven by the AGN
\citep[see ][and references therein]{veilleux05a}. Examples in the
local universe include NGC~3079 \citep{cecil01a,cecil02a} and NGC~4388
\citep{veilleux03a}. In fact, the current data point to exactly this
occuring in the AGN in our sample, if these outflows are in fact
AGN-driven. NIR IFS with adaptive optics would be able to probe the
physics of launching and collimation at smaller scales.

\subsubsection{Comparison of Mass, Momentum, and
  Energy} \label{sec:comp-mpe}

We computed the mass, momentum, and energy in the outflows in our
sample using the SRFW and BSB models (\S\,\ref{sec:models} and
Table~\ref{tab:mpe}). The mass and momentum in the wind do not differ
significantly between galaxies with and without AGN. However, the
energy outflow rates in the galaxies with AGN are a factor $\sim$$4$
higher than those with no AGN, primarily because of higher velocities
(\S\,\ref{sec:comp-vel}).

The fraction of the momentum and energy in the ionized phase (compared
to the neutral phase) rises to over half in two of the AGN, and is
negligible in the rest of our sample (\S\,\ref{sec:multiphase-wind}
and Table~\ref{tab:mperats_in}). Again, this is due to the high
velocities reached in the ionized gas components of these two systems.

Table~\ref{tab:mperats_host} shows the energy outflow rate in the wind
compared to the mechanical luminosity of the starburst in each
system. The mechanical luminosity follows from the continuous
starburst models of \citet{leitherer99a}, with a correction applied
for a Salpeter IMF with lower mass cutoff of 0.1~\msun. In the model
of starburst winds driven by mechanical energy from stellar winds and
supernovae, there is sufficient mechanical energy from the starburst
in each system to power the outflow. For the galaxies without AGN, the
efficiency with which this mechanical energy must be converted into
the kinetic energy of the outflow is low: 3\%\ to 13\%. However, for
the AGNs in our sample, it is much higher: 100\%\ for F08572$+$3915
and Mrk~231, and $\ga$20\%\ for Mrk~273 (with only the ionized phase
accounted for in the latter case). A high thermalization efficiency is
inferred for M82 \citep{strickland09a}, though much of the kinetic
energy in this system may reside in the hot, X-ray emitting wind fluid
rather than the ionized and neutral gas components. The fraction of
the starburst mechanical energy that goes into the neutral and ionized
phases is certainly less than the total thermalization efficiency, and
quite possibly much less \citep{strickland00a}. The starburst
thermalization efficiencies required for the winds in the AGN in our
sample are thus uncomfortably high.

\begin{figure}
  %\plotone{f53.eps}
  \centering \includegraphics[width=6.5in]{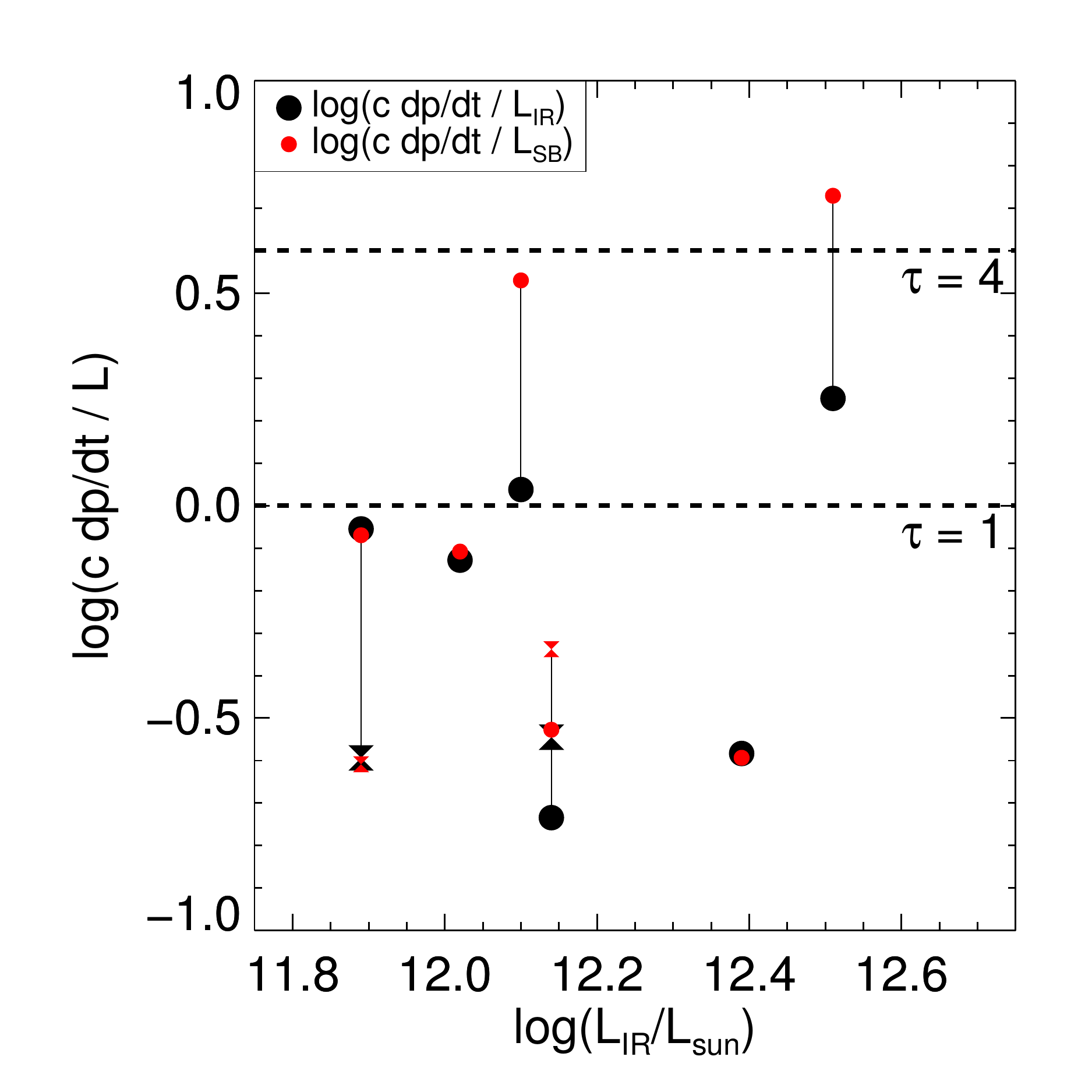}
  \caption{Momentum outflow rate normalized to infrared luminosity and
    starburst luminosity vs. infrared luminosity. Large, black symbols
    represent $c~dp/dt / L_{IR}$, and small, red symbols represent
    $c~dp/dt / L_{SB}$. Circles (hourglasses) are results from the
    SRFW (BSB) model. The dashed lines represent required optical
    depths of the outflow to photons if all of $L_{IR}$ or $L_{SB}$ is
    scattered by the wind. Note that powering the outflows with the
    starburst alone requires higher optical depths in some cases.}
  \label{fig:dpdt_v_lir}
\end{figure}

If instead the wind is driven by radiation pressure from the high
luminosity in these systems \citep[e.g.,][]{king03a,murray05a}, then
we can compare the available momentum injection rate ($\tau\lir/c$;
\citealt{murray05a}) to the wind's momentum outflow
rate. Table~\ref{tab:mperats_host} and Figure~\ref{fig:dpdt_v_lir}
show that these numbers are comparable -- i.e., $\tau$ must be near
unity for these outflows to be driven by radiation pressure. Given
that most of the luminosity of these systems emerges in the
far-infrared, almost every photon emerging from the nuclear power
sources interacts at least once with surrounding interstellar dusty
gas. Thus, radiative driving with $\tau = 1$ is a natural, though
certainly not the only, explanation for the powerful outflows in these
systems. The luminosity is injected at very different scales for the
starburst and AGN, however, since the MIR sizes of the AGN in our
sample are much smaller than the systems without AGN
\citep{soifer00a,imanishi11a}.

Any radiation pressure on the wind results from both a starburst and
any AGN. Thus, in cases where both are present, it is important to
understand how much each is contributing to the
outflow. Figure~\ref{fig:dpdt_v_lir} shows that simply intercepting
the radiation from the starburst in these systems requires the wind to
be quite optically thick ($\tau \ga 4$) in at least two cases. In the
radiation pressure model, it is simpler to assume that the entire
luminosity of the system, from both starburst and AGN, are powering
the outflow.

If an AGN drives the highest velocity flows in our sample, then the
energy outflow rate in the wind is a small fraction of the AGN's
luminosity ($0.1-0.4$\%), comfortably within the requirements of
recent models of AGN feedback
\citep{hopkins10a,fauchergiguere12a}. Table \ref{tab:mperats_host}
shows the ratio of the energy outflow rate in the wind to the AGN
luminosity in each system in our sample.

In summary, we reach several conclusions by studying the mass,
momentum, and energy in these winds. (1) The outflow mass in our
sample is independent of the galaxy's power source, within the
errors. (2) The starbursts in our sample are consistent with driving
outflows via radiation pressure or mechanical energy from stellar
winds and supernovae. (3) The data are consistent with AGN playing a
significant role in driving the outflow in those systems where an AGN
is present (though the mass, momenta, and energy do not conclusively
show that an AGN is contributing). The outflows in the AGN in our
sample are overall more energetic, have a more energetic ionized
phase, and require uncomfortably large thermalization efficiencies if
they are driven solely by mechanical energy from stellar winds and
supernovae. In terms of momentum and energy, however, they are
consistent with radiative driving by both the starburst {\it and} AGN
or with recent models of AGN feedback.

\subsubsection{Comparison of Velocities} \label{sec:comp-vel}

The energetic winds in the AGN in our sample are largely a result of
their large velocities compared to the
starbursts. Table~\ref{tab:vels} shows the spatially-averaged and
maximum values of \vfifty\ and \vtsig\ for each gas phase and each
galaxy, and Figure~\ref{fig:vel_v_lagn} shows how these values depend
on the AGN luminosity. We show the AGN luminosity computed from the
average AGN fraction of the bolometric luminosity from six MIR
diagnostics, as well as plausible uncertainty in this value using the
range of luminosities computed from individual MIR diagnostics
\citep{veilleux09a}. We assume that the AGN luminosities are upper
limits for the galaxies in our sample without evidence for AGN at
other wavelengths.

Both galaxies with and without an AGN host fast outflows. In systems
{\it without} an AGN, we find spatially-averaged \vfifty\ (\vtsig)
values of $-$250 ($-$450 to $-$650) \kms\ in the neutral gas and
$-$100 ($-$500) \kms\ in the ionized gas. The highest velocities
(maximum \vtsig) in each system (considering all spaxels and phases)
range from $-$900~\kms\ to $-$1250~\kms.

In systems {\it with} an AGN, we find spatially-averaged \vfifty\
(\vtsig) values are $-$100 to $-$400 ($-$450 to $-$900) \kms\ in the
neutral gas and $-$200 to $-$1500 ($-$700 to $-$2800) \kms\ in the
ionized gas.  The highest velocities (maximum \vtsig) reached in any
spaxel in each system range from $-$1450~\kms\ to $-$3350~\kms.

We conclude from this analysis that {\bf the presence of a QSO in
  major mergers yields higher outflow velocities.} The two systems
with the highest neutral gas velocities (Mrk~231 and F08572$+$3915)
contain QSOs, and the two systems with the highest ionized gas
velocities (Mrk~273 and F08572$+$3915) also contain QSOs. Though there
is clearly overlap between the velocity distributions of the galaxies
with and without AGN, only those with an AGN reach outflow velocities
exceeding $-$1200~\kms, and every system with an AGN exceeds this
velocity in at least one gas phase. Furthermore, {\bf only the ionized
  gas phase reaches velocities of $\ga$2000~\kms, and only in mergers
  hosting a QSO}. The same trend of outflow velocity with AGN
luminosity is seen in the molecular phase, again with velocities lower
than 2000~\kms\ (\citealt{sturm11a}; Veilleux et al. 2013, in prep.),
providing independent confirmation of this result.

How does deprojection affect these conclusions? In Mrk~273 and
VV~705:NW, we can deproject in some areas of our data using the BSB
model.  Figure~\ref{fig:v50_v_lagn} shows how this deprojection
affects the derived average velocities. In VV~705:NW, the system with
the highest observed velocities in a non-AGN, the velocities change
very little because the superbubble is observed face-on. In Mrk~273,
however, the superbubble is highly inclined, and the derived
velocities double. Thus, in these two sources, at least, deprojection
only strengthens the above conclusions.

This conclusion differs from that of \citet{rupke05c} and
\citet{krug10a}, who argued that the AGN in LIRGs and ULIRGs did not
accelerate the wind to higher velocities. However, these previous
studies relied on single-aperture spectra and a single gas phase
(neutral), as well as incomplete information about the AGN content of
individual systems. It is clear that combining spatially-resolved,
multiphase information, as in the current study, with more precise AGN
diagnostics is key to fully understanding the power sources of winds
in these galaxies.

The maximum velocities we observe in the neutral phase of ULIRG winds
are also higher than previously reported
\citep{heckman00a,rupke02a,rupke05c,rupke05b,martin05a,martin06a}. It
is clear from our spatially resolved data that this is because the
highest velocities are often not found along the line of sight to the
optical galaxy nucleus.

The {\it total} luminosity of the system does not appear to determine
the wind velocities in our sample (Figure~\ref{fig:vel_v_lir}). The
range of infrared luminosities in our sample spans only a factor of
four, and the two highest velocity systems are in the middle of the
luminosity range. This is consistent with previous results
\citep{rupke05b,rupke05c}.

The high velocities, masses, and energies we observe in ULIRGs with
luminous AGN are comparable to the properties of outflows inferred to
arise on kpc scales in other radio-quiet QSOs
\citep{moe09a,dunn10a,borguet13a}. The spatial constraints in other
radio-quiet QSOs depend on detailed photoionization modeling and are
thus not as direct as the measurements presented here. If they are
correct, the similarity of these two results suggests that such
QSO-driven flows are a general property of radio-quiet QSOs, whether
found in mergers or otherwise. A more detailed census of outflows in
radio-quiet QSOs of different types is clearly in order.

Outflow velocities up to $1500-2000$~\kms\ are also seen in what
appear to be purely starburst objects: so-called Lyman-break analogs,
and compact starbursts at $z\sim0.6$
(\citealt{tremonti07a,heckman11a,diamondstanic12a}; though obscured
AGN have not been ruled out in these systems). The highest outflow
velocities from these starbursts apparently require very compact star
formation \citep{overzier09a,heckman11a,diamondstanic12a}. ULIRGs also
contain compact star formation, with star formation rate surface
densities of $100-1000$~\msun~yr$^{-1}$ (based on their infrared
luminosities and molecular disk radii; \citealt{downes98a}),
comparable to Lyman break analogs and the $z\sim0.6$ compact
starbursts \citep{overzier09a,diamondstanic12a}. However, the compact
star formation in ULIRGs is heavily obscured and occurs amid massive
reservoirs of dense gas, while the Lyman-break analogs and $z\sim0.6$
compact starbursts are UV-luminous and relatively unobscured
\citep{heckman05a,overzier09a,diamondstanic12a}. Our data shows that
an AGN is required in ULIRGs to achieve the highest velocities. We
speculate that the higher velocities observed in UV-luminous compact
starbursts compared to starburst ULIRGs result from the outflow not
having to transfer energy and momentum into a dusty, dense, massive
interstellar medium around the starburst.

\subsection{Feedback in Major Mergers} \label{sec:feedback}

An important outstanding question is the impact of these massive,
high-velocity flows on their host galaxies and nuclear black
holes. These galaxies retain a central molecular disk
\citep{downes98a}, as well as rapid star formation and active galactic
nuclei, so any negative feedback has yet to be felt.

In Table~\ref{tab:mperats_host}, we show that the mass outflow rate,
while uncertain, is a significant fraction of the star formation
rate. Thus, the outflow evacuates gas at a rate similar to the star
formation rate as it propagates through the galaxy. The outflow mass
will only increase if we include the molecular phase
\citep{feruglio10a,sturm11a}. Furthermore, the outflow is highly
mass-loaded, meaning that most of its mass comes from gas it has
entrained while moving through the
galaxy. Table~\ref{tab:mperats_host} compares the mass outflow rate
(measured) with the mass input from the starburst itself (from models;
\citealt{leitherer99a}). These numbers suggest that, on average, about
75\% of the wind mass comes from entrained material, and perhaps more
when other gas phases are included.

As we argued in \S\,\ref{sec:wind-structure-ion}, a comparison of the
dynamical timescales of the wind ($\la$10~Myr) and merger
($\sim$100~Myr) suggests that the duty cycle of winds is high and that
these outflows are relatively continuous. Similarly, the timescale for
star formation to deplete the nuclear molecular gas reservoir in these
systems ($4-6\times10^9$~\msun\ for the four galaxies with CO
interferometric data; \citealt{downes98a}) is of order $30-40$~Myr
(the average star formation rate in our sample is 143~\smpy). The
timescale for the wind to deplete the nuclear gas reservoir is
slightly larger: $\sim$100~Myr, based on the mass outflow rates we
calculate here. Thus, the gas consumption, mass outflow, and merger
dynamical timescales are all similar, and all three processes will be
important in regulating star formation.

The overall picture of star formation self-regulation is a complex
one, but it is clear that gas outflows do play a role in governing
star formation. Molecular gas is part of the dusty, neutral wind we
detect here, and these outflows will serve to end star formation in
particular regions on a shorter timescale than would the exhaustion of
the gas reservoir by star formation itself. This is certainly one
means of negative feedback on star formation.

Feedback to AGN accretion, however, is difficult to probe on these
scales. Observations at smaller scales, comparison to simulations, and
population studies may all be necessary to better constrain whether
the fast AGN outflows in these systems are true feedback on the AGN
itself, and on what timescales.

\section{SUMMARY} \label{sec:summary}

In this study, we have presented deep integral field spectroscopy of
the optical emission and absorption lines in one luminous and five
ultraluminous infrared galaxies at $z\leq0.06$. Three of these
galaxies have dust-enshrouded QSOs, and three have no apparent AGN. By
fitting the \nad\ absorption line and multiple Gaussian components to
strong emission lines, we perform a spatial and spectral decomposition
of the data. We are able to separate rotating from outflowing or
tidally-disturbed gas, and have detected neutral and ionized outflows
on kpc scales in all six galaxies. This detailed data set has enabled
us to study the multiphase structure and power sources of the winds in
these systems.

We reach a number of major conclusions.

\begin{figure}
%  \plotone{f17.eps}
  \centering \includegraphics[width=6.5in]{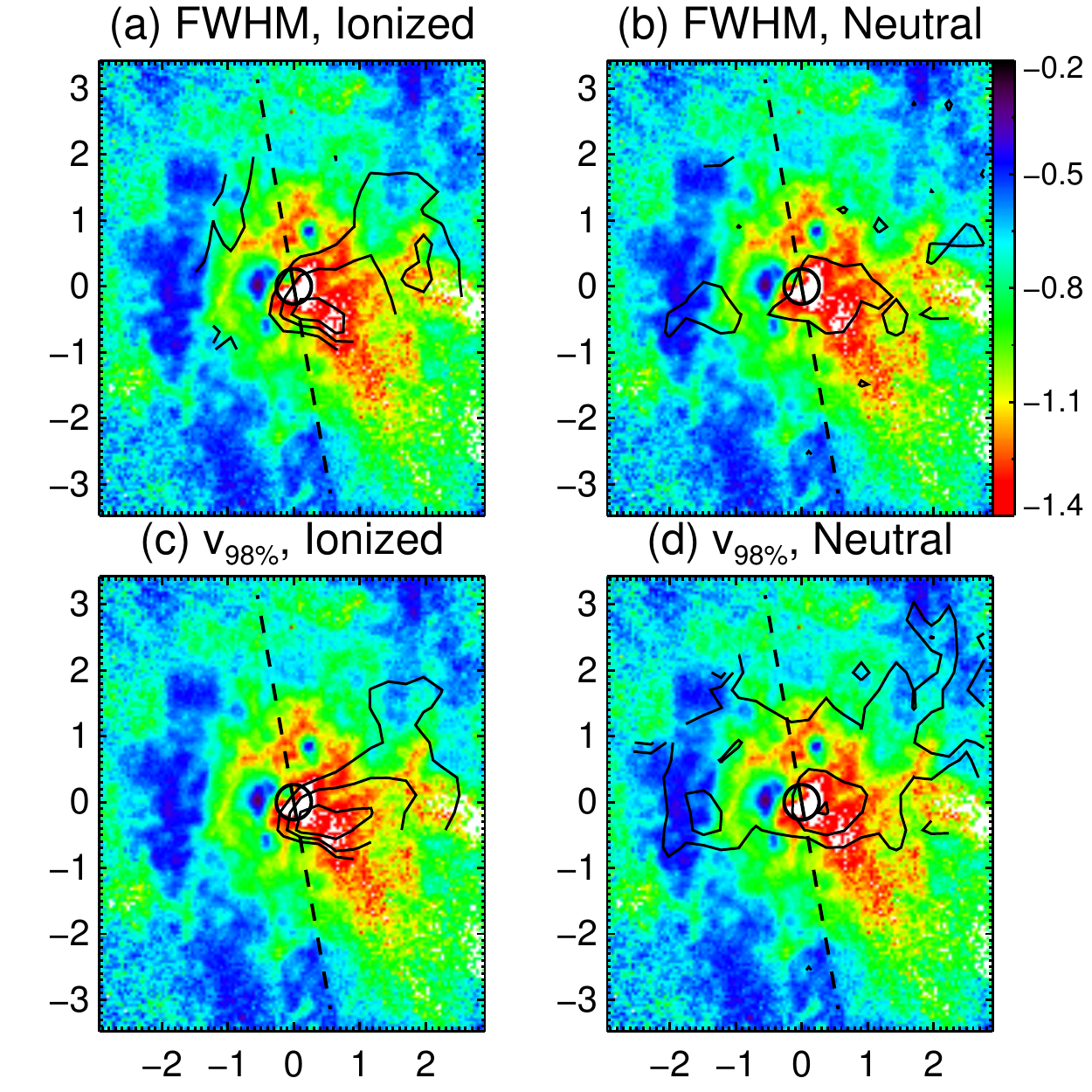}
  \caption{Color maps of the \hst\ continuum flux ratio,
    log(F435W/F814W), in F10565$+$2448, with velocity contours
    overlaid: (a) ionized gas FWHM (contours: 500, 600, 700~\kms); (b)
    neutral gas FWHM (contours: 400~\kms); (c) ionized gas \vtsig\
    (contours: -800, -700, -600~\kms); and (d) neutral gas \vtsig\
    (contours: -700, -600, -500~\kms). The dashed line represents the
    position angle of the CO disk \citep{downes98a}, and the open
    circle the GMOS continuum peak. The relative color and morphology
    of the blue clumps to the left (N) are consistent with unobscured
    star forming regions in the near side of a galaxy disk, while the
    dusty, high-velocity gas to the right (S) is almost certainly a
    minor-axis outflow. The minor-axis outflow reveals dusty filaments
    and unresolved motions of hundreds of \kms, consistent with dusty,
    shocked gas. If this outflow is emerging along the minor axis of
    the CO disk, it is inclined $\sim$20\arcdeg\ away from the line of
    sight \citep{downes98a}. (See \S\,\ref{sec:f10565_kin} for more
    details.)}
  \label{fig:dustpeak_f10565}
\end{figure}

\begin{figure}
%  \plotone{f29.eps}
  \centering \includegraphics[width=5.5in]{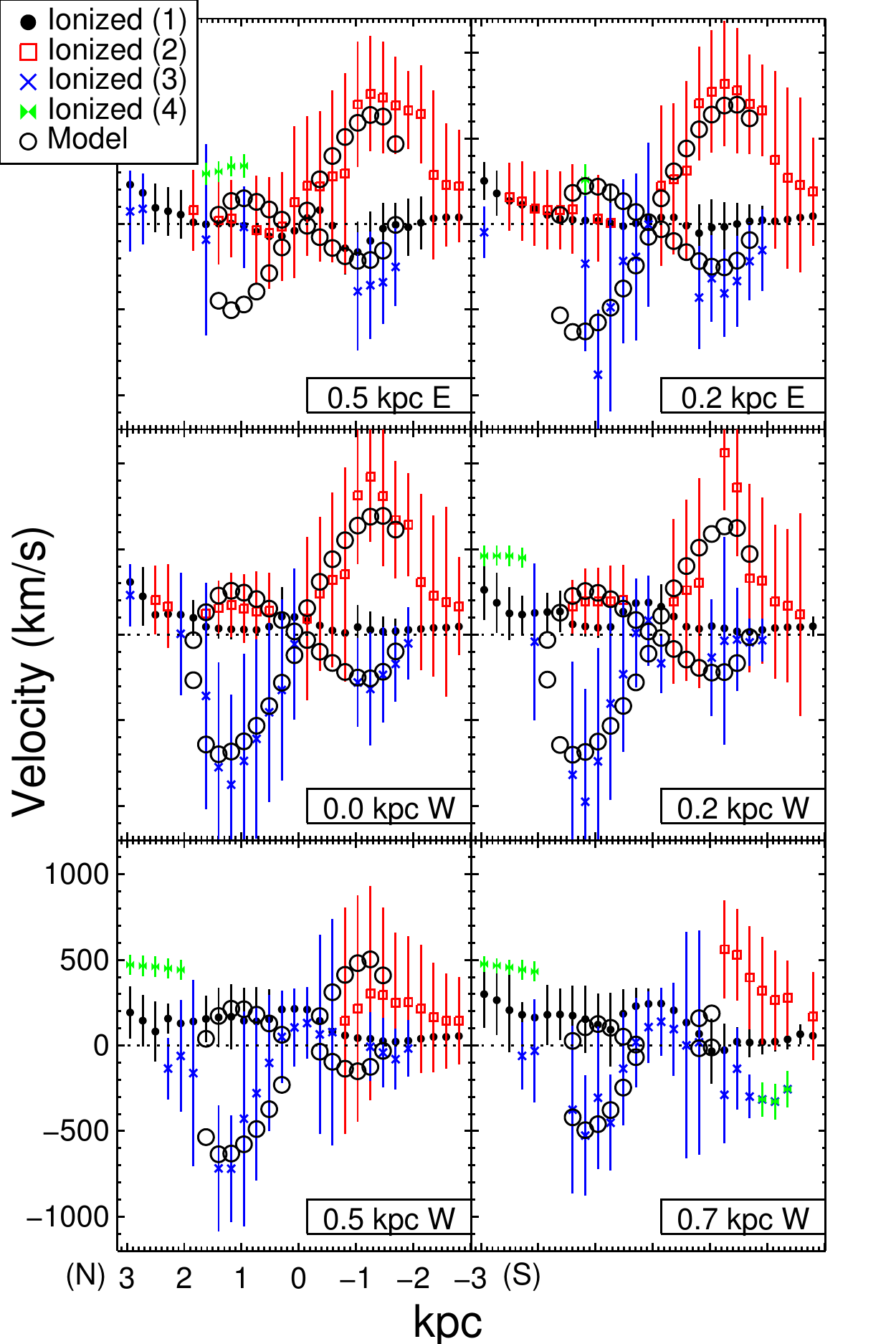}
  \caption{\scriptsize Position-velocity diagrams of the ionized gas
    in Mrk~273.  Six columns are shown, with horizontal distance from
    the outflow center in kpc overlaid on each. Each ionized gas
    velocity component is shown (1/black solid circles $=$ narrow,
    rotating; 2/red open squares $=$ broad, redshifted; 3/blue crosses
    $=$ broad, blueshifted; and 4/green bowties $=$ narrow,
    non-rotating; \S\,\ref{sec:mrk273-kinematics}), as well as the
    $n=2$ bipolar superbubble model fit to the data (open black
    circles). The vertical lines emerging from each point represent
    the spread in the velocity distribution ($\pm$1$\sigma$). Model
    parameters are given in Table~\ref{tab:modbbl}. (See
    \S\S\,\ref{sec:bsb} and \ref{sec:mrk273-bsb} for more details.)}
  \label{fig:pv_e_mrk273}
\end{figure}

\begin{figure}
  %\plotone{f41.eps}
  \centering \includegraphics[width=5.5in]{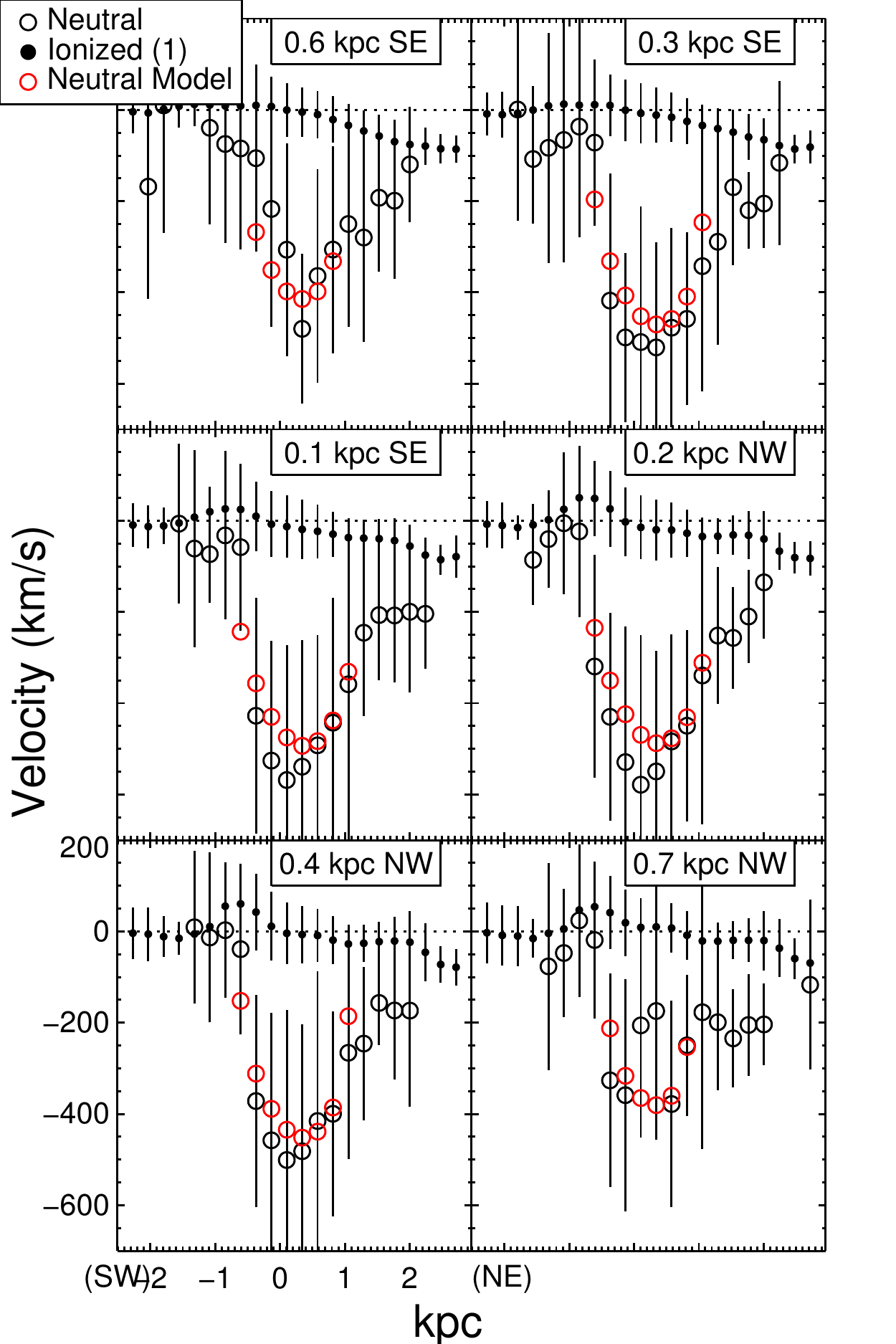}
  \caption{Position-velocity diagrams of the neutral gas in VV~705:NW.
    Six columns are shown, with horizontal distance from the outflow
    center in kpc overlaid on each. The neutral gas \vfifty\ values
    are shown (black open circles), as well as the $n=2$ bipolar
    superbubble model fit to the data (red open circles) and the
    narrow ionized gas component (small black points). The vertical
    lines emerging from each point represent the spread in the
    velocity distribution ($\pm$1$\sigma$). Model parameters are given
    in Table~\ref{tab:modbbl}. (See \S\,\ref{sec:vv705-kinematics} for
    more details.)}
  \label{fig:pv_a_vv705nw}
\end{figure}

\begin{figure}
  %\plotone{f49.eps}
  \centering \includegraphics[width=6.5in]{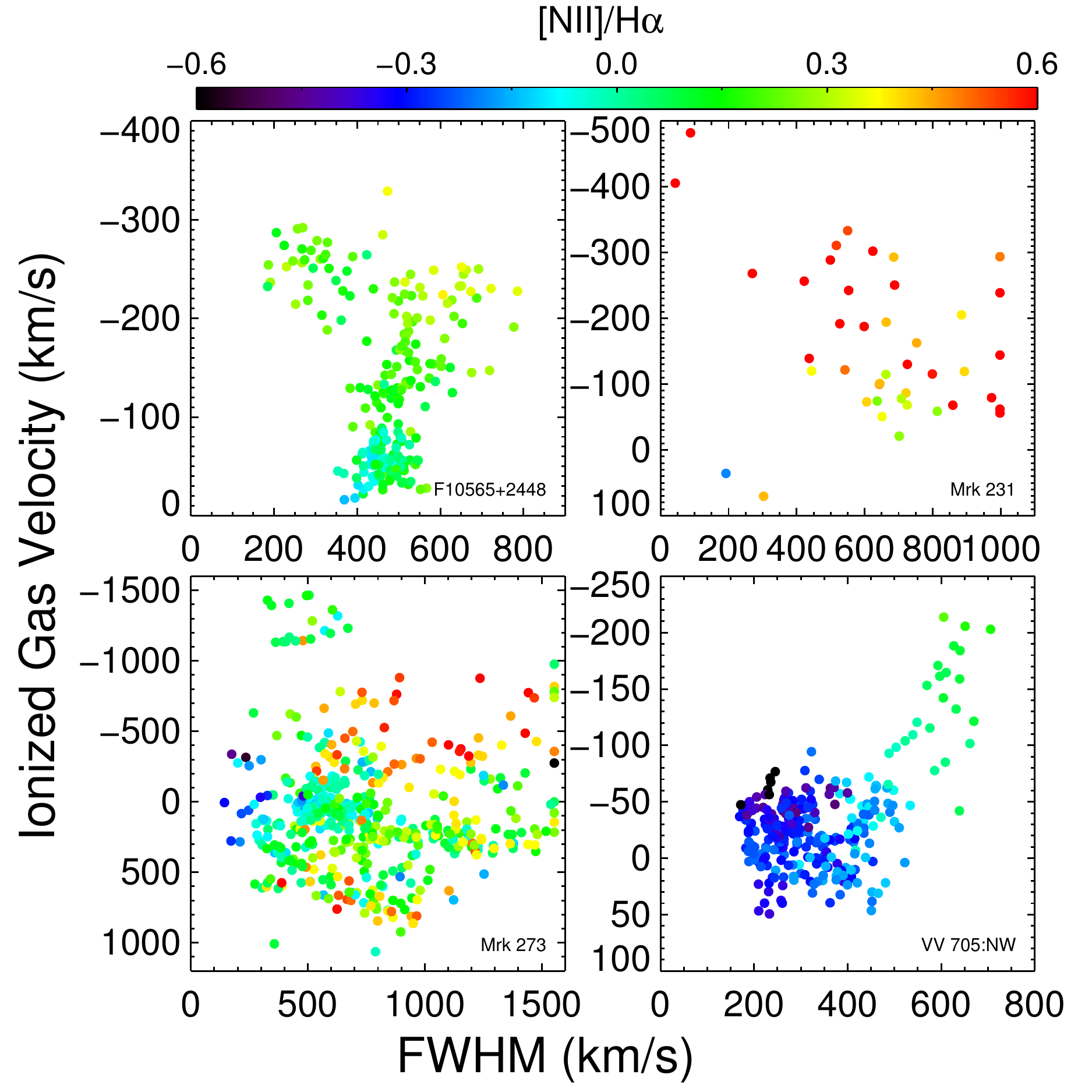}
  \caption{Ionized gas velocity (\vfifty) vs. FWHM for the broad
    emission line component, in the four nuclei in our sample with
    ionized gas outflows and resolved \nt/\ha\ data. The color of each
    point represents its \nt/\ha\ flux ratio. Within a given source,
    there is a clear trend for spaxels of higher velocity with respect
    to systemic and higher FWHM to have higher gas excitation. This
    suggests that the wind is being shock-excited, though there may be
    a contribution from AGN photoionization in those systems where an
    AGN is present, as well as a contribution from stellar
    photoionization at the lowest velocities
    (\S\,\ref{sec:gas-excitation}).}
  \label{fig:vel_v_fwhm}
\end{figure}

\begin{figure}
  %\plotone{f50.eps}
  \centering \includegraphics[width=6.5in]{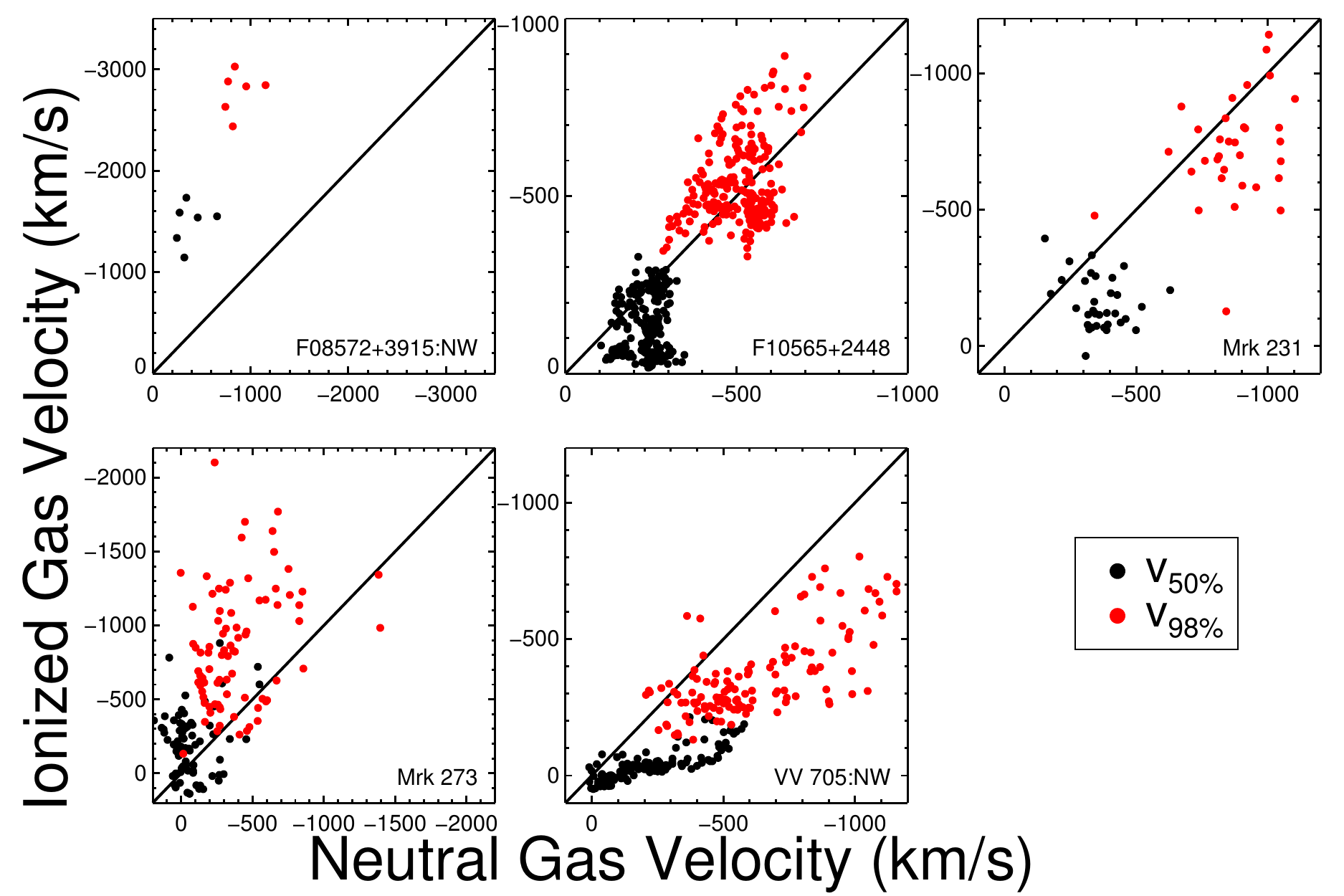}
  \caption{Ionized vs. neutral gas velocity, for the 5 nuclei in our
    sample with both ionized and neutral winds. Each point represents
    a spaxel with outflowing gas. The black points show \vfifty, and
    the red points \vtsig. The solid black line denotes
    equality. Within at least three systems (F10565$+$2448, Mrk~273,
    and VV~705:NW), there is a clear correlation between the
    velocities of the two gas phases
    (\S\,\ref{sec:multiphase-wind}). It is unclear, however, what
    causes the deviations from equality, and why these deviations
    differ from object to object.}
  \label{fig:vel_a_v_e}
\end{figure}

\begin{figure}
  %\plotone{f51.eps}
  \centering \includegraphics[width=6.5in]{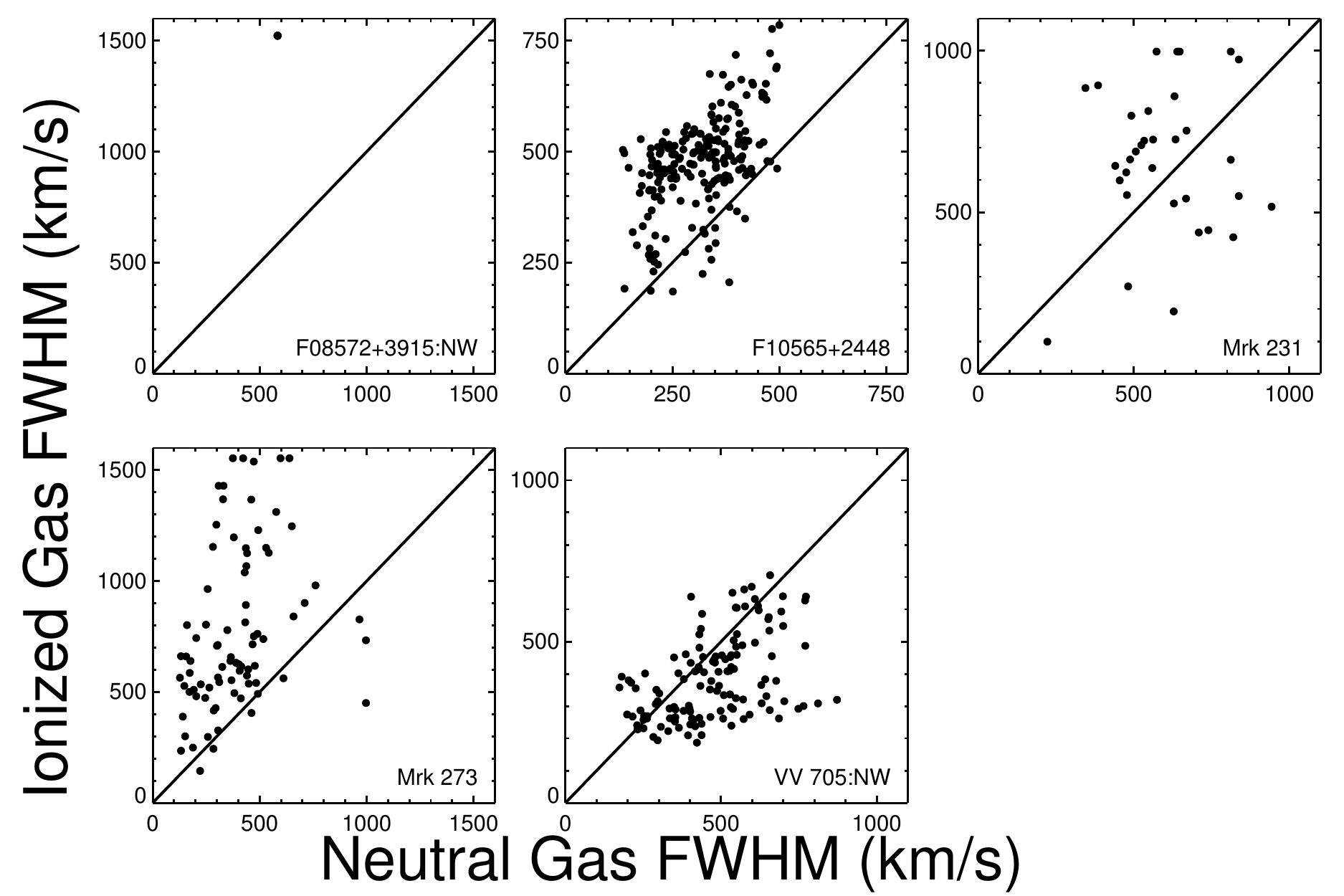}
  \caption{Ionized vs. neutral gas FWHM, for the 5 nuclei in our
    sample with both ionized and neutral winds. Each point represents
    a spaxel with outflowing gas. The solid black line denotes
    equality. Within at least three systems (F10565$+$2448, Mrk~273,
    and VV~705:NW), there is a clear correlation between the
    velocities of the two gas phases
    (\S\,\ref{sec:multiphase-wind}). It is unclear, however, what
    causes the deviations from equality, and why these deviations
    differ from object to object.}
  \label{fig:fwhm_a_v_e}
\end{figure}

\begin{figure}
  %\plotone{f52.eps}
  \centering \includegraphics[width=6.5in]{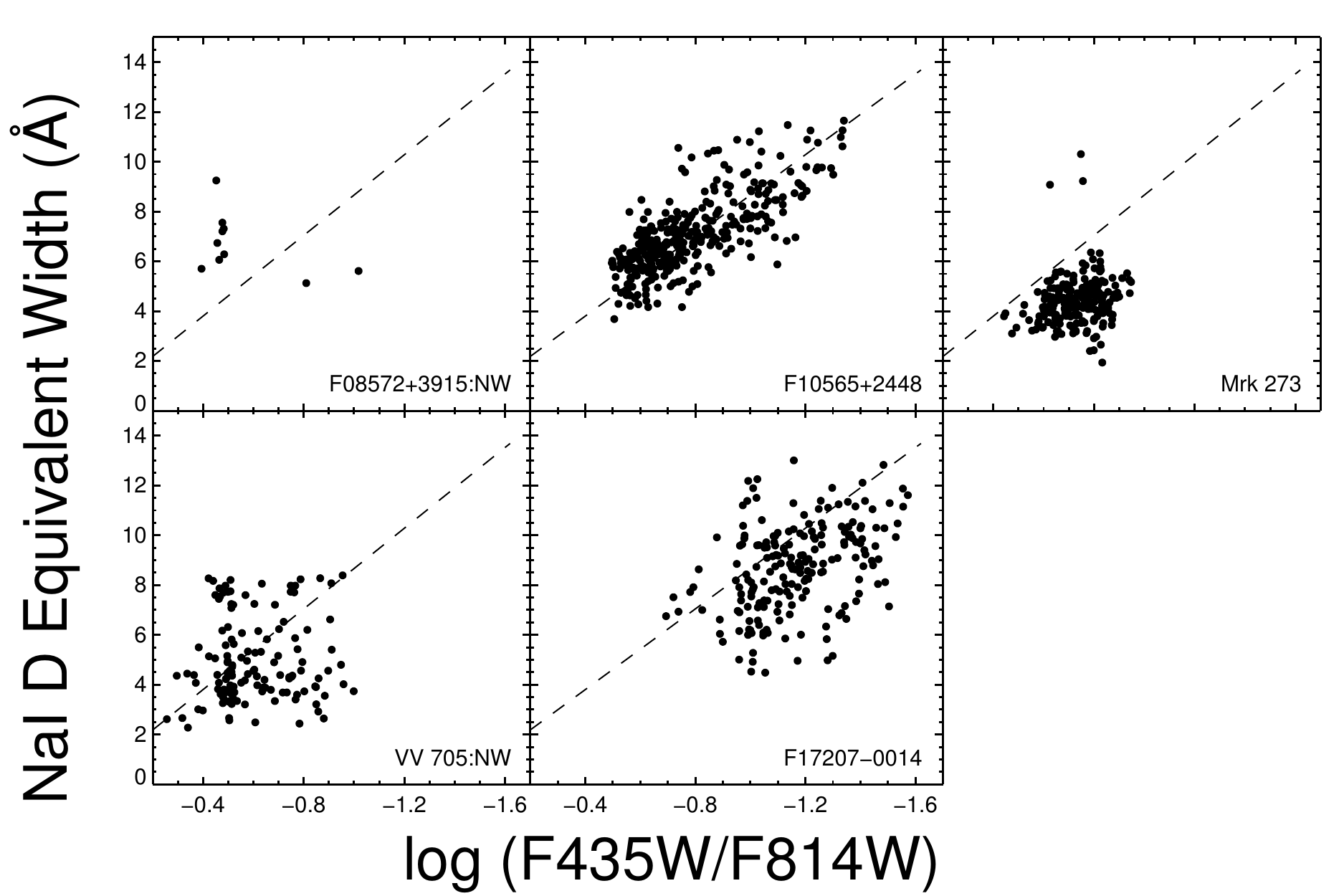}
  \caption{Rest-frame equivalent width of \nad\ vs. \hst\ color. The
    colors are binned and aligned to correspond to GMOS spaxels. The
    dashed line is a fit to F10565$+$2448. The correlations for
    F10565$+$2448 and F17207$-$0014 suggest that the dusty, neutral
    wind is attenuating the galaxy's continuum light. Though they have
    smaller dynamic range, the other systems fall near this
    correlation; these winds may also be dusty
    (\S\,\ref{sec:multiphase-wind}).}
  \label{fig:weq_v_hstcol}
\end{figure}

\begin{figure}
  %\plotone{f54.eps}
  \centering \includegraphics[width=6.5in]{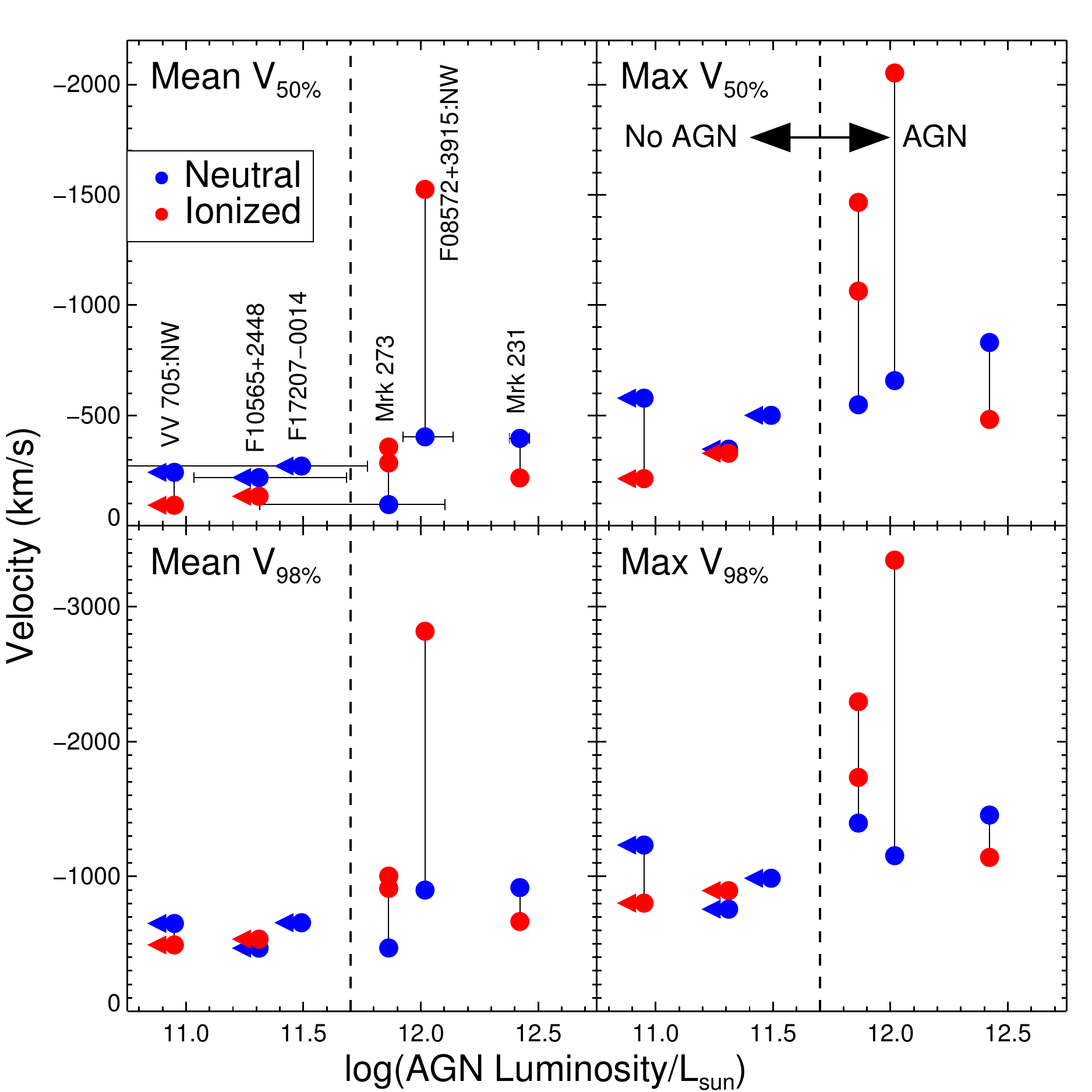}
  \caption{Velocity vs. AGN luminosity. Spatial averages and maximal
    values are computed from the projected velocities of outflowing
    components in all spaxels with outflowing gas. Blue (red) symbols
    represent neutral (ionized) gas. Two ionized gas symbols are shown
    for Mrk~273, representing the approaching and receding sides of
    the outflow. AGN luminosity is computed from the infrared
    luminosity and the average of six mid-infrared (MIR) diagnostics
    \citep{veilleux09a}; the horizontal error bars represent the range
    of AGN luminosities determined from individual MIR
    diagnostics. The AGN luminosities for systems without clear
    multiwavelength evidence for an AGN are shown as upper limits. As
    labeled by the arrows in the upper right panel, the dashed line
    divides systems with and without clear evidence for an AGN. It is
    clear that galaxies with a luminous AGN (or QSO; F08572$+$3914:NW,
    Mrk~231, and Mrk~273) reach higher outflow velocities (up to
    $1450-3350$~\kms\ in each object with a QSO;
    \S\,\ref{sec:comp-vel}). }
  \label{fig:vel_v_lagn}
\end{figure}

\begin{figure}
  %\plotone{f55.eps}
  \centering \includegraphics[width=6.5in]{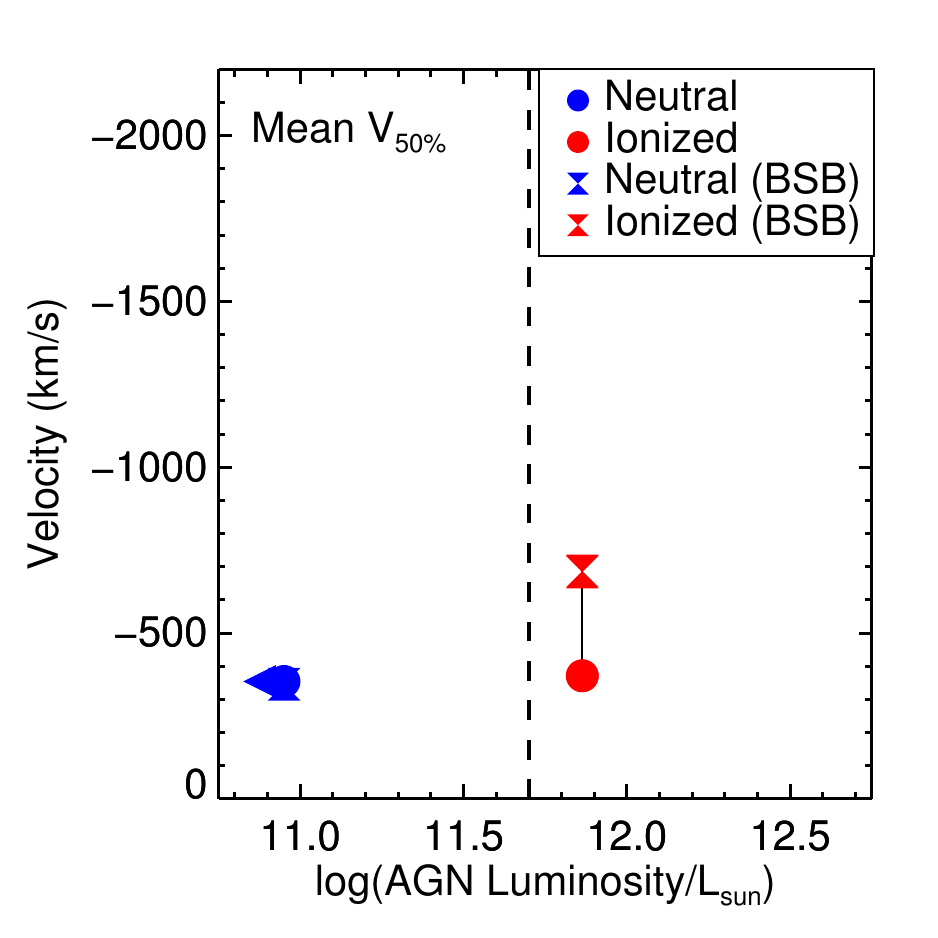}
  \caption{Velocity (\vfifty) vs. AGN luminosity for the two systems
    with a bipolar superbubble (Mrk~273 and VV~705:NW), to illustrate
    projection corrections to Figure~\ref{fig:vel_v_lagn}. Hourglasses
    show average deprojected velocities from the BSB model, and
    circles show average projected velocities. To enter these
    averages, spaxels must fall within the superbubble boundary and
    contain significant absorption or emission. The average velocities
    in VV~705:NW change little on deprojection, since it is a face-on
    bubble, while those in the highly inclined Mrk~273 bubble roughly
    double (\S\,\ref{sec:comp-vel}).}
  \label{fig:v50_v_lagn}
\end{figure}

\begin{figure}
  %\plotone{f56.eps}
  \centering \includegraphics[width=6.5in]{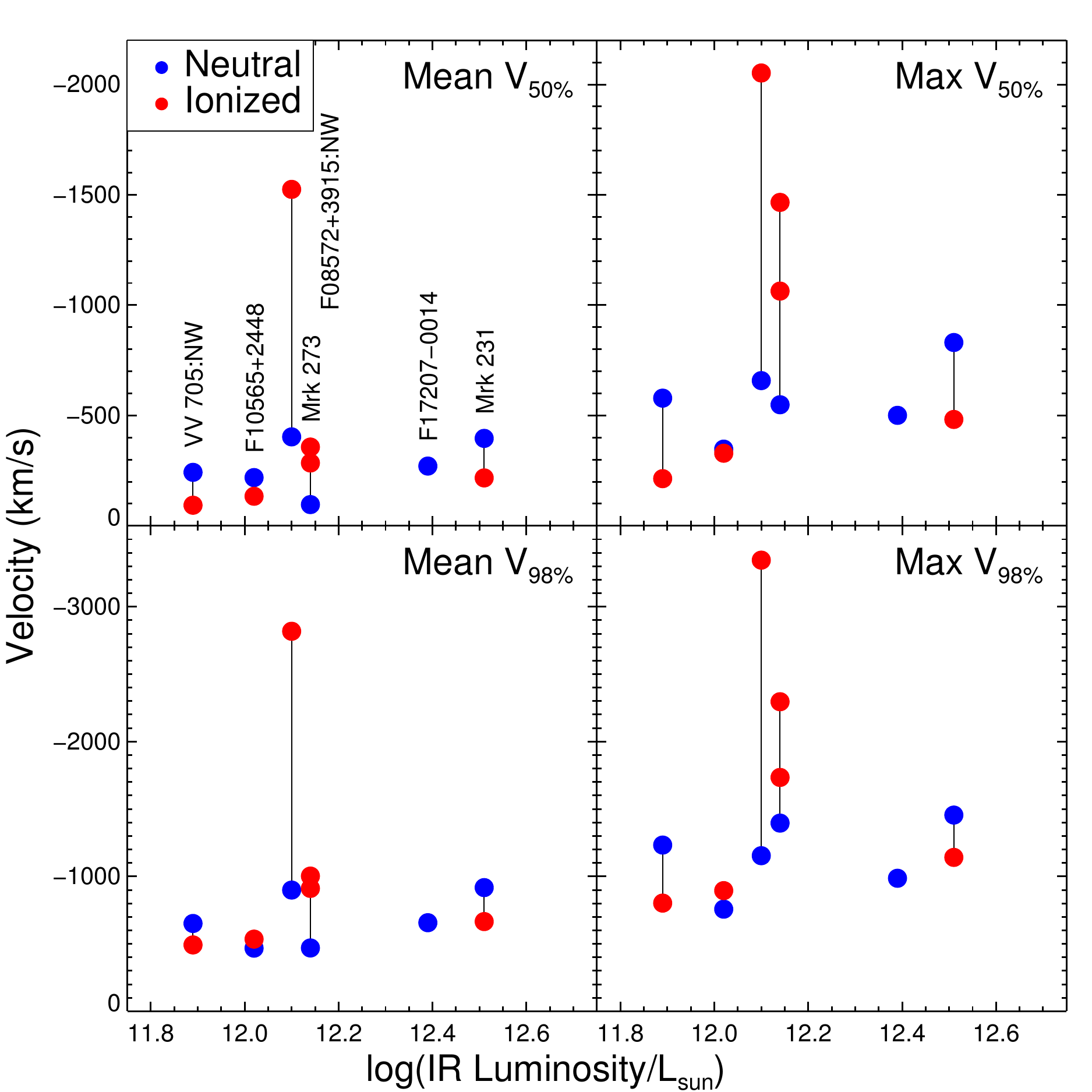}
  \caption{Velocity vs. total infrared luminosity (see
    Figure~\ref{fig:vel_v_lagn} for more details). There is no obvious
    relationship between velocity and infrared luminosity. The dynamic
    range in luminosity is only a factor of four, and the starburst
    and starburst$+$AGN systems are mixed; the sources with AGN are
    not the brightest.}
  \label{fig:vel_v_lir}
\end{figure}

\begin{enumerate}
\item{Where CO maps of the rotation of a nuclear disk are available (4
    of the 6 galaxies in our sample), the narrow component of the
    ionized gas matches the rotation of the molecular gas in those
    regions where they are coincident
    (\S\,\ref{sec:wind-structure-ion}). On larger scales, the narrow
    component of the ionized gas combines a continuation of the
    rotation of the nuclear disk with tidally-induced kinematics.}
\item{In 5 of 6 galaxies, the dynamics and structure of at least one
    gas phase at projected galactocentric radii of $1-2$~kpc are
    consistent with a bipolar outflow collimated by the nuclear
    molecular disk (\S\,\ref{sec:wind-structure-ion}). In two galaxies
    (Mrk~273 and VV~705:NW), this bipolar outflow takes the form of a
    bipolar superbubble that is detected on the basis of clear
    kinematic signatures (Figures~\ref{fig:pv_e_mrk273} and
    \ref{fig:pv_a_vv705nw}). On larger scales, the winds are less
    collimated. The prevalence of minor axis flows on small scales,
    and their short timescales compared to the merger dynamical time,
    suggests they are a somewhat continuous phenomenon.}
\item{Consistent with the results of \citet{martin06a}, the winds in
    our systems reach projected galactocentric radii of at least
    $2-5$~kpc, with the detectable size of the wind limited in most
    cases by either the size of our FOV or sensitivity.}
\item{Consistent with the results of \citet{sharp10a},
    \citet{rich11a}, and \citet{soto12b}, we find that the excitation
    of the broad component of the ionized gas is due primarily to
    wind-driven shocks, since emission line ratios in this component
    increase with both velocity with respect to systemic and with line
    width (Figure~\ref{fig:vel_v_fwhm}). A contribution from AGN
    photoionization is also likely in those galaxies with a QSO.}
\item{Within several systems, we find strong correlations between the
    velocities of the ionized and neutral gas phases on a
    spaxel-by-spaxel basis (Figures~\ref{fig:vel_a_v_e} and
    \ref{fig:fwhm_a_v_e}). The velocities are not often identical,
    however, and it is not yet clear what determines the relative
    velocities of the two phases.}
\item{We calculate that most of the mass, momentum, and energy in the
    wind is contained in the neutral gas phase
    (Table~\ref{tab:mperats_in}). The exceptions are the galaxies with
    the highest ionized gas velocities, which divide the momentum and
    energy roughly evenly between the ionized and neutral
    phases. However, the contribution from the ionized phase may be
    underestimated because of our inability to make extinction
    corrections, and uncertainties in the neutral gas values remain
    because of uncertain ionization corrections.}
\item{Spatially-resolved comparisons between the neutral phase of the
    wind and the optical continuum (Figures~\ref{fig:dustpeak_f10565}
    and \ref{fig:weq_v_hstcol}) show that the outflows are dusty in at
    least some cases, and thus partly responsible for obscuring the
    optical continuum emission in these systems on kpc scales.}
\item{In those systems without an AGN, the properties of the wind are
    consistent with being driven by radiation pressure or supernova
    mechanical energy. In those systems with a QSO, however, the
    required thermalization efficiencies appear too high to explain
    the energetics on the basis of supernova mechanical energy
    alone. Furthermore, if radiation pressure from a starburst alone
    drove the outflow in the QSOs, the wind optical depth would need
    to exceed unity by a factor of $\sim$4. In the systems with a QSO,
    the data are consistent with the wind being driven by a
    combination of radiation pressure and/or mechanical energy from
    the starburst {\it and} QSO (\S\,\ref{sec:comp-mpe}).}
\item{The galaxies with QSOs have the highest outflow velocities
    (Figure~\ref{fig:vel_v_lagn}), reaching projected velocities of at
    least 1450~\kms\ in each system with a QSO, and up to 3350~\kms\
    in one (F08572$+$3915:NW). Galaxies without AGN reach maximum
    projected velocities of $\sim$1000~\kms. We conclude that the QSO
    plays an important role in accelerating the wind in those systems
    in which a QSO is present.}
\item{The highest gas velocities ($2000-3000$~\kms) are reached in the
    ionized gas of two buried QSOs (Mrk~273 and F08572$+$3915).  Gas
    velocities this high are not seen in the neutral gas.}
\item{When the ionized, neutral, and molecular phases are combined,
    these merger outflows are massive enough to evacuate star-forming
    gas from the galaxy at a rate comparable to the star formation
    rate, suggesting that negative feedback is occurring
    (\S\,\ref{sec:feedback}).}
\end{enumerate}

Besides these scientific conclusions, the results of our study also
show that single aperture nuclear spectra, or even long slit spectra,
do not adequately capture the structural complexity of winds in major
mergers. In particular, IFS data is required to disentangle gas
motions due to rotation, tidal motions, and outflow. Furthermore, peak
velocities measured in nuclear apertures, or along slits in a random
direction, are not likely to probe the peak velocities in the outflow,
which in our data often occur outside the nuclear line of sight.

Though we are able to draw important conclusions from this sample, it
is clear that its scope is limited by its size. Larger IFS samples of
ULIRGs will put on firmer footing, or modify, the conclusions reached
here. Observations at higher spatial resolution (with, e.g., NIR IFS
adaptive optics observations) will further illuminate the launching
region and mechanism of these winds, particularly in those systems
where the wind is poorly resolved (the ionized phases in F08572$+$3915
and Mrk~231) or where there are multiple, closely-spaced nuclei
(Mrk~273). In QSO-powered systems, studying the inner regions will
also help to determine the physical mechanism by which the AGN helps
to power the outflow. Finally, observations of other emission lines
(e.g., \hb\ and \oth) will enable us to better constrain the
connection of the dust to the ionized gas phase, the mass of the
ionized gas phase, and the ionization mechanism of the emission line
gas.

\acknowledgments The authors are pleased to thank Alex Piazza for
creating the HST color maps; Vivian U and Anne Medling for sharing
their data and conclusions on Mrk~273 prior to publication; Nahum
Arav, Mike Crenshaw, Aleks Diamond-Stanic, Claude-Andre
Faucher-Giguere, Travis Fischer, Crystal Martin, Kurt Soto, and
Christy Tremonti for helpful conversations; the organizers of the 2011
Ringberg, AGN Winds in Charleston, and NRAO Jets and Outflows
meetings; and Jenny Shih for her work in prior stages of this
project. We also thank David Fanning and Craig Markwardt for their
excellent IDL software packages. Finally, we are grateful for the
detailed comments of the anonymous referee, whose input significantly
improved the paper. D.~S.~N.~R. was supported by NASA through a
Herschel data analysis award, by the Rhodes College Faculty
Development Endowment, and by a Cottrell College Science Award.

This work was based primarily on observations obtained at the Gemini
Observatory (program IDs GN-2007A-Q-107 and GN-2010A-Q-41), which is
operated by the Association of Universities for Research in Astronomy,
Inc., under a cooperative agreement with the NSF on behalf of the
Gemini partnership: the NSF (United States), the Science and
Technology Facilities Council (United Kingdom), the National Research
Council (Canada), CONICYT (Chile), the Australian Research Council
(Australia), Minist\'{e}rio da Ci\^{e}ncia, Tecnologia e
Inova\c{c}\~{a}o (Brazil) and Ministerio de Ciencia, Tecnolog\'{i}a e
Innovaci\'{o}n Productiva (Argentina).

The \hst\ observations described herein were obtained from the Hubble
Legacy Archive, which is a collaboration between the Space Telescope
Science Institute (STScI/NASA), the Space Telescope European
Coordinating Facility (ST-ECF/ESA) and the Canadian Astronomy Data
Centre (CADC/NRC/CSA).

Small portions of this work are based on observations made with the
Spitzer Space Telescope, obtained from the NASA/ IPAC Infrared Science
Archive, both of which are operated by the Jet Propulsion Laboratory,
California Institute of Technology under a contract with NASA.

Aspects of this work have made use of the SDSS. Funding for the SDSS
and SDSS-II has been provided by the Alfred P. Sloan Foundation, the
Participating Institutions, the NSF, the U.S. Department of Energy,
NASA, the Japanese Monbukagakusho, the Max Planck Society, and the
Higher Education Funding Council for England. The SDSS Web Site is
{\tt http://www.sdss.org/}.

\bibliography{apj-jour,dsr-refs}

\clearpage
  
\clearpage

\clearpage

\clearpage

\clearpage

\clearpage

\clearpage

\begin{deluxetable}{lclcccrlcc}
%  \rotate
  \tablecaption{Sample\label{tab:sample}}
  \tabletypesize{\scriptsize}
  \tablewidth{0pt}

  \tablehead{\colhead{Galaxy} & \colhead{Other name} & \colhead{$z$} &
    \colhead{log(\lir/\lsun)} & \colhead{$f_{25}/f_{60}$} &
    \colhead{AGN Frac.} & \colhead{SFR} &
    \colhead{Spec. Type} & \colhead{Merger Class} & \colhead{Nuc. Sep.} \\
    \colhead{(1)} & \colhead{(2)} & \colhead{(3)} & \colhead{(4)} &
    \colhead{(5)} & \colhead{(6)} & \colhead{(7)} & \colhead{(8)} &
    \colhead{(9)} & \colhead{(10)}}

  \startdata
  F08572$+$3915 & \nodata & 0.0584 (NW) & 12.10   & 0.241   & 0.72    &      61 & H/C (NW) & IIIb    & 5.0 \\
  \nodata       & \nodata & 0.0585 (SE) & \nodata & \nodata & \nodata & \nodata & H (SE) & \nodata & \nodata \\
  F10565$+$2448 & \nodata & 0.0431      & 12.02   & 0.105   & 0.17    &     150 & C       & triple  & triple \\
  Mrk~231       & F12540$+$5708 & 0.0422& 12.51   & 0.287   & 0.71    &     162 & S1      & IVb     & 0.0 \\
  Mrk~273       & F13428$+$5608 & 0.0373& 12.14   & 0.105   & 0.46    &     128 & S2      & IIIb    & 1.0 \\
  % VV~705        & F15163$+$4255, Mrk~848, I~Zw~107 & 0.04035 (NW) & 11.89 & 0.157 & 0.10 & 120 & C (NW) & IIIb & 7.5 \\
  VV~705        & F15163$+$4255, & 0.04035 (NW) & 11.89 & 0.157 & 0.10 & 120 & C (NW) & IIIb & 7.5 \\
  ~        & Mrk~848, I~Zw~107 & ~ & ~ & ~ & ~ & ~ & ~ & ~ & ~ \\
  \nodata       & \nodata & 0.0407 (SE) & \nodata & \nodata & \nodata & \nodata & C (SE)  & \nodata & \nodata \\
  F17207$-$0014 & \nodata & 0.0430      & 12.39   & 0.050   & 0.11    &     376 & H       & IVa     & 0.0
  \enddata

  \tablecomments{Col. 1: Galaxy name. Names starting with ``F'' are
    from the {\it Infrared Astronomical Satellite (IRAS)} Faint Source
    Catalog. Col. 2: Other common names. Col. 3: Redshift
    (geocentric), from the current study. Col. 4: $8-1000$\micron\
    infrared luminosity, based on {\it IRAS} or {\it Spitzer Space
      Telescope} fluxes from, in order of preference:
    \citet{veilleux09a}; \citet{surace04a}; \citet{sanders03a}; or
    NED. Col. 5: 25-to-60 \micron\ flux density ratio. Col. 6: ``AGN
    fraction'', or fraction of the bolometric luminosity of the galaxy
    due to an AGN. This value is determined from MIR diagnostics
    \citep{veilleux09a} except in the case of VV~705, for which we
    assume a value of 10\%\ (\S\,\ref{sec:vv705}). For ULIRGs,
    $L_{bol}\sim1.15\lir$ \citep{sanders96a,veilleux09a}. Col. 7: Star
    formation rate in \smpy, as determined from the AGN fraction, the
    infrared luminosity, and the prescription of
    \citet{kennicutt98a}. Col. 8: Optical spectral type, from
    \citet{yuan10a} except for F08572$+$3915; we determine types for
    this system in \S\,\ref{sec:f08572_ps}. H $=$
    \ion{H}{2}-region-like, C $=$ composite, S1/2 $=$ Seyfert
    1/2. Col. 9: Merger class based on system from
    \citet{veilleux02a}. Classes are taken from this paper for
    F08572$+$3914 and Mrk~231. For other sources: F10565$+$2448 class
    was changed from IIIa to triple based on archival \hst\ images;
    Mrk~273 class was changed from IVb to IIIb based on
    \citet{scoville00a} and \citet{iwasawa11a}; and VV~705 and
    F17207$-$0014 were newly classified. Col. 10: Nuclear separation
    in kpc, from \citet{scoville00a} except for F10565$+$2448 and
    VV~705. The separation for VV~705 is based on archival \hst\
    data.}
  
\end{deluxetable}

\begin{deluxetable}{lclrl}
  \tablecaption{Observations\label{tab:obs}}
  \tablewidth{0pt}

  \tablehead{\colhead{Galaxy} & \colhead{Dates} & \colhead{$t_{exp}$}
    & \colhead{PA} & \colhead{Dim} \\
    \colhead{(1)} & \colhead{(2)} & \colhead{(3)} & \colhead{(4)} &
    \colhead{(5)}}

  \startdata
  F08572$+$3915:NW & 10apr14, 10may11, 10may19 & $3\times1800$s & 81$\arcdeg$ & $16\times22$ \\
  F08572$+$3915:SE & 10apr03, 10apr14, 10may19 & $4\times1800$s & 81$\arcdeg$ & $16\times24$ \\
  F10565$+$2448    & 07feb12                   & $7\times1800$s & 90$\arcdeg$ & $22\times26$ \\
  Mrk~231          & 07jul19                   & $5\times~900$s & 35$\arcdeg$ & $21\times25$ \\
  Mrk~273          & 10jun13                   & $4\times1800$s &  0$\arcdeg$ & $20\times27$ \\
  VV~705:NW        & 10apr14, 10apr20, 10apr22 & $5\times1800$s & 35$\arcdeg$ & $16\times22$ \\
  VV~705:SE        & 10apr14, 10apr20, 10apr22 & $4\times1800$s & 35$\arcdeg$ & $16\times22$ \\
  F17207$-$0014    & 10may09, 10may10, 10jun03 & $4\times1800$s & 90$\arcdeg$ & $22\times30$
  \enddata

  \tablecomments{Col. 1: Galaxy name.  Col. 2: UT dates of
    observations (in YYmmmDD format). Col. 3: Number of exposures
    $\times$ length of each exposure, in seconds. Col. 4: Position
    angle, E of N, of long axis of final data cube. Col. 5: Size of
    final data cube, in units of 0\farcs3 spaxels.}
  
\end{deluxetable}

\begin{deluxetable}{lcrrrrrr}
%  \rotate
  \tablecaption{Outflow Velocity Statistics\label{tab:vels}}
  \tablewidth{0pt}

  \tablehead{\colhead{Galaxy} & \colhead{Phase} &
    \colhead{$\langle$FWHM$\rangle$} & \colhead{Max. FWHM} &
    \colhead{$\langle \vfifty \rangle$} & \colhead{Max. \vfifty} &
    \colhead{$\langle \vtsig
      \rangle$} & \colhead{Max. \vtsig} \\
    \colhead{(1)} & \colhead{(2)} & \colhead{(3)} & \colhead{(4)} &
    \colhead{(5)} & \colhead{(6)} & \colhead{(7)} & \colhead{(8)} }

  \startdata
 F08572+3915:NW  &        neutral  &     583  &     583  &    -403  &    -658  &    -898  &   -1153 \\
        \nodata  &        ionized  &    1522  &    1522  &   -1524  &   -2052  &   -2817  &   -3345 \\
\hline
    F10565+2448  &        neutral  &     293  &     544  &    -218  &    -347  &    -468  &    -757 \\
        \nodata  &        ionized  &     472  &     785  &    -133  &    -328  &    -535  &    -894 \\
\hline
        Mrk 231  &        neutral  &     613  &     996  &    -396  &    -830  &    -917  &   -1455 \\
        \nodata  &        ionized  &     527  &     997  &    -217  &    -482  &    -665  &   -1141 \\
\hline
        Mrk 273  &        neutral  &     439  &     996  &     -96  &    -548  &    -469  &   -1395 \\
        \nodata  & ionized (blue)  &     760  &    1553  &    -356  &   -1465  &   -1002  &   -2295 \\
        \nodata  &  ionized (red)  &     737  &    1553  &     284  &    1063  &     911  &    1734 \\
\hline
      VV 705:NW  &        neutral  &     481  &     872  &    -242  &    -578  &    -651  &   -1232 \\
        \nodata  &        ionized  &     469  &     705  &     -93  &    -213  &    -492  &    -802 \\
\hline
    F17207-0014  &        neutral  &     454  &     835  &    -270  &    -500  &    -656  &    -986 \\
\hline
\hline
  Starburst ULIRGs\tablenotemark{a} & neutral & 330 & ~ & $-$170 & ~ & $-$450 & ~
  \enddata

  \tablecomments{Col. 2: Gas phase. Col. 3$-$4: Mean and maximum of
    full width at half maximum of velocity distribution, in
    \kms. Col. 5$-$6: Mean and maximum of velocity at center of
    (Gaussian) distribution. Col. 7$-$8: Mean and maximum of $\vtsig
    \equiv \vfifty - 2\sigma$.}

  \tablenotetext{a}{Computed from starburst-driven ULIRGs in
    \citet{rupke05b}. This is a sample average, as opposed to the
    spatial averages in the current sample.}

\end{deluxetable}

\begin{deluxetable}{lcccc}
  \tablecaption{Single Radius Free Wind
    Parameters \label{tab:modsrfw}} \tablewidth{0pt}

  \tablehead{\colhead{Galaxy} & \colhead{Phase} & \colhead{Radius} &
    \colhead{$\Omega/4\pi$} \\
    \colhead{(1)} & \colhead{(2)} & \colhead{(3)} & \colhead{(4)}}
  
  \startdata
 F08572+3915:NW  &        neutral  & 2.0  & 0.06/0.02 \\
        \nodata  &        ionized  & 2.0  &      0.18 \\
\hline
    F10565+2448  &        neutral  & 5.0  & 0.17/0.12 \\
        \nodata  &        ionized  & 5.0  &      0.11 \\
\hline
        Mrk 231  &        neutral  & 3.0  & 0.24/0.14 \\
        \nodata  &        ionized  & 3.0  &      0.06 \\
\hline
        Mrk 273  &        neutral  & 4.0  & 0.05/0.03 \\
        \nodata  &        ionized  & 4.0  &      0.16 \\
\hline
      VV 705:NW  &        neutral  & 3.0  & 0.14/0.08 \\
        \nodata  &        ionized  & 3.0  &      0.07 \\
\hline
    F17207-0014  &        neutral  & 5.0  & 0.10/0.06
  \enddata

  \tablecomments{Col. 2: Gas phase. Col. 3: Fixed wind radius, in
    kpc. Col. 4: Integrated covering factor, expressed as a fraction
    of 4$\pi$. The first number for the neutral phase does not include
    a correction for the fitted covering factor of the absorption line
    in each spaxel; the second does (see \S\,\ref{sec:SRFW}).}

\end{deluxetable}

\begin{deluxetable}{lcccrrrc}
  \tablecaption{Bipolar Superbubble Parameters\label{tab:modbbl}}
  \tablewidth{0pt}

  \tablehead{\colhead{Galaxy} & \colhead{Phase} & \colhead{$R_z$} &
    \colhead{$R_{xy}$} & \colhead{$i$} & \colhead{PA} &
    \colhead{$v_{max}$} & \colhead{$n$} \\
    \colhead{(1)} & \colhead{(2)} & \colhead{(3)} & \colhead{(4)} &
    \colhead{(5)} & \colhead{(6)} & \colhead{(7)} & \colhead{(8)} }

  \startdata
  Mrk~273   & ionized & 1.0$\pm$0.1 & 0.8$\pm$0.1        & 75$\pm$5\arcdeg & $-5\pm10$\arcdeg  & $1600\pm500$ & 2\tablenotemark{a} \\
  \nodata   & \nodata & 1.0$\pm$0.1 & 0.8$\pm$0.1        & 75$\pm$5\arcdeg & $-5\pm10$\arcdeg  & $2100\pm500$ & 3\tablenotemark{a} \\
  VV~705:NW & neutral & 2$\pm$1     & 1\tablenotemark{a} & 6$\pm$6\arcdeg  & $35\pm40$\arcdeg & $500\pm150$  & 2\tablenotemark{a} \\
  \nodata   & \nodata & 3$\pm$2     & 1\tablenotemark{a} & 4$\pm$4\arcdeg  & $35\pm40$\arcdeg & $600\pm150$  & 3\tablenotemark{a}
  \enddata

  \tablecomments{Col. 2: Gas phase. Col. $3-4$: Semi-principal axes of
    the ellipsoid bubble, in kpc. The $z$ axis is the common axis
    shared by the two counterpropagating bubbles in each
    system. Col. 5: The inclination of the bubble's $z$ axis with
    respect to the line of sight (for $i=0\arcdeg$, the $z$ axis and
    LOS are parallel). Col. 6: The position angle of the bubble's $z$
    axis (for PA $=0\arcdeg$, the $z$ axis of the near side bubble is
    directed N). Col. 7: The (maximum) velocity of the bubble along
    the $z$ axis, in \kms. Col. 8: The power-law index of the bubble's
    velocity field: $v = v_z (r/r_z)^n$, where $r$ is the distance
    from the center of the wind, and $v_z$ and $r_z = 2R_z$ are the
    velocity and distance along the $z$ axis.}

  \tablenotetext{a}{This parameter is fixed.}

\end{deluxetable}

\begin{deluxetable}{lcccccccc}
  \tablecaption{Mass, Momentum, and Energy\label{tab:mpe}}
  \tablewidth{0pt}

  \tablehead{\colhead{Galaxy} & \colhead{phase} & \colhead{model} &
    \colhead{$M$} & \colhead{$dM/dt$} & \colhead{$p$} &
    \colhead{$c\times dp/dt$} & \colhead{$E$} & \colhead{$dE/dt$}\\
    \colhead{~} & \colhead{~} & \colhead{~} & \colhead{\msun} &
    \colhead{\msun/yr} & \colhead{dyne~s} & \colhead{\lsun} &
    \colhead{erg} & \colhead{erg/s} \\
    \colhead{(1)} & \colhead{(2)} & \colhead{(3)} & \colhead{(4)} &
    \colhead{(5)} & \colhead{(6)} & \colhead{(7)} & \colhead{(8)} &
    \colhead{(9)} }

  \startdata
 F08572+3915:NW  &neutral  & SRFW  &  7.88  &  1.39  & 48.98  & 11.95  & 56.69  & 42.78 \\
        \nodata  &ionized  & SRFW  &  6.93  &  0.88  & 48.47  & 11.84  & 56.58  & 43.05 \\
        \nodata  &  total  & SRFW  &  7.92  &  1.51  & 49.10  & 12.20  & 56.94  & 43.24 \\
\hline
    F10565+2448  &neutral  & SRFW  &  9.10  &  1.81  & 49.80  & 11.95  & 57.17  & 42.42 \\
        \nodata  &ionized  & SRFW  &  7.88  &  0.16  & 48.15  & 10.00  & 56.06  & 40.93 \\
        \nodata  &  total  & SRFW  &  9.12  &  1.82  & 49.81  & 11.95  & 57.20  & 42.44 \\
\hline
        Mrk 231  &neutral  & SRFW  &  8.92  &  2.25  & 50.02  & 12.82  & 57.74  & 43.68 \\
        \nodata  &ionized  & SRFW  &  6.97  & -0.10  & 47.67  & 10.29  & 55.50  & 41.17 \\
        \nodata  &  total  & SRFW  &  8.92  &  2.25  & 50.02  & 12.82  & 57.74  & 43.69 \\
\hline
        Mrk 273  &neutral  & SRFW  &  8.23  &  0.90  & 48.79  & 11.07  & 56.50  & 41.87 \\
        \nodata  &ionized  & SRFW  &  8.07  &  0.87  & 48.77  & 11.25  & 56.76  & 42.27 \\
        \nodata  &  total  & SRFW  &  8.46  &  1.19  & 49.08  & 11.47  & 56.95  & 42.41 \\
        \nodata  &ionized  &  BSB  &  7.27  &  1.18  & 48.35  & 11.66  & 56.36  & 42.77 \\
\hline
      VV 705:NW  &neutral  & SRFW  &  8.53  &  1.59  & 49.36  & 11.89  & 56.97  & 42.61 \\
        \nodata  &ionized  & SRFW  &  7.66  &  0.28  & 48.05  & 10.13  & 55.92  & 41.09 \\
        \nodata  &  total  & SRFW  &  8.59  &  1.61  & 49.38  & 11.90  & 57.01  & 42.62 \\
        \nodata  &neutral  &  BSB  &  8.22  &  1.33  & 48.86  & 11.36  & 56.55  & 42.16 \\
\hline
    F17207-0014  &neutral  & SRFW  &  8.63  &  1.53  & 49.52  & 11.87  & 57.05  & 42.52
  \enddata

  \tablecomments{Col. (2): Gas phase. Col. (3): Outflow model. SRFW =
    Single radius, mass conserving free wind. BSB = Superbubble.
    Col. (4)$-$(9): Logarithms of the mass, mass outflow rate,
    momentum, momentum outflow rate, energy, and energy outflow rate
    contained in the wind (\S\,\ref{sec:models}). The ionized gas
    values depend on electron density as $n_e^{-1}$, and we assume
    $n_e = 10$~cm$^{-3}$. The neutral gas values depends on ionization
    fraction $y$ as $(1-y)^{-1}$, and we assume $y = 0.9$. Note that
    correcting the ionized gas for reddening will increase the values
    in this phase, as well as the total.}

\end{deluxetable}

\begin{deluxetable}{lccccccc}
  \tablecaption{Percent of Mass, Momentum, and Energy in Neutral
    Phase\label{tab:mperats_in}} \tablewidth{0pt}

  \tablehead{\colhead{Galaxy} & \colhead{model} & \colhead{$M$} &
    \colhead{$dM/dt$} & \colhead{$p$} &
    \colhead{$dp/dt$} & \colhead{$E$} & \colhead{$dE/dt$}\\
    \colhead{(1)} & \colhead{(2)} & \colhead{(3)} & \colhead{(4)} &
    \colhead{(5)} & \colhead{(6)} & \colhead{(7)} & \colhead{(8)} }

  \startdata
 F08572+3915:NW  & SRFW  &  89  &  76  &  76  &  56  &  56  &  34 \\
    F10565+2448  & SRFW  &  94  &  97  &  97  &  98  &  92  &  96 \\
        Mrk 231  & SRFW  &  98  &  99  &  99  &  99  &  99  &  99 \\
        Mrk 273  & SRFW  &  59  &  51  &  51  &  39  &  35  &  28 \\
      VV 705:NW  & SRFW  &  88  &  95  &  95  &  98  &  91  &  97
  \enddata

  \tablecomments{Col. 2: Outflow model. SRFW = Single radius, mass
    conserving free wind. BSB = Superbubble.  Col. 3$-$8: Percent of
    the mass, mass outflow rate, momentum, momentum outflow rate,
    energy, and energy outflow rate in the neutral phase of the wind
    (compared to the total in the neutral and ionized phases). Note
    that correcting the ionized gas for reddening will decrease these
    values.}

\end{deluxetable}

\begin{deluxetable}{lccrrrrr}
%  \rotate
  \tablecaption{Power Sources and Feedback \label{tab:mperats_host}}
  \tablewidth{0pt}

  \tablehead{
    \colhead{Galaxy} & \colhead{phase} & \colhead{model} &
    \colhead{$c\times dp/dt/$} & \colhead{$dE/dt/$} &
    \colhead{$dE/dt/$} &
    \colhead{$dM/dt/$} & \colhead{$dM/dt/$} \\
    \colhead{~} & \colhead{~} & \colhead{~} & 
    \colhead{$\lir$} & \colhead{$dE/dt_{SB}$} &
    \colhead{$L_{AGN}$} &
    \colhead{$SFR$} & \colhead{$dM/dt_{SB}$} \\
    \colhead{(1)} & \colhead{(2)} & \colhead{(3)} & \colhead{(4)} &
    \colhead{(5)} & \colhead{(6)} & \colhead{(7)} & \colhead{(8)}
  }

  \startdata
 F08572+3915:NW  &neutral  & SRFW  &-0.21  &-0.45  &$>$-2.88  &-0.39  & 0.58 \\
        \nodata  &ionized  & SRFW  &-0.32  &-0.17  &$>$-2.61  &-0.90  & 0.08 \\
        \nodata  &  total  & SRFW  & 0.04  & 0.01  &$>$-2.42  &-0.28  & 0.70 \\
\hline
    F10565+2448  &neutral  & SRFW  &-0.13  &-1.19  &   -2.53  &-0.36  & 0.62 \\
        \nodata  &ionized  & SRFW  &-2.08  &-2.68  &   -4.02  &-2.01  &-1.03 \\
        \nodata  &  total  & SRFW  &-0.13  &-1.18  &   -2.52  &-0.35  & 0.63 \\
\hline
        Mrk 231  &neutral  & SRFW  & 0.25  & 0.03  &$>$-2.38  & 0.04  & 1.02 \\
        \nodata  &ionized  & SRFW  &-2.28  &-2.48  &$>$-4.90  &-2.31  &-1.33 \\
        \nodata  &  total  & SRFW  & 0.25  & 0.04  &$>$-2.38  & 0.05  & 1.03 \\
\hline
        Mrk 273  &neutral  & SRFW  &-1.13  &-1.68  &$>$-3.64  &-1.21  &-0.23 \\
        \nodata  &ionized  & SRFW  &-0.95  &-1.28  &$>$-3.24  &-1.23  &-0.25 \\
        \nodata  &  total  & SRFW  &-0.73  &-1.14  &$>$-3.09  &-0.92  & 0.06 \\
        \nodata  &ionized  &  BSB  &-0.54  &-0.78  &$>$-2.74  &-0.93  & 0.05 \\
\hline
      VV 705:NW  &neutral  & SRFW  &-0.06  &-0.91  &   -1.99  &-0.49  & 0.49 \\
        \nodata  &ionized  & SRFW  &-1.82  &-2.44  &   -3.51  &-1.80  &-0.82 \\
        \nodata  &  total  & SRFW  &-0.05  &-0.90  &   -1.97  &-0.47  & 0.51 \\
        \nodata  &neutral  &  BSB  &-0.59  &-1.36  &   -2.44  &-0.75  & 0.23 \\
\hline
    F17207-0014  &neutral  & SRFW  &-0.58  &-1.49  &   -2.61  &-1.05  &-0.07
%\hline
  \enddata

  \tablecomments{Col. 2: Gas phase. Col. 3: Outflow model. SRFW =
    Singe radius, mass conserving free wind. BSB = Superbubble.
    Col. 4: Logarithm of the momentum outflow rate divided by the
    momentum input rate if all of the galaxy's infrared luminosity
    accelerates the wind and each photon is intercepted once. Col. 5:
    Logarithm of the energy outflow rate divided by the mechanical
    energy production rate from a continuous starburst
    \citep{leitherer99a}. Col. 6: Logarithm of the energy outflow rate
    divided by the AGN luminosity. Col. 7: Logarithm of mass outflow
    rate normalized to the star formation rate. Col. 8: Logarithm of
    the mass outflow rate divided by the hot gas mass production rate
    expected from a continuous starburst \citep{leitherer99a}.}

\end{deluxetable}

\end{document}